\documentclass{article}

\usepackage{graphicx} 
\usepackage{amsmath}
\usepackage{amssymb}
\usepackage{hyperref}
\usepackage{placeins}
\usepackage{array}
\usepackage{booktabs}
\usepackage{float}
\usepackage{natbib}
\usepackage{comment}
\usepackage{xspace}

\usepackage{xcolor} 
\usepackage[colorinlistoftodos]{todonotes}

\usepackage{graphicx}
\usepackage{subcaption}

\usepackage{float}

\usepackage{array}
\newcolumntype{C}[1]{>{\centering\arraybackslash}p{#1}}

\newcommand{\Uniform}[2]{\text{U}\left(#1,#2\right)}

\newcommand{\Beta}[2]{\text{Beta}\left(#1,#2\right)}
\newcommand{\Poisson}[1]{\text{Poisson}\left(#1\right)}

\newcommand{\Prob}[1]{\text{Pr}\left(#1\right)}
\newcommand{\E}[2][]{\text{E}_{#1}\!\left(#2\right)}

\newcommand{\KLdiv}[2]{\text{KL}\left( #1 \middle\| #2 \right)}
\newcommand{\juneold}{\texttt{JUNE21}\xspace}
\newcommand{\junegen}{$\text{JUNE}(\mathbf{X},\theta)$\xspace}
\newcommand{\junepi}{\(\pi_{\text{JUNE}}\)\xspace}
\newcommand{\junenew}{\texttt{JUNE26}\xspace}

\DeclareMathOperator*{\argmin}{argmin}

\newcommand{\dataT}[0]{\mathcal{D}^{(t)}}

\definecolor{red1}{rgb}{1.0, 0.9, 0.9}   
\definecolor{red5}{rgb}{0.8, 0.0, 0.0}   

\definecolor{green1}{rgb}{0.9, 1.0, 0.9} 
\definecolor{green5}{rgb}{0.0, 0.5, 0.0} 

\definecolor{mygreen}{RGB}{0,150,0}
\definecolor{myred}{RGB}{220,20,60}

\definecolor{purple1}{rgb}{0.949, 0.941, 0.969}
\definecolor{purple5}{rgb}{0.329, 0.153, 0.561}

\definecolor{redlaura}{RGB}{219, 43, 57}
\definecolor{bluelaura}{RGB}{111, 138, 183}

\definecolor{darkorangejoel}{RGB}{209, 95, 2}

\newcommand{\matrixt}[1]{\mathbf{#1}^{(t)}}

\usepackage{main_style}

\usepackage[utf8]{inputenc} 
\usepackage[T1]{fontenc}    
\usepackage{hyperref}       
\usepackage{url}            
\usepackage{booktabs}       
\usepackage{amsfonts}       
\usepackage{nicefrac}       
\usepackage{microtype}      
\usepackage{lipsum}
\usepackage{fancyhdr}       
\usepackage{graphicx}       
\graphicspath{{media/}}     

\usepackage{orcidlink}

\title{\emph{GENIE}: Generative Neural Inference for Epidemics} 

\author{
  Laura M Guzm\'{a}n-Rinc\'{o}n$^{*}$\orcidlink{0000-0003-4592-2896} \\
  Medical Research Council Biostatistics Unit \\
  University of Cambridge 
  \And
  George R E Bradley$^{*}$ \orcidlink{0000-0003-2504-5152}\\
  Medical Research Council Biostatistics Unit \\
  University of Cambridge 
  \And
  Joel Kandiah$^{*}$\orcidlink{0000-0001-8314-0620} \\
  Medical Research Council Biostatistics Unit \\
  University of Cambridge
  \And
  Kyriakos Flouris\orcidlink{0000-0001-7952-1922} \\
  Medical Research Council Biostatistics Unit \\
  University of Cambridge
  \And
  Pietro Li\`{o}\orcidlink{0000-0002-0540-5053} \\
  Department of Computer Science and Technology \\
  University of Cambridge
  \And
  Paul J. Birrell\orcidlink{0000-0001-8131-4893} \\
  UK Health Security Agency, United Kingdom \\
  Medical Research Council Biostatistics Unit \\
  University of Cambridge
  \And
  Alexander E. Zarebski\orcidlink{0000-0003-1824-7653} \\
  Medical Research Council Biostatistics Unit \\
  University of Cambridge
  \And
  Daniela De Angelis \orcidlink{0000-0001-6619-6112}\\
  Medical Research Council Biostatistics Unit \\
  University of Cambridge \\
  \texttt{dd227@cam.ac.uk}
}

\begin{document}
\maketitle

\begin{abstract}

The SARS-CoV-2 pandemic highlighted the ongoing risk infectious diseases pose to society and the value of reliable information on the likely future burden. 
When forecasting an epidemic at fine spatial resolution, traditionally used mechanistic compartmental model struggle to capture highly complex granular transmission dynamics, resulting in inaccurate and overconfident forecasts. However, detailed Agent-Based Models (ABMs), are challenging to calibrate and are too computationally expensive to use in real-time.
Amortized simulation-based inference promises to overcome this difficulty by exploiting the power of machine learning (ML) to perform approximate forecasting at near-real-time using arbitrarily complex models of epidemics.
In this work we introduce Generative Neural Inference for Epidemics (GENIE), a spatio-temporal ML-based framework for high-resolution forecasting of the burden of respiratory pathogens. 
GENIE is designed to reflect two key characteristics of outbreaks: (i) shared biological mechanisms across locations and (ii) location-specific characteristics affecting transmission dynamics. 
This results in the model architecture having two modules: (i) a Local Infection Encoder - which learns to represent disease dynamics shared across all locations and (ii) a Local Profile Encoder - which learns location-specific representations.
Using simulations from a high-resolution spatio-temporal ABM, GENIE is trained to generate samples from an approximate posterior predictive distribution of future epidemic trajectories.
Benchmarked against established statistical and ML models, GENIE demonstrates superior performance across a range of measures including the timing and magnitude of peak hospitalisations.

\end{abstract}

*these authors contributed equally. \\


\section{Introduction}

The SARS-CoV-2 pandemic highlighted the burden infectious diseases impose on healthcare systems. Throughout the crisis, mathematical modelling was an integral component of the scientific evidence used to guide government response \cite{brooks-pollock_modelling_2021}, with a crucial focus on predicting future epidemic health burden. In the UK, several models were central to producing real-time forecasts aimed at quantifying various measures of such burden, such as hospitalisations and mortality \cite{Birrell2025,jewell_bayesian_2023,keeling_predictions_2021,funk_short-term_2020}. Beyond the recent pandemic, extensive research e.g. explored forecasting for diverse pathogens using a broad range of approaches \cite{gomez_holistic_2025}. These include mechanistic, phenomenological/statistical and machine learning models
\cite{keshavamurthy_predicting_2022,banholzer_comparison_2023}, with each approach presenting distinct trade-offs between interpretability, computational demand, and predictive robustness across different phases of an epidemic.

Mechanistic compartmental models offer the advantage of incorporating domain-specific epidemiological knowledge, allowing for high levels of interpretability and the capability to evaluate counterfactual scenarios \cite{keshavamurthy_predicting_2022,banholzer_comparison_2023,ye_integrating_2025}. 
However, they make rigid assumptions on the mechanism of transmission of infection and can be difficult to adapt to rapidly changing circumstances.
Furthermore, incorporating population heterogeneity through additional compartments quickly leads to high-dimensional models that are difficult to calibrate while still lacking the detail to inform effective policymaking
\cite{keshavamurthy_predicting_2022}.

Alternatively, agent-based models (ABMs) represent interactions between individuals, where these granular, micro-level behaviours lead to large-scale emergent properties (see e.g. \cite{JUNE}).
While they offer a more realistic representation of human behaviour and heterogeneity, this comes at the cost of significant computational expense and challenges in calibration for routine forecasting.
Statistical models are a further popular alternative that provide a robust framework for short-term forecasting with limited data \cite{banholzer_comparison_2023}.
While these models are effective when underlying statistical assumptions hold, they often lack epidemiological interpretability and may fail to extrapolate when those assumptions no longer apply to the evolving situation \cite{keshavamurthy_predicting_2022}.

In recent years, there has been a surge in the application of deep learning in epidemic inference and forecasting \cite{ye_integrating_2025}.
Many neural network are trained on real-world surveillance data, and employ spatial neural networks such as graph-based \cite{kapoor2020,gao_stan_2021} or attention-based recurrent architectures \cite{Jung2022,Jiao2025} to capture dependencies between geographical locations.
However, these models are typically constrained by their dependence on historical outbreak data, limiting their ability to generalise to novel scenarios.
Training models on simulation-based training has also been proposed as a way to address that data scarcity \cite{ye_integrating_2025, dudley2025mantis,zarebski2026amortized,OutbreakFlow,wang2019defsi,Cape2026}.
A benefit of training on simulated data is that it allows models to learn to predict quantities that cannot be measured and hence would be difficult to learn from real-world data, e.g. the effective reproduction number~\cite{zarebski2026amortized}.
Another benefit of this approach is that, while simulating data and training the neural network can be expensive, once this is done, evaluating a trained network is typically extremely fast.
This means that if multiple datasets need to be analysed, or the forecasts generated multiple times, the average cost per evaluation decreases, a property that has led to these methods being referred to as ``amortized inference'' \cite{zammit-mangion_neural_2025}.

Many of these neural network models are trained on data simulated with ABMs or compartmental models but remain primarily temporal or do not provide explicit probabilistic forecasts \cite{wang2019defsi,Cape2026,Awais2025-sg,rodriguez2023einns}.
There is a need for an approach that integrates probabilistic simulation-based training with flexible and granular spatial modelling to enable forecasting across pathogens and spatial resolutions.
To address this gap we introduce the Generative Neural Inference for Epidemics (GENIE) architecture.
Our objective is to generate spatio-temporal probabilistic forecasts of infectious disease burden, that are trained on realistic ABM-generated scenarios, and account for spatial heterogeneity in disease spread.
Conceptually, GENIE is designed to reflect two key characteristics of an outbreak: 
\textit{biological mechanisms}, such as how a pathogen typically spreads;
and \textit{location-specific characteristics}, such as local demographics.
By using a graph-based representation of the geography, the model learns how infections in one area propagate into socio-demographically related areas.
The model handles forecasting through a two-stage approach.
First, next-day forecasts of the entire geography are approximated by a family of parametric distributions, with parameters estimated via our neural architecture.
Then, to project further into the future, the model generates long-term probabilistic forecasts by sampling from the approximate predictive posterior autoregressively.

Rather than attempting to learn these dynamics from real-world data, which are scarce and often subject to observational lags and noise, GENIE is formulated as an amortized inference model trained on synthetic epidemics generated with the JUNE agent-based model \cite{JUNE}.
JUNE simulates individual-level interactions, household compositions, and commuting patterns in a customised geography, allowing us to produce highly granular data on the infection process and its burden. 

By extending the JUNE framework, we generated a dataset of simulated epidemics that was used to train GENIE to demonstrate its ability to produce accurate short and long-term forecasts of daily hospitalisations, deaths, unobserved infections and the effective reproduction number.

The remainder of this paper is structured as follows: 
Section~\ref{sec:methods}, presents the methodology, providing information on the JUNE ABM and the generation of our synthetic epidemic dataset; a description of the approximate Bayesian inference methodology used (Section~\ref{sec:methods_bayesian_inference}); the choice of the neural network model (Section~\ref{sec:network_architecture}); the selection of measures and experiments to understand the performance of GENIE (Sections \ref{sec:methods_evaluation}, \ref{sec:methods_ablation} and\ref{sec:methods_peak_performance}).
Section \ref{sec:results} presents the results of our computational experiments.
Finally, in Section \ref{sec:discussion} we comment on the implications of this work for the development of high-resolution spatio-temporal epidemic forecasting.

\section{Methods}\label{sec:methods}

We propose a forecasting tool GENIE, a framework designed around a core neural network which makes one day ahead forecasts. This neural network is trained on data generated by the JUNE simulator.
First, we describe the modified JUNE agent-based model (\texttt{JUNE26}) used to generate a synthetic dataset of realistic outbreak trajectories at a fine grained geographic scale (Section~\ref{sec:data}).
Then we formulate this as an amortized Bayesian inference problem, i.e. approximating the intractable predictive posterior distribution (Section~\ref{sec:methods_bayesian_inference}). 
We then detail the GENIE architecture, specifically covering the Graph Neural Networks to capture spatial and socio-demographic structure (Section~\ref{sec:network_architecture}). 
Finally we outline the experiments, baseline models, ablation studies and evaluation measures used to assess GENIE's predictive capabilities (Sections~\ref{sec:methods_evaluation}-\ref{sec:methods_peak_performance}).

\subsection{The JUNE ABM}\label{sec:data}

JUNE is an agent-based model, developed during the SARS-Cov2 pandemic \cite{JUNE} that describes how an infection spreads within a population across a defined geography.
We refer to the version of this model published in 2021  as \juneold. 
By default, 
\juneold uses the full English geography 
calibrated to multiple sources of data published by the UK Office of National Statistics (ONS). 
The model subdivides the English population down to the Output Area level (100–625 residents) and tracks movement and infection events at the MSOA level (5,000–15,000 residents) \cite{ONS_StatisticalGeographies}. By integrating local socio-demographic and mobility patterns, \juneold models the movement of  individuals and their interactions as random events. 
The framework also allows for the customisation of the characteristics of a pathogen such as infectivity, severity, and infection progression. 
At each timestep,  movements, contact events and progression through health-states such as infection, recovery, hospitalisation and death of each individual in a defined geography are modelled probabilistically, resulting in highly heterogeneous stochastic epidemic trajectories.

\begin{figure}[h!]
    \centering
    \includegraphics[width=0.90\textwidth]{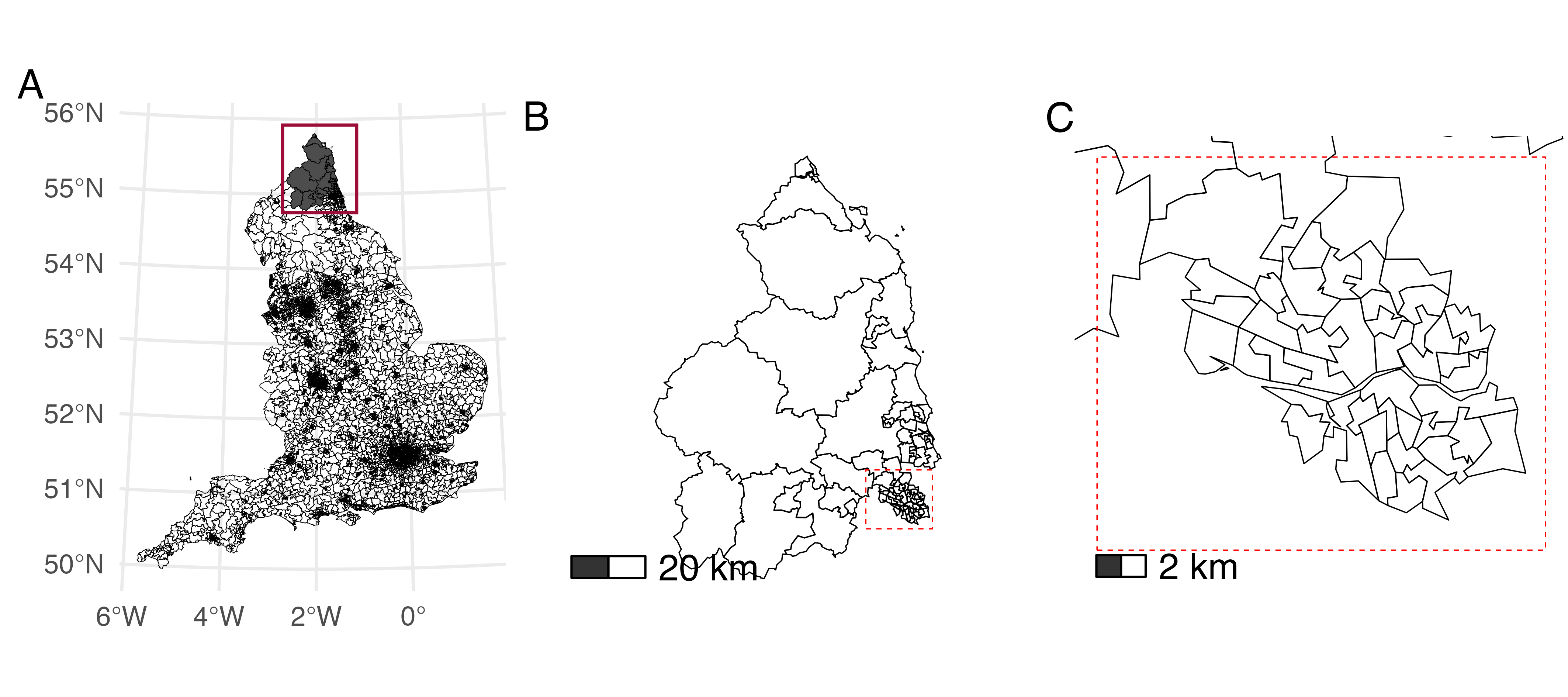}
    
    \caption{\textbf{JUNE simulation geography} (Panel A) Map of MSOAs in England, where the geography used for the JUNE simulations is shaded in grey. 
    (Panel B) Geography used for the JUNE simulations. 
    Each MSOA was defined by the UK census to represent a similar population size; hence, larger areas correspond to lower population density. (Panel C) Detailed view of the most densely populated area in the geography, which includes the cities of Newcastle upon Tyne and Gateshead.}
    \label{fig:geography}
\end{figure}

Here, we configure \juneold to simulate an epidemic spreading in the North East of England (Figure \ref{fig:geography}B), covering 84 MSOAs and approximately 700,000 inhabitants.
This region includes urban, suburban and rural MSOAs, thus providing heterogeneity in population structure, mobility and behavioural patterns and serving as a suitable setting for a proof-of-concept evaluation of GENIE.
A selection of 15 location-specifc features derived from the ONS data (see Table \ref{tab:static_features} are incoporated into GENIE as inputs to provide contextual information for each MSOA, as described in Section \ref{sec:network_architecture}. 

\begin{table}[ht]
\centering
\begin{tabular}{@{}l p{3.5cm} p{8.5cm} p{2.2cm}@{}}
\toprule
\textbf{No.} & \textbf{Feature} & \textbf{Description} & \textbf{Type / Units} \\
\midrule
1 & Number of care-home residents & Total number of individuals residing in registered care homes within the MSOA. & Count \\
\midrule
2--9 & Households (size 1--8) & Number of households of size 1 through 8. & Count \\
\midrule
10 & Index of Multiple Deprivation (IMD) centile & Socio-economic deprivation score \newline (0 = most deprived, 1 = least deprived). & Ordinal \newline (centile) \\
\midrule
11 & Number of students & Total number of individuals registered as students in the MSOA. & Count \\
\midrule
12 & Number of residents & Total population residing in the MSOA. & Count \\
\midrule
13 & Population density & Residents per square kilometre of land area. & Residents/km$^2$ \\
\midrule
14 & Latitude & Latitude of the MSOA centroid (degrees). & Continuous \\
\midrule
15 & Longitude & Longitude of the MSOA centroid (degrees). & Continuous \\
\bottomrule
\end{tabular}
\caption{{List of the 15 static location-specific features for each of the MSOAs in the simulation geography. The values are based on data obtained from the ONS.} 
}
\label{tab:static_features}
\end{table}

To generate a range of outbreak scenarios, we modified the original \texttt{JUNE21} by introducing additional disease and immunity parameters, such as immunity waning rates and location-specific severity, and by adding heterogeneity to existing parameters (see Table \ref{tab:june_parameters} in the Appendix for the list of parameters used).
Incorporating immunity waning is particularly important to produce reinfection dynamics and more realistic long-term outbreak trajectories. 
We refer to the resulting modified version of the model as \texttt{JUNE26}.
The chosen ranges for these parameters ensure that the simulations capture a wider spectrum of realistic respiratory outbreaks that both include and diverge from SARS-CoV-2-like dynamics (see Appendix Section \ref{sec:appendix_data_parameters}).

We generated 1,000 outbreak simulations across the 84 MSOAs of the selected geography, by sampling from the parameter ranges, producing ensemble of trajectories that carry  both parametric uncertainty and intrinsic uncertainty due to stochastic dynamics. The number of simulations was limited by the computational cost.

For each MSOA, three time-varying measures of healthcare burden: daily infections, hospitalisations, and deaths, were generated, along with an epidemiological metric, the effective reproduction number ($R_t$) derived from \texttt{JUNE26}'s output (see Appendix Figure \ref{fig:simulations} for a randomly selected subset of simulations across all burdens aggregated over the entire geography). To contextualise the simulations with known epidemiological characteristics, the basic reproduction number ($R_0$) and the mean generation time ($GT$) were derived for each simulation from the transmission tree extracted from the \texttt{JUNE26} output (see Section \ref{sec:appendix_data_R0} in the Appendix for details on these derivations). Figure \ref{fig:parameter_histograms} shows the histogram of the resulting $R_0$ and $GT$ for all simulations in the dataset, compared to estimates for the wild \cite{billah_reproductive_2020,lau_joint_2021} and the Omicron variants of SARS-CoV-2 \cite{liu_effective_2022,manica_intrinsic_2022}, the respiratory syncytial virus \cite{reis_simulation_2018,cohen_incidence_2024}, and seasonal influenza \cite{biggerstaff_estimates_2014,chan_estimating_2025}.

\begin{figure*}[htb]
    \centering
    \includegraphics[width=0.90\textwidth]{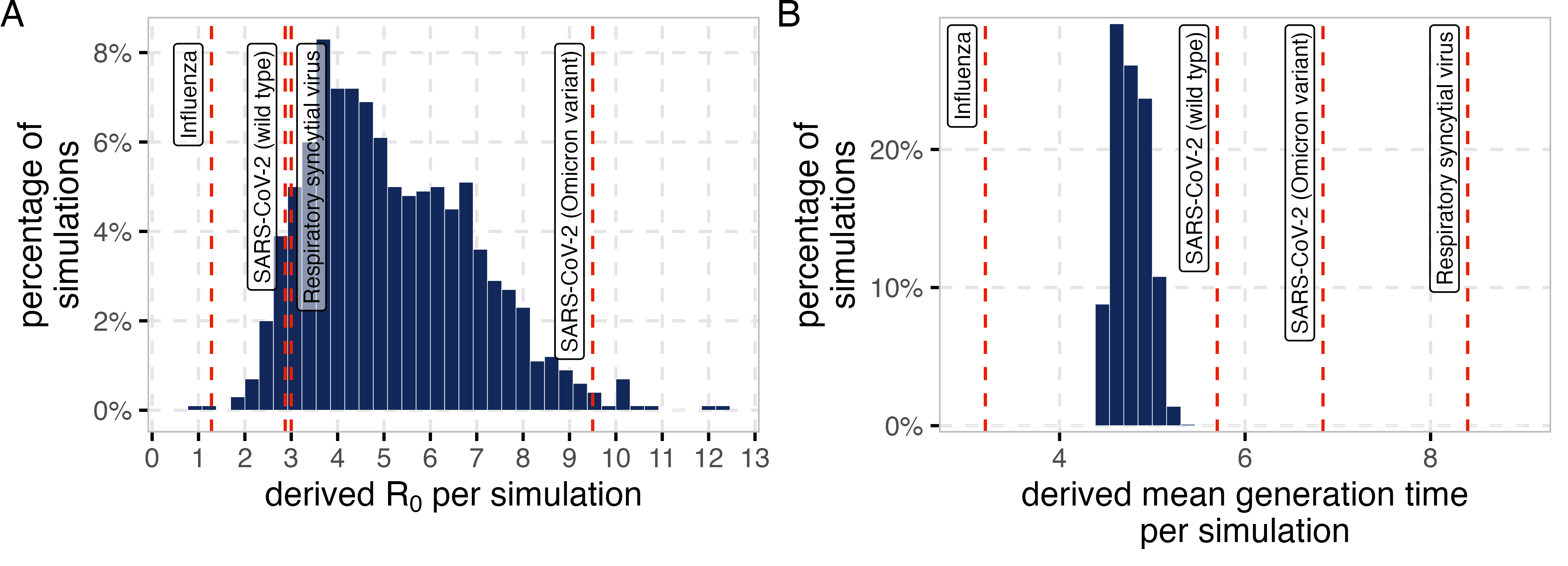}
    \caption{\textbf{Distribution of key epidemiological parameters in \texttt{JUNE26} simulations.} Histograms of (Panel A) the basic reproduction number, and (Panel B) generation times, across 1,000 simulations, compared to published estimates for various diseases.}
    \label{fig:parameter_histograms}
\end{figure*}

The observations from a realisation of JUNE form two matrix-valued time series.
In the first, the matrix of observations on day \(t\) is \( \matrixt{Y} \in \mathbb{R}^{M \times 2}\), where there is a row for each of the \(M=84\) MSOAs in the geography, and one column for each of the two values of interest: the daily number of hospitalisations, and deaths.
The second is a matrix \(\matrixt{W}\in\mathbb{R}^{M\times 2}\) of quantities of fundamental interest but which cannot be observed in practice, the daily number of infections, and the effective reproduction number, all measured at the level of MSOA.

JUNE models workplace and school interactions, which leads to a noticeable ``day of the week'' effects on the incidence of infection.
As a consequence, it it useful to record which day of the week day \(t\) falls on.
To do this, we use \(\matrixt{D}\), a one-hot-encoding \citep{bishop2023deep} of the day of the week of day \(t\), i.e. a vector of length seven where there is a one in the \(d\)th entry if \(t\) is the \(d\)th day of the week; all other entries are zero.

We consider simulations of 180 days, starting from the first seeding event.
The matrix \(\mathbf{X} \in \mathbb{R}^{M \times S}\) contains static socio-demographic features associated with each MSOA in the geography, as described in Section \ref{sec:data} for \junenew. In what follows we will denote the epidemiological parameters governing a \junenew simulation by $\theta$ and the joint distribution over $\{(\matrixt{Y},\matrixt{W},\matrixt{D}): t\leq180\}$, generating outputs of a \junenew simulation, by \junegen.

\subsection{Approximate Bayesian Inference of the Forecasting Distribution Function}\label{sec:methods_bayesian_inference}

We now consider the task of generating probabilistic forecasts of the epidemics by sampling from the posterior predictive distribution \(p(\matrixt{Y},\matrixt{W}|\dataT{})\) of \(\matrixt{Y}\) and \(\matrixt{W}\) given the data, \(\dataT{}:=(\{\mathbf{Y}^{(\tau)}:\tau<t\},\mathbf{X},\matrixt{D})\) that have been observed by day \(t\) 
under the stochastic process defined by JUNE.
Sampling from this \emph{conditional} distribution directly is intractable.
Instead we will approximate it
with a distribution \(q(\cdot,\kappa)\) parameterised by \(\kappa\), chosen to minimise the forwards Kullback-Leibler divergence (KL-divergence) between the true predictive distribution and its approximation. 
For a specific set of observations \(\dataT{}\), the optimal posterior predictive approximation is obtained by
\[
\kappa^* = \argmin_{\kappa}\KLdiv{p(\matrixt{Y},\matrixt{W}\mid \dataT{})}{q(\matrixt{Y},\matrixt{W};\kappa)}.
\] 
Rather than solving this optimisation problem for a particular instance of the data \(\dataT{}\), we can generalise this to choosing an optimal function \(\kappa^{*}(\cdot)\) which takes any observations \(\dataT{}\) and returns the optimal parameters \(\kappa^{*}\) for that particular dataset. 
We can cast this as the following optimisation problem for the optimal function \(\kappa^*(\cdot)\), among some family of functions \(\mathcal{K}\):
\begin{align}\label{eq:min_loss}
  \kappa^*(\cdot) &= \argmin_{\kappa(\cdot)\in\mathcal{K}} \E[\dataT{}, \matrixt{Y},\matrixt{W}]{\KLdiv{p(\matrixt{Y},\matrixt{W}\mid \dataT{})}{q(\matrixt{Y},\matrixt{W};\kappa(\dataT{}))}}\notag \\
  &= \argmin_{\kappa(\cdot)\in\mathcal{K}} \left[-\E[\dataT{},\matrixt{Y},\matrixt{W}]{\ln\left(q(\matrixt{Y},\matrixt{W};\kappa(\dataT{}))\right)}\right]
\end{align}
\noindent
where the expectation is over the joint distribution of \(\dataT{}\), \(\matrixt{Y}\) and \(\matrixt{W}\)~\citep{zammit-mangion_neural_2025}.

We cannot evaluate the expectation in Equation~\eqref{eq:min_loss} directly, but we can sample from the \emph{joint} distribution of \(\dataT{},\matrixt{Y},\matrixt{W}\), which allows us to make a Monte-Carlo approximation of the integral.
To sample from this joint distribution (see Section \ref{sec:data}), we first draw epidemiological parameters \(\theta\) from a prior distribution, \(\pi_{\text{JUNE}}\), and then sample the remaining variables from JUNE, \(\{\mathbf{Y}^{(\tau)},\mathbf{W}^{(\tau)}:\tau\leq t\},\matrixt{D}\sim\text{JUNE}(\mathbf{X},\theta)\).
The population characteristics \(\mathbf{X}\) are fixed across the simulations here as we are primarily interested in a single geography; alternatively, this might be viewed as putting a degenerate prior on \(\mathbf{X}\).

Equation~\eqref{eq:min_loss} describes an optimisation problem, over a family of functions \(\mathcal{K}\), for approximating posterior predictive distributions.
The generative process \junegen and the prior distribution of its parameters, \junepi are described above, we now turn our attention towards the choice the approximating distributions, \(q\) (and some standardisation of our input data \(\dataT{}\)), and the architectural choices behind the neural networks in \(\mathcal{K}\). 

The family of functions \(\mathcal{K}\) is a set of functions arising from a fixed neural network architecture (described in Section~\ref{sec:network_architecture}) where the weights and biases of the neural network index the different functions \(\kappa(\cdot)\).
The computation then is optimising the parameters of the neural network with respect to the objective in Equation~\eqref{eq:min_loss}.

Rather than working with a set of matrices, \(\{\mathbf{Y}^{(\tau)}:\tau<t\}\), we instead consider, for each $t$ a \emph{context} matrix \(\matrixt{C}\in\mathbb{R}^{M\times 2N}\).
Each row of \(\matrixt{C}\) contains all the observations for a given MSOA in preceding \(N\) days, i.e. the set \(\{\mathbf{Y}^{(\tau)}:t-N\leq\tau<t\}\). 
Concatenating the \(\matrixt{C}\)  matrices into a tensor in \(\mathbb{R}^{M\times 2\times N}\) 
and then flattening them over the time and observation-type dimensions to get a matrix.
For \(t\leq0\) we set \(\matrixt{Y}=\mathbf{0}\), the zero matrix, as prior to the start of the epidemic, there are no hospitalisations or deaths observed (due to the pathogen). 
The value of \(N\) controls how much memory the context matrix contains. 

In constructing our approximations of the predictive distribution \(q\), we make the simplifying assumption of conditional independence among the variables and MSOAs.
We defined \(q\) as follows:
\begin{equation}\label{eq:distribution_q}
\begin{aligned}
    q(\matrixt{Y},\matrixt{W};\kappa(\dataT{})) := \prod_{m\in[M]}\Bigg [  &q_{\text{hosp}}((\matrixt{Y})_{m,1}; \kappa_{\text{hosp}}(\tilde{\mathcal{D}}^{(t)}))  q_{\text{death}}((\matrixt{Y})_{m,2}; \kappa_{\text{death}}(\tilde{\mathcal{D}}^{(t)}))
    \times\\
    &\quad q_{\text{inc}}((\matrixt{W})_{m,1}; \kappa_{\text{inc}}(\tilde{\mathcal{D}}^{(t)})) q_{R}((\matrixt{W})_{m,2}; \kappa_{R}(\tilde{\mathcal{D}}^{(t)})) \Bigg ].
\end{aligned}
\end{equation}
\noindent
where \((\matrixt{Y})_{m,b}\) and \((\matrixt{W})_{m,b}\) denote the \(m,b\)th element of the respective matrix.
The tuple \(\tilde{\mathcal{D}}^{(t)}=(\matrixt{C},\mathbf{X},\matrixt{D})\) is the truncation of the data to the context length of \(N\). 
The \(q_{\text{hosp}}\), \(q_{\text{death}}\), \(q_{\text{inc}}\) are chosen to be negative binomial distributions, because they model the count data of number of hospitalisations, deaths and incidence of infection.
The \(q_{R}\) is a log-normal distribution to model the reproduction number.
The functions \(\kappa_{\cdot}(\cdot)\) are defined by the neural networks described below which map the input data to parameters of the \(q_{\cdot}\)
Since the negative binomial and log-normal distributions can be parameterised in terms of a mean and dispersion parameter, the functions \(\kappa_{\cdot,\cdot}(\cdot)\) return values in \(\mathbf{R}^{2}\).

\subsubsection{Sampling from posterior predictive distribution}\label{sec:forecast-rollout}

So far we have considered approximating the posterior predictive distribution of \(\matrixt{Y},\matrixt{W}\) given \(\dataT{}\), the observations up until day \(t-1\).
To extend this to forecasts over a horizon of \(H\) days, i.e. \(\mathbf{Y}^{(t:t+H)},\mathbf{W}^{(t:t+H)}\), consider the following:
\begin{align}
    p(\mathbf{Y}^{(t:t+H)},\mathbf{W}^{(t:t+H)}\mid\dataT{})&=\prod_{i=0}^{H} p(\mathbf{Y}^{(t+i)},\mathbf{W}^{(t+i)}\mid \mathbf{Y}^{(t:t+i-1)},\mathbf{W}^{(t:t+i-1)},\dataT{}) \notag \\
    &=\prod_{i=0}^{H} p(\mathbf{Y}^{(t+i)},\mathbf{W}^{(t+i)}\mid \mathcal{D}^{(t+i)},\mathbf{W}^{(t:t+i-1)}) \label{eq:fx-defn} \\
    &\approx\prod_{i=0}^{H} p(\mathbf{Y}^{(t+i)},\mathbf{W}^{(t+i)}\mid \mathcal{D}^{(t+i)}) \label{eq:fx-approx} \\
    &\approx\prod_{i=0}^{H} q(\mathbf{Y}^{(t+i)},\mathbf{W}^{(t+i)}; \kappa(\tilde{\mathcal{D}}^{(t+i)})) \label{eq:fx-qtrunc}
\end{align}
where~\eqref{eq:fx-defn} follows from the definition of \(\dataT{}\) and~\eqref{eq:fx-approx} is a simplifying approximation equivalent to a conditional independence assumption in which \(\mathcal{D}^{(t+i)}\) contains all the information in \(\{\mathbf{W}^{\tau}: \tau<t+i\}\).
In~\eqref{eq:fx-qtrunc}, we include the approximation by the distribution \(q\) and make explicit that we are truncating the history (by using \(\tilde{\mathcal{D}}^{(t+i)}\) instead of \(\mathcal{D}^{(t+i)}\)) to only include observations from the previous \(N\) days.

The factorisation of \(p(\mathbf{Y}^{(t:t+H)},\mathbf{W}^{(t:t+H)}\mid\dataT{})\) above suggests a way to draw samples from an approximation of the posterior predictive distribution: autoregressively sample from our approximation \(q\) in Equations~\eqref{eq:distribution_q} for each day in the forecast horizon and combine the samples.
In practice, we generate probabilistic forecasts of \(\mathbf{Y}^{(t:t+H)},\mathbf{W}^{(t:t+H)}\mid\dataT{}\) by drawing a sample \(\{(\mathbf{Y}^{(t:t+H)}_{j},\mathbf{W}^{(t:t+H)}_{j})\}_{j\in[J]}\) using the following steps:
\begin{enumerate}
    \item draw \(J\) samples, \(\{(\mathbf{Y}^{(t)}_{j},\mathbf{W}^{(t)}_{j})\}_{j\in[J]}\), from the distribution \(q(\mathbf{Y}^{(t)},\mathbf{W}^{(t)};\kappa(\dataT{}))\);
    \item for each of the \(\mathbf{Y}^{(t)}_{j}\) extend the context to \(\mathcal{D}^{(t+1)}_{j}\) to form \(\{\mathcal{D}^{(t+1)}_{j}\}_{j\in[J]}\);
    \item for each of the \(\mathcal{D}^{(t+1)}_{j}\) draw a single sample from the distribution \(q(\mathbf{Y}^{(t+1)},\mathbf{W}^{(t+1)};\kappa(\mathcal{D}^{(t+1)}_{j}))\) to form \(\{(\mathbf{Y}^{(t+1)}_{j},\mathbf{W}^{(t+1)}_{j})\}_{j\in[J]}\);
    \item repeat steps 2 and 3, incrementing time appropriately, until the forecast horizon is reached.
\end{enumerate}
The result of these steps is \(J\) draws from the approximate posterior predictive distribution which can be combined to form the predictive sample \(\{(\mathbf{Y}^{(t:t+H)}_{j},\mathbf{W}^{(t:t+H)}_{j})\}_{j\in[J]}\).
These draws can then be used to form Monte-Carlo estimates of the functionals described below.
All the forecasts presented here were generated using a context of \(N=30\) days. 
Each forecast consists of \(J=150\) sampled trajectories.

\subsection{Graph neural networks for spatio-temporal forecasting}\label{sec:network_architecture}

Equation~\eqref{eq:min_loss} presents the task of sampling from a posterior predictive distribution as an optimisation problem over a family of functions \(\mathcal{K}\) with an unspecified distribution \(q\). Equation~\eqref{eq:distribution_q} then presents a specific instance of \(q\) related to the specific quantities being predicted. IN this section we focus on defininga suitable family of neural networks \(\mathcal{K}\) which act to provide our estimating functions \(\kappa\in\mathcal{K}\), appropriate to the specific outbreak prediction problem.

\begin{figure}[h]
\centering
\includegraphics[width=1.0\textwidth]{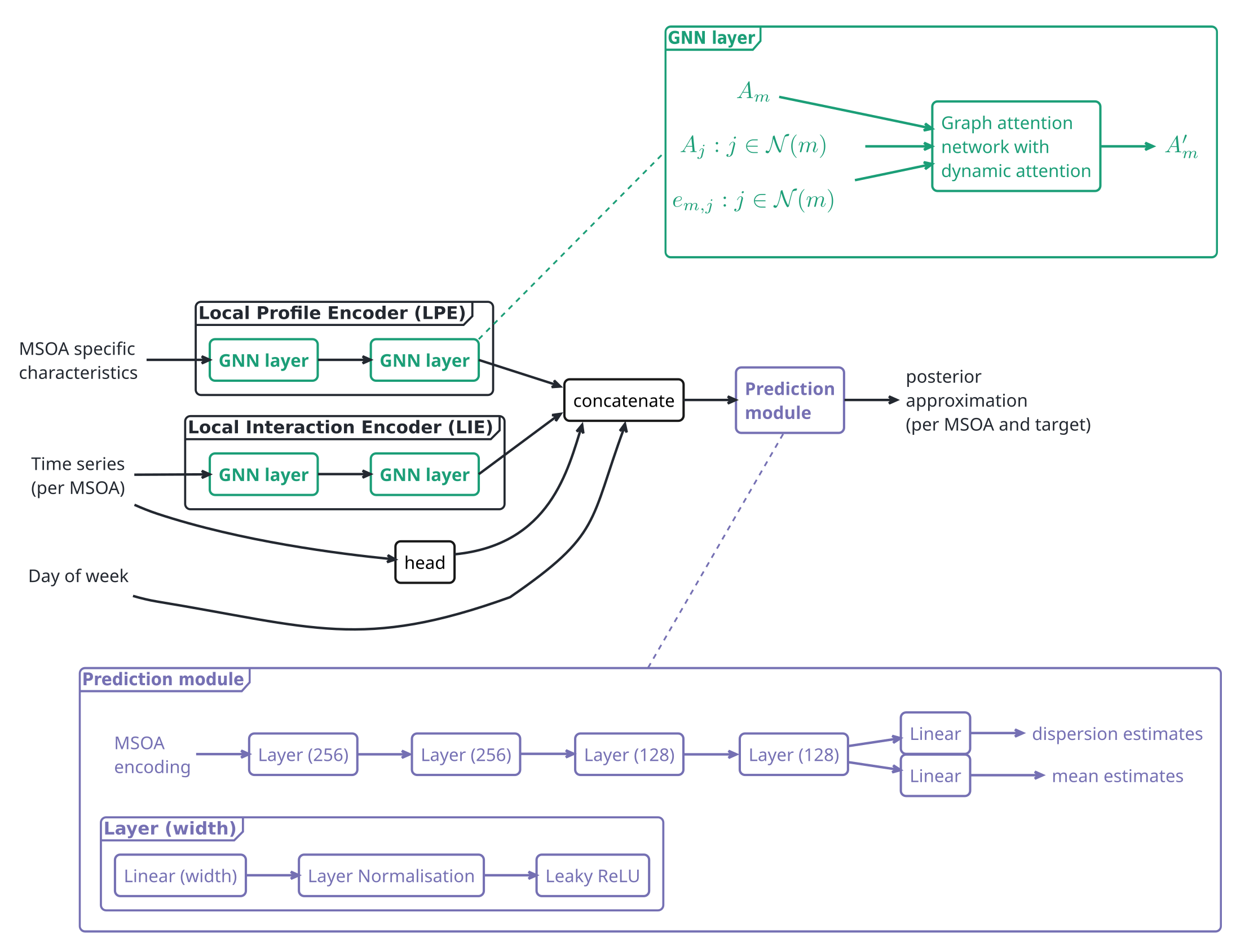}
\caption{\label{fig:genie-architecture}The Local Interaction Encoder and the Local Profile Encoder use graph neural networks (GNNs) to produce a representation of the spatially indexed data that can be passed into the prediction module to approximate the posterior predictive distribution. The input to this neural network are graphs with annotated nodes and edges; the node annotations are either the time series of observations in the rows of the context matrix \(\matrixt{C}\) or the rows of the MSOA characteristics matrix, \(\mathbf{X}\). The output of the neural network is the parameters of the posterior approximation givnen by Equation~\eqref{eq:distribution_q}.
Each GNN layer (see inset) uses dynamic attention (i.e. GATv2) to form a representation \(A_{m}'\) given the node's initial representation \(A_m\) (which starts as a row of either \(\matrixt{C}\) or \(\mathbf{X}\)) and the representation of its neighbours, \(\mathcal{N}(m)\), and the edge weights, \(e_{mj}\).
The prediction module is shared across each of the MSOAs when predicting the parameters of the posterior approximation. The details of the prediction module are shown in the inset.
}
\end{figure}

GENIE is designed to leverage the knowledge that aspects of the dynamics of a given outbreak are invariant across geographical location. This means that the \textit{biological mechanism} is shared, while \textit{location-specific characteristics}, such as local demographics and contact patterns are additional influences on transmission dynamics. Consequently, GENIE is architected with three principal components (see Figure \ref{fig:genie-architecture}): (i) a \emph{Local Profile Encoder} (LPE), which processes the socio-demographic features of an MSOA, ii) a \emph{Local Interaction Encoder} (LIE), which processes the time series of observations from an MSOA, and (ii) a \emph{Prediction Module} (PM), which uses the encodings produced by the LPE and LIE to approximate the posterior predictive distribution as given by Equation~\eqref{eq:distribution_q}.
The LPE and LIE use graph neural networks (GNNs) to form an encoding of the data for each MSOA in a way that accounts for the data from neighbouring MSOAs.

The neighbour relationship between MSOAs comes from representing the MSOAs as the nodes of a graph \(\mathcal{G}=(\mathcal{V},\mathcal{E})\) with labeled nodes and edges (as illustrated in Figure~\ref{fig:Graph_Consruction}).
The edge set \(\mathcal{E}=\mathcal{E}_{\text{geo}}\cup\mathcal{E}_{\text{socio}}\), where the subsets of edges capture geographical proximity, or proximity in sociodemographic features.
The edges in \(\mathcal{E}_{\text{geo}}\) come from linking each MSOA to its nine geographically closest neighbours.
The geographical distance is the Haversine distance, the Earth's great-circle distance, between the centroids of the MSOAs.
The edges in \(\mathcal{E}_{\text{socio}}\) come from linking each MSOA to its nine closest neighbours measured as the Euclidean distance between their feature vectors in \(\mathbf{X}\).

The edges in \(\mathcal{E}\) are undirected. 
For an edge between MSOAs \(i\) and \(j\) the weight on that edge is \((d_{ij}-d_{\min})/(d_{\max}-d_{\min})\) where \(d_{ij}\) is the Haversine distance between the centroids of MSOAs \(i\) and \(j\) with \(d_{\max}\) and \(d_{\min}\) being the maximum and minimum of these distances between any pair of nodes in the geography.

The nodes of \(\mathcal{G}\) may be annotated with the rows of either \(\matrixt{C}\) or \(\mathbf{X}\).
When computing an encoding with the LIE, we annotate the nodes of \(\mathcal{G}\) with the rows of \(\matrixt{C}\).
The GNN layers of the LIE then compute an encoding (updated node annotations) defining the matrix \(\matrixt{H}_{\text{LIE}}\in\mathbb{R}^{M\times8}\), where the \(m\)th row contains the encoding of the time series data for the \(m\)th MSOA.
Similarly, the GNN layers in the LPE produce a matrix \(\mathbf{H}_{\text{LPE}}\in\mathbb{R}^{M\times8}\) encoding the sociodemographic features of each MSOA in relation to their neighbours.

For each MSOA, the prediction module (PM) returns a point estimate of the mean and dispersion parameter for each of distributions on the right hand side of Equation~\ref{eq:distribution_q}. 
When approximating the posterior predictive distribution for MSOA \(m\) on day \(t\), the PM takes as input the vector \([(\mathbf{H}_{\text{LPE}})_{m} | (\matrixt{H}_{\text{LIE}})_{m} | \matrixt{D} | (\mathbf{Y}^{(t-1)})_{m}]\in\mathbb{R}^{8+8+7+2}\), where \((\mathbf{M})_{m}\) indicates the \(m\)th row of matrix \(\mathbf{M}\), and returns a matrix \(\matrixt{\Phi}_{m}\in\mathbb{R}^{4\times2}\).
The number of rows in the matrix \(\matrixt{\Phi}_{m}\) come from there there being four variables across the matrices \(\matrixt{Y}\) and \(\matrixt{W}\), the number of hospitalisations, deaths, infections and the reproduction number. 
The number of columns in \(\matrixt{\Phi}_{m}\) come from each of the approximating distributions \(q_{\text{hosp}}\), \(q_{\text{death}}\), \(q_{\text{inc}}\) and \(q_{R}\) each taking two parameters each (a mean and dispersion).

As shown in Figure~\ref{fig:genie-architecture}, the LIE and LPE are implemented as two-layer GNNs \citep{velickovic2018graph}, specifically the GATv2 active attention architecture of \cite{brody2021how}, with 128 hidden units in the first layer and 8 in the second layer, each with two attention heads per layer.

The PM is implemented as a multi-layer perceptron consisting of four hidden layers with dimensions \([256, 256, 128, 128]\), each followed by layer normalisation and a LeakyReLU activation. The output of the final hidden layer branches into two independent linear heads to estimate the mean and dispersion separately across all targets, i.e. each head produces one of the columns of \(\matrixt{\Phi}_{m}\).

\subsubsection{Training}\label{sec:methods_training}

The synthetic dataset comprised 1,000 simulated outbreaks across 84 MSOAs per burden type.
This dataset was partitioned into training (70\%), validation (15\%), and testing (15\%) subsets by randomly selecting 300 simulations without replacement (150 each for validation and testing), retaining 700 simulations for training.

The network is trained to minimise the loss function defined by Equation~\ref{eq:min_loss} using stochastic mini-batching with AdamW~\cite{pytorchAdamWx2014}.
A OneCycleLR scheduler~\cite{onecyclelr} updates the learning rate once per batch across 3,000 steps (300 iterations of 10 batches). 
Under this scheduler the learning rate warms up from a fraction of the maximum learning rate, before cosine-annealing over the remaining steps.
A dropout rate of 0.2 is applied between each internal layer of the prediction module during training to attempt to reduce overfitting.
The training and optimisation parameters are summarised in Table~\ref{tab:training_hyperparameters}.

Mini-batches are formed by randomly sampling days from the simulated dataset with replacement.
In each iteration of training, $10\times 64$ days are sampled (along with the preceding days needed as context) and divided into 10 mini-batches of size 64.
For each sampled day the full set of MSOAs is included.

After every iteration, the validation loss is computed on a randomly sampled batch from the validation set.
The final model parameters are those with the lowest validation loss (i.e., early stopping with infinite patience).

\begin{table}[h!]
\centering
\begin{tabular}{p{4.5cm} p{7.0cm}@{}}
\toprule
\textbf{Parameter} & \textbf{Value} \\
\toprule
Optimiser & AdamW \\
Weight decay & $1 \times 10^{-4}$ \\
Learning rate scheduler & OneCycleLR \\
Maximum learning rate & $1 \times 10^{-3}$ \\
Scheduler steps & 3,000 (10 batches/iteration $\times$ 300 iterations) \\
Dropout rate & 0.2 \\
Epochs & 300 \\
Batch size & 64 sampled timesteps \\
Samples per iteration & $10 \times 64$ timesteps \\
Dataset split (Train / Validation / Test) & 70\% / 15\% / 15\% (700 / 150 / 150 simulations) \\
\bottomrule
\end{tabular}
\caption{Training and optimisation parameters for the GENIE model.}
\label{tab:training_hyperparameters}
\end{table}

\subsection{Model Evaluation}\label{sec:methods_evaluation}
We evaluate the forecasting performance of GENIE across the full epidemic.
Assessing performance across different settings provides insight into the model's ability to perform using sparse, early data as well as more stable transmission dynamics.
The evaluation is structured to answer four key questions:

\begin{samepage}
\begin{description}
    \item[(i)] How does GENIE perform across the entire disease timeline?
    \item[(ii)] How accurately does GENIE forecast during periods of disease emergence?
  \item[(iii)] Is GENIE able to forecast the timing and magnitude in the peak number of hospitalisations?
    \item[(iv)] What is the importance of location specific information to forecasting performance? What is the relevance of using a GNN in the model structure?
\end{description}
\end{samepage}

To evaluate GENIE's forecasting performance for questions (i) and (ii), we select two competitive baseline models representing distinct modelling paradigms: Mantis, which is a neural network-based probabilistic time series forecasting model trained on simulated data \cite{dudley2025mantis}; and \texttt{hhh4}, a statistical spatio-temporal forecasting model \cite{meyer2017spatio}.
Ideally, a spatial neural network-based model would have been included, but to our knowledge, no such model exists that has not been trained on real-world data (and hence might be expected to generalise to our context).
    
Mantis is a foundation model for time-series forecasting of epidemics, trained on a large number of mechanistic simulations of epidemics; it outperformed all models in the US Centers for Disease Control and Prevention's COVID-19 Forecast Hub~\cite{dudley2025mantis}.
Mantis is a suitable baseline for comparison, however there are two aspects of it which limit the extent to which it can be compared to GENIE.
The first is that it is a temporal model only, so the forecasts for each MSOA are made independently.
We evaluate the forecasting performance of GENIE against Mantis at the individual MSOA and burden level using the Continuous Ranked Probability Score (CRPS), a scoring rule for univariate probabilistic forecasting that jointly assesses calibration and sharpness \cite{gneiting_strictly_2007}.
The second limitation is that Mantis generates quantiles of the predictive distribution, rather than individual samples.
The lack of samples of individual trajectories precludes the peak analysis described below.

To evaluate univariate MSOA-level predictive distributions, we employ the \emph{continuous ranked probability score} (CRPS) \cite{gneiting_strictly_2007}.
For an true value $y \in \mathbb{R}$, the CRPS of a predictive CDF $F$ can be expressed in its kernel representation as
\begin{equation}
\label{eq:crps_kernel}
\text{CRPS}(F,y) = \mathbb{E}_{X \sim F}[|X - y|] - \frac{1}{2}\mathbb{E}_{X, X' \sim F}[|X - X'|],
\end{equation}
where $X$ and $X'$ are independent draws from $F$.
In practice, we approximate both expectations in Equation~\eqref{eq:crps_kernel} with a Monte Carlo estimate (as detailed in Appendix Section~\ref{sec:appendix_model_comparison}).

\texttt{hhh4} is a spatio-temporal statistical model for infectious disease surveillance which decomposes disease burden counts into three components: endemic, autoregressive, and spatio-temporal .
It is primarily used to model disease dynamics and transmission between regions, while accounting for seasonality, covariates, and spatial connectivity.
Since \texttt{hhh4} is a spatio-temporal model, we assess performance at two levels: at the level of individual MSOAs for each burden metric using the CRPS; and across all MSOAs jointly using the Energy Score \cite{gneiting_strictly_2007}.

The \emph{energy score} (ES) is a strictly proper mutli-variate generalisation of the CRPS \cite{gneiting_strictly_2007}, which enables it to measure the predictive performance of the joint predictive distribution over all the MSOAs.
For an true vector $\mathbf{y}\in\mathbb{R}^M$ the ES of a multivariate predictive CDF $\mathbf{F}$ is defined by the following:
\begin{equation}
\text{ES}(\mathbf{F}, \mathbf{y}) = \mathbb{E}_{\mathbf{X} \sim \mathbf{F}}[\|\mathbf{X} - \mathbf{y}\|_2] - \frac{1}{2}\mathbb{E}_{\mathbf{X}, \mathbf{X}' \sim \mathbf{F}}[\|\mathbf{X} - \mathbf{X}'\|_2],
\end{equation}
where $\mathbf{X}$ and $\mathbf{X}'$ are independent draws from $\mathbf{F}$.
Analogous to the univariate CRPS, the ES evaluates the joint accuracy and multivariate calibration against the spatial dispersion and cross-region uncertainty of the ensemble.

The comparison between GENIE, Mantis and \texttt{hhh4} is applied only to observed measures of burden: daily hospitalisations and deaths forecasts.
Details of Mantis and \texttt{hhh4} configurations are in the Appendix Section \ref{sec:appendix_model_comparison}, as well as the definition of the CRPS and energy score.

\subsection{Ablation Study}\label{sec:methods_ablation}

To evaluate the contribution of the LPE encoder and spatial structure to GENIE's forecasting performance, and address question (iii) above, we performed a couple of ablation studies:
\begin{description}
    \item[LPE ablation] We compare the performance of GENIE against a version without the LPE, to test the importance of the MSOA characteristics for forecast performance.
    \item[Graph ablation] We compare the performance of GENIE against a version in which the graph structure is absent, i.e. \(\mathcal{E}=\emptyset\). To do this, we replace the GNN layers and PM with a fully-connected MLP consisting of three hidden layers with dimensions [128, 64, 32], each followed by Layer Normalisation, LeakyReLU, activation, and dropout ($p = 0.2$ during training), and a final linear output layer. 
\end{description}
In each case, both variants are evaluated on the same training/testing/validation splits as the full model, and with all other components unchanged.

\subsection{Predictive Peak Performance}\label{sec:methods_peak_performance}

Scoring rules such as the CRPS provide a generic evaluation of a whole predictive distribution . 
In reality, we may be primarily interested in specific aspects of a forecast trajectory, such as the timing and magnitude of the peaks and troughs in specific burden, e.g. the number of hospitalisations.
To investigate the ability of models to predict peaks across a large set of realisations, we need a formal definition for peak (and trough) occurrence and magnitude. It is key to evaluate this, as previous work has shown that even if a model minimises daily forecast error it may not be the best model for forecasting peaks evaluated on timing and height \cite{Morbey2023}.

At the level of MSOAs, the daily number of hospitalisations is small, so the turning points 
of the time series are obscured.
We aggregate the hospitalisation time series geographically to the level of Lower Tier Local Authority (LTLA), which divides the chosen geography into three LTLAs (115,000–316,000 residents), and temporally to ISO week \citep{ISO.8601-1}.
An ISO week begins on a Monday and ends on a Sunday.
Such a definition is common in epidemic modelling \citep{Birrell2025, Mook2020}. 

For ISO week \(w\), let \(m_{l}^{(w)}\) be the median (across the days of the ISO week \(w\)) of the daily total number of hospitalisations across the MSOAs in LTLA \(l\).
Let \(s_{l}^{(w)}\) denote the sum of the daily total number of hospitalisations across the MSOAs in LTLA \(l\).

Let \(\sigma_{l}^{(w)}\) track the sign of the change in the weekly (median) number of hospitalisations:
\begin{equation*}
    \sigma_{l}^{(w)}=
    \begin{cases}
        \text{sign}\left( m_{l}^{(w)} - m_{l}^{(w-1)} \right), & m_{l}^{(w)}\neq m_{l}^{(w-1)},\\
        \sigma_{l}^{(w-1)}, & m_{l}^{(w)}=m_{l}^{(w-1)},
    \end{cases}
\end{equation*}
so constant values take the direction associated with the previous step and a turning point is placed at the last week of a plateau.

We define a \emph{turning point} for an LTLA \(l\) as the ISO week \(w\) where the \(\sigma_{l}^{(w)}\) changes. 
The turning point is a \emph{peak} if \(\sigma_{l}^{(w+1)}=-1\) and \(\sigma_{l}^{(w)}=+1\).
The turning point is a \emph{trough} if \(\sigma_{l}^{(w+1)}=+1\) and \(\sigma_{l}^{(w)}=-1\).
To reduce the sensitivity to stochasticity in the time series, we can extend this to have the requirement that the preceding \(\rho\) weeks had the same sign.
I.e. for a peak, we would require \(\sigma_{l}^{(w+1)}=-1\) and \(\sigma_{l}^{(w-p)}=+1\) for \(p=0,1,\ldots,\rho-1\), and similarly for a trough.
For our main results we use \(\rho=1\), however, we also consider \(\rho=2\) in a sensitivity analysis in Appendix Section~\ref{sec:appendix_peaks_sensitivity}.

The definition of turning points above depends on values prior to the forecast horizon.
To each aggregate forecast of \(m_{l}^{(w)}\), we prepend the observed number of hospitalisations so that we can produce estimates of all the peaks and troughs from the start of the time series up until the end of the forecast horizon.
Tracking this history explicitly, allows predictions of a given peak/trough to be identified as the prediction of the \(p\)th peak/trough in the time series.

Using the sampling strategy in \S\ref{sec:forecast-rollout}, we draw \(J\) epidemic trajectories from the (approximate) posterior predictive distribution (as described in Section~\ref{sec:forecast-rollout}).
For each of these draws, we can apply the peak definition given above to get samples of potential peaks within the forecast horizon of \(H\) days.
Let \(\mathcal{J}_{p}^{+}\) denote the subset of the epidemic trajectories in which there is a \(p\)th peak during the forecast horizon.
In evaluating the peak predictions we focus on two key measures:
\begin{description}
    \item [\(\Prob{\text{correct week}}\)] To measure the prediction of peak timing, we use the posterior probability associated with the true peak week in the prediction. This is estimated by the proportion of all of the $J$ sampled trajectories in which the $p$-th peak falls in the correct ISO week of the peak.
    \item [Mean relative error (\(\text{MRE}\)) in peak magnitude] To measure the prediction of peak magnitude, we use the MRE conditioning on there being a peak.
    This is estimated by the mean relative (absolute) error in the predicted peak hospitalisations ($s_l^{(w)}$), using the subset \(\mathcal{J}_{p}^{+}\) of trajectories in which there is a \(p\)th peak.
\end{description}

Further formal details for these measures are provided in the Appendix Section~\ref{sec:methods_peak_definition}, along with some additional metrics including to test the sensitivity to these choices: the probability of detecting a peak, the mean absolute error (MAE) in predicted peak magnitude, and the bias in the predicted peak week.
To measure the predicted peak existence we use the posterior probability, i.e. \(\Prob{\text{peak in horizon}}\), associated with there being a peak in the forecast horizon.
This is estimated by the overall proportion of the $J$ sampled trajectories that contain a $p$-th peak anywhere within the forecast horizon.
To gain an absolute measure of peak magnitude, we use the MAE conditioning on there being a peak.
This is estimated by the mean absolute error in the predicted peak hospitalisations ($s_l^{(w)}$), using the subset \(\mathcal{J}_{p}^{+}\) of trajectories in which there is a \(p\)th peak.
We measure the bias in the peak timing under the posterior distribution conditioning on there being a peak.
This is estimated by mean difference between the predicted and true peak weeks, across the subset $\mathcal{J}_p^+$.

\section{Results}\label{sec:results}

\subsection{Example forecasts}

\begin{figure*}[h]
    \centering
    \includegraphics[width=0.8\textwidth]{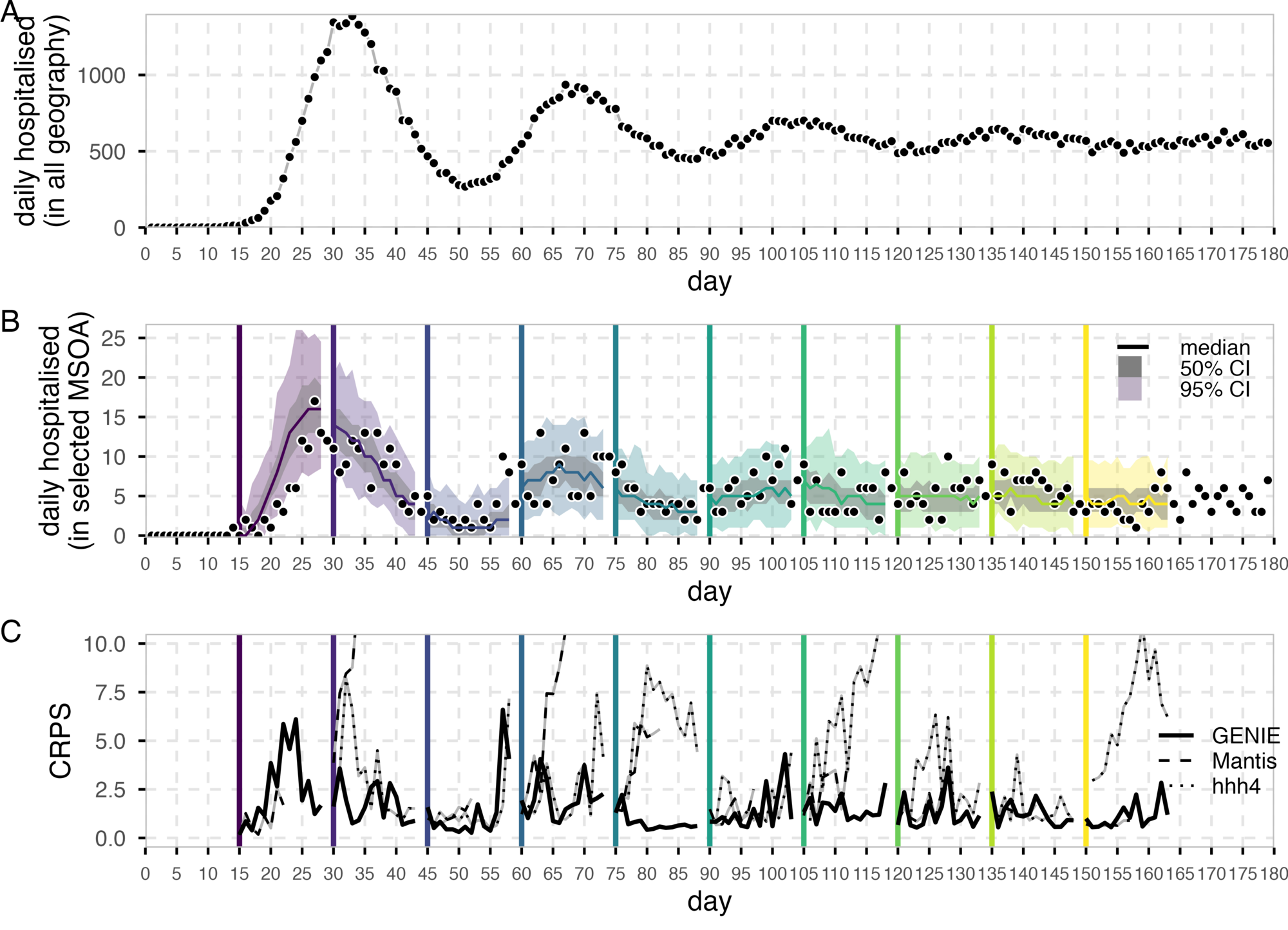}
    \caption{\textbf{Example GENIE forecasts} Comparison of GENIE model forecast against observed hospitalisation data. (Panel A) Total daily hospitalisations on the whole geography, providing a reference for the overall epidemic state. (Panel B) Daily hospitalisations for a sample MSOA. Includes 14-day GENIE forecasts from 10 starting points (days 15–150). Shaded areas represent 50\% and 95\% credible intervals (CIs). (Panel C) Daily CRPS for GENIE, Mantis and \texttt{hhh4}. Lower scores indicate higher accuracy and precision.}
    \label{fig:overview_simA_hospitalisations}
\end{figure*}

Figure \ref{fig:overview_simA_hospitalisations}A shows the daily number hospitalisations aggregated over the whole geography (shown in Figure~\ref{fig:geography}) in one of the simulated epidemics.
Figure \ref{fig:overview_simA_hospitalisations}B also shows an example of forecasts produced with GENIE of daily hospitalisations in a selected MSOA.
The forecasts are started at ten points spaced 15 days apart, with the first one day after the first hospitalisation in the simulation. 
The 50\% and 95\% credible intervals (CIs) come from the empirical quantiles of the simulated trajectories.
Figure \ref{fig:overview_simA_hospitalisations}C shows a comparison of the CRPS (described in Section~\ref{sec:methods_evaluation}) for forecasts generated using GENIE, and two alternative models: Mantis and \texttt{hhh4}.
For the majority of the points in the comparison, GENIE has a lower CRPS --- lower is better --- than the alternatives.

\begin{figure*}[h]
    \centering
    \includegraphics[width=0.9\textwidth]{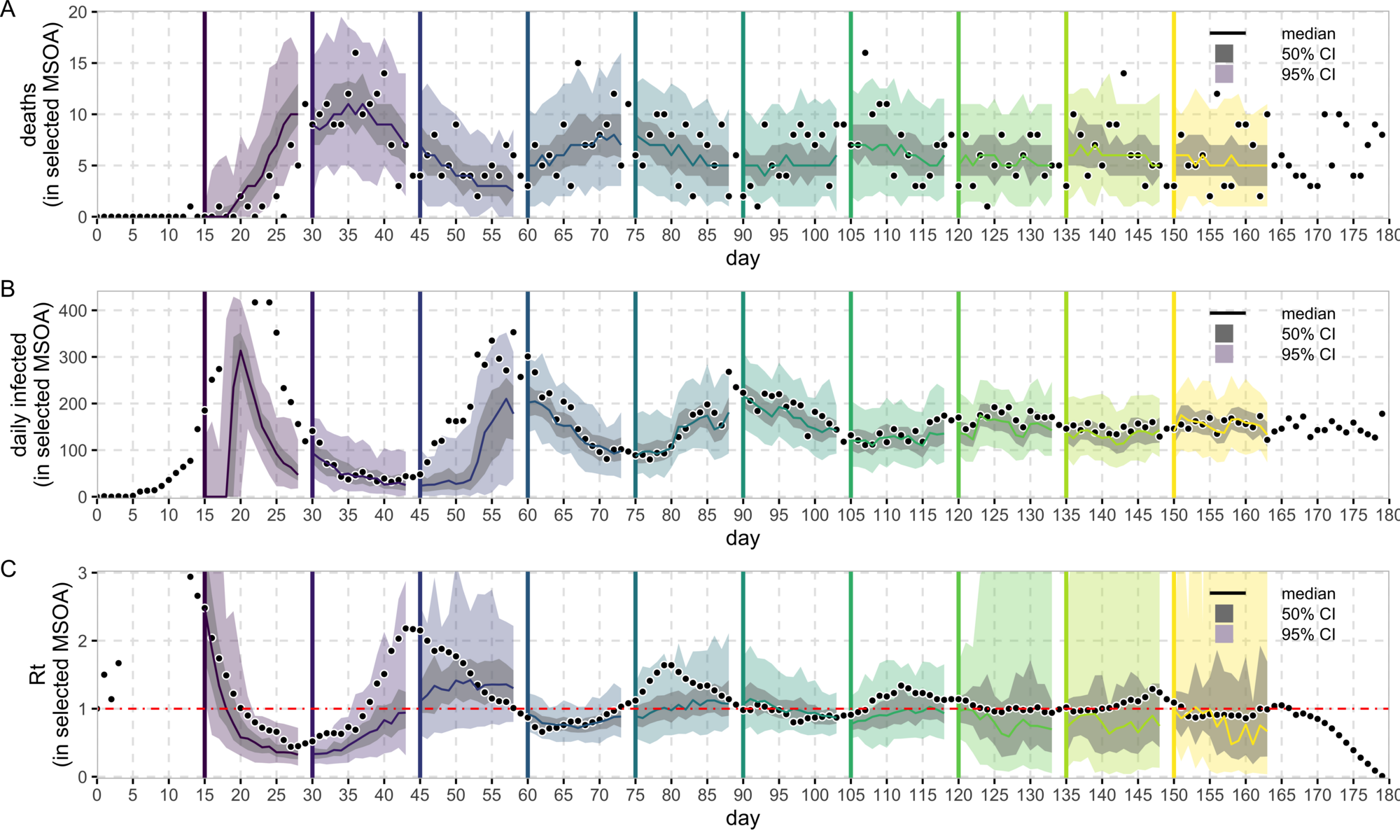}
    \caption{\textbf{Forecasts of additional measures} Evaluation of GENIE model performance for deaths, infection rates, and the case reproduction number $R_t$. (Panel A) Observed death timeline with 14-day GENIE forecasts initiated at 10 points (days 15-150). (Panel B) Forecasted daily infections compared against \texttt{JUNE26} simulation ground truth. (Panel C) Forecast case reproduction number ($R_t$) compared against ground truth derived from the \texttt{JUNE26} model.}
    \label{fig:overview_simA_other_burdens}
\end{figure*}

Figure~\ref{fig:overview_simA_other_burdens} shows the forecasts for deaths, daily number of infections, and the effective reproduction number $R_t$ for the same simulation shown in Figure~\ref{fig:overview_simA_hospitalisations}.
In each panel, black points indicate the ground truth extracted from the \texttt{JUNE26} simulation, as described in \ref{sec:appendix_data_R0}, the shaded regions denote the 50\% and 95\% CIs of the predicted distribution.
Our model produces a good fit to the ground truth, with improved agreement after the first month of the epidemic (once more data is available in the context).
For infections, early forecasts do not fully capture the initial peak, reflecting the earlier rise infections relative to hospitalisations and deaths and the limited amount of available context when forecasting begins.
For $R_t$, the model follows the general trend of the ground truth but underestimates peak magnitudes.

\FloatBarrier

\subsection{Forecast performance}

To answer Question (i) from Section~\ref{sec:methods_evaluation}, we compare the forecast performance of GENIE relative to two alternatives, Mantis and \texttt{hhh4}, across all test simulations using CRPS (and ES in the case of \texttt{hhh4}), for multiple forecast initialisation times, forecast horizons, and simulations.
These comparison reveal that GENIE performs better than these alternative models in the majority of test cases considered.

Raw score values are sensitive to signal magnitude, so direct comparison across simulations or timesteps is not meaningful.
Instead, models are compared by counting how often GENIE outperforms Mantis for a given forecast starting point (counted from the day the first hospitalisation is recorded) and window horizon.
The first row of Table~\ref{tab:results_comparison} shows the percentage of forecasts in which GENIE produces better forecasts, one comparison per forecast start time per MSOA per simulation.
GENIE outperformed Mantis 69\% of the time when forecasting daily hospitalisations and 61\% of the time in forecasting deaths.

When comparing GENIE with \texttt{hhh4} we repeat the analysis of how often GENIE produces a better forecast using either CRPS or the ES. 
Forecast accuracy, calibration, and sharpness are evaluated at each MSOA individually using the CRPS score, while a multivariate generalisation, the ES, is used to assess these properties jointly across locations.
The second and third rows of Table~\ref{tab:results_comparison} report that GENIE outperforms \texttt{hhh4} in the majority of forecasts.
Further details of these results are visualised in Figure \ref{fig:comparison_models_hosp} and Figure \ref{fig:comparison_models_death} of the Appendix.

\begin{table}[h]
\centering
\begin{tabular}{@{}l C{3.5cm} C{2cm} @{}}
\toprule
\textbf{} & \textbf{Daily hospitalisations} & \textbf{Deaths} \\
\midrule
Mantis (CRPS) & 69.78\% & 61.86\% \\
\midrule
\texttt{hhh4} (CRPS) & 58.54\% & 56.91\% \\
\midrule
\texttt{hhh4} (Energy) & 78.12\% & 79.02\% \\
\bottomrule
\end{tabular}
\caption{{Percentage of simulations where GENIE achieved the best score across different forecasting parameters compared to either Mantis or \texttt{hhh4}.} 
}
\label{tab:results_comparison}
\end{table}

\subsection{Emergence Phase}\label{sec:results_emergence}

To answer Question (ii) from Section~\ref{sec:methods_evaluation}  about GENIE's performance during disease emergence phase in greater detail, we considered the ES of short-term hospitalisation forecasts.
Specifically, we generated forecasts with a 14 day horizon starting on each day from day 0 to day 30 following the first recorded hospitalisation.
Supplementary Figure~\ref{fig:score_emergence_comparison} shows the percentage of instances in which GENIE produces a better forecast (than either Mantis or \texttt{hhh4}) by forecast initialisation date.
In each case, we find that GENIE produces better forecasts for the majority of initiation dates, although Mantis produces better forecasts for the initial 12 days when there is very little data available to inform forecasts.
These comparison reveal that GENIE performs better than these alternative models in the majority of emergence phase test forecasts, although in the very early stages Mantis outperforms GENIE.

\subsection{Predictive Peak Performance}\label{sec:results_peaks}

To answer Question (iii) from Section~\ref{sec:methods_evaluation}  about GENIE's ability to predict peak hospitalisation we compared it to \texttt{hhh4} using the measures described in Section~\ref{sec:methods_peak_performance} for the first three peaks using the \(\rho=1\) peak definition.
We do not include Mantis in this comparison as it makes quantile-based forecasts which precludes the evaluation of measures that apply at the level of individual samples of the epidemic trajectory.
The results below are reported by the offset (defined in Equation~\ref{eq:days_after_peak}) between the day a forecast was issued and the midpoint of the peak week, so that negative values correspond to forecasts issued in advance.

\begin{figure*}[h]
    \centering
    \includegraphics[width=0.90\textwidth]{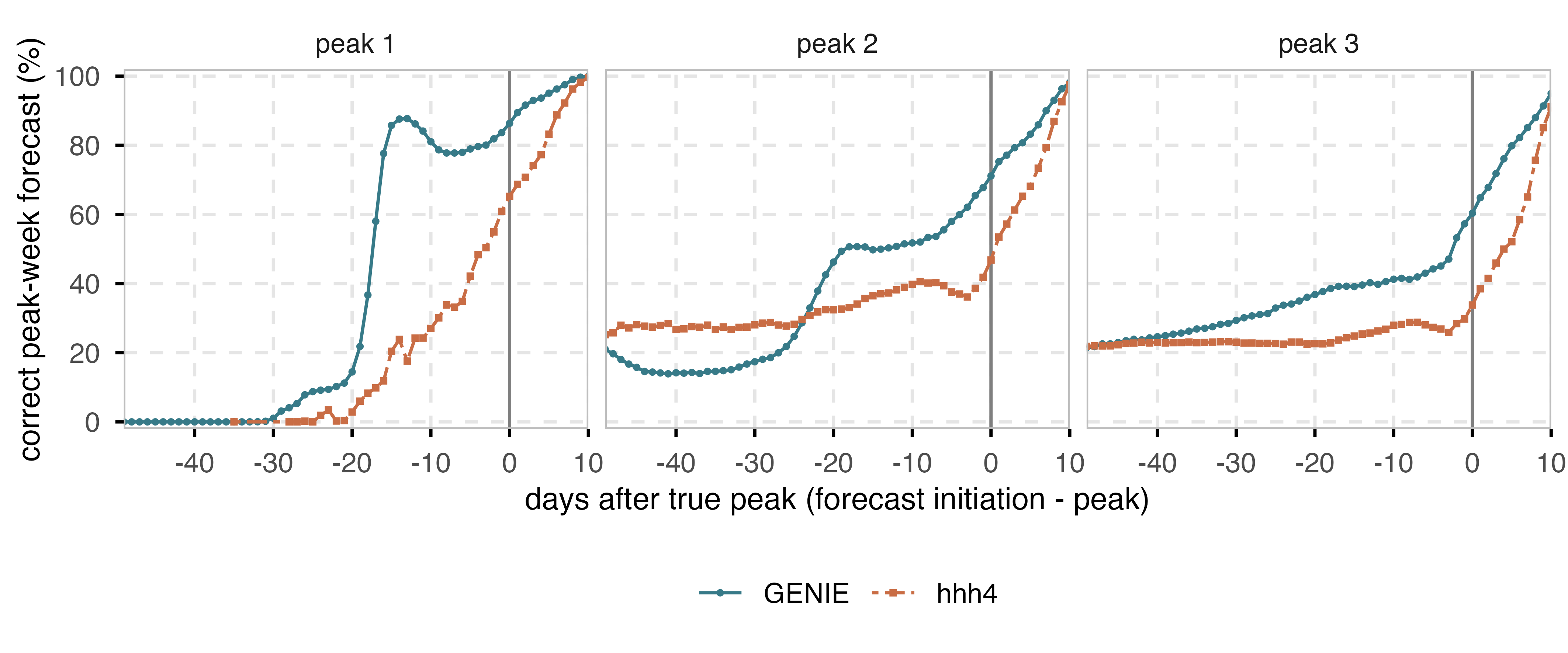}
    \caption{\textbf{Accuracy of the predicted peak week.} Percentage of forecast samples placing peak $p$ in the correct ISO week (as defined in Equation~\eqref{eq:peak_week_accuracy}) for GENIE (solid blue) and \texttt{hhh4} (dashed orange), for the first three peaks. The horizontal axis is the offset (defined by Equations~\eqref{eq:days_after_peak}) between the forecast initiation day and the day of the peak; the vertical line marks the true peak timing. Samples containing no $p$-th peak count as incorrect. With the exception of some very early forecasts for peak 2, GENIE outperforms \texttt{hhh4}. The \texttt{hhh4} curve for the first peak begins 35 days ahead of it, the earliest lead time at which an \texttt{hhh4} forecast of that peak exists (Appendix Section~\ref{sec:appendix_peaks_availability}).}
    \label{fig:peak_week_accuracy}
\end{figure*}

Figure~\ref{fig:peak_week_accuracy} shows the average percentage of forecast trajectories in which the peak falls in the correct week for both GENIE and \texttt{hhh4}.
Note that while accuracy in predicting peak timing broadly increases as the peak approaches, it is not until at least one week after the peak that it can be said to have occurred with certainty.
Neither model identifies the week of the first peak more than a month in advance, but GENIE exceeds $50\%$ accuracy 17 days ahead of it and reaches $87.6\%$ two weeks ahead, where \texttt{hhh4} attains $23.8\%$ and does not itself exceed $50\%$ until three days before the peak.
On the day of the peak GENIE leads by $24.3$ and $26.5$ percentage points for the second and third peaks; only at lead times beyond 24 days on the second peak are the two comparable. The timing errors show the same pattern (Appendix Section~\ref{sec:appendix_peaks}).

The \texttt{hhh4} curves for the first peak begin 35 days before it. This is a property of the benchmark: \texttt{hhh4} is refitted at each initiation day and requires a median of 22 days of data before it converges, and the first peak follows its first forecast by at most 35 days, so no earlier forecast of that peak exists. From the second peak onwards both models cover all offsets (Appendix Section \ref{sec:appendix_peaks_availability}).

\begin{figure*}[h]
    \centering
    \includegraphics[width=0.95\textwidth]{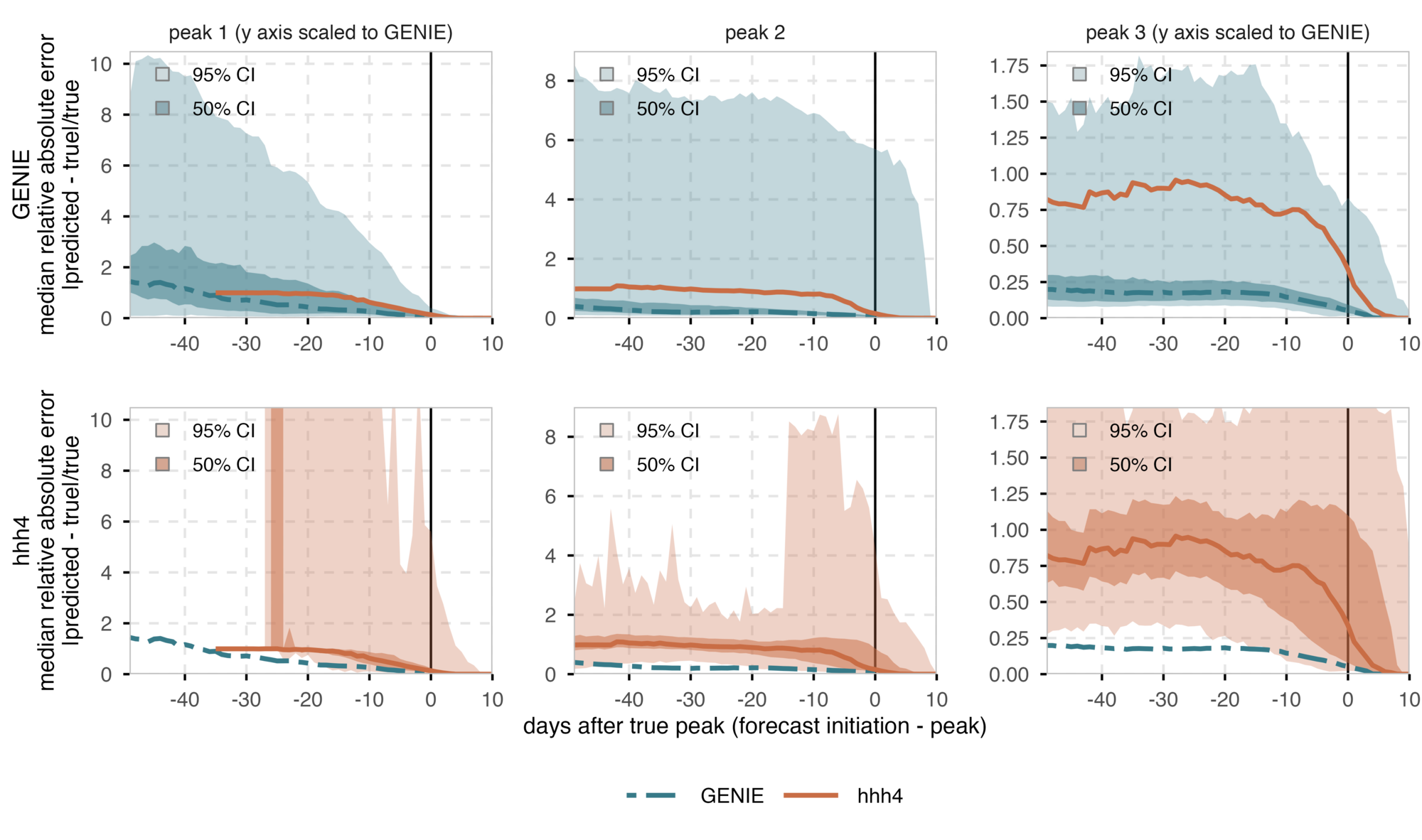}
    \caption{\textbf{Relative error in the predicted peak height.} Median MRE \eqref{eq:mre_height} in the height of peak $p$, i.e. as this is relative to the true peak height, an error of $1$ is as large as the peak itself. Both rows show the median MRE of each model (GENIE solid blue, \texttt{hhh4} dashed orange) with the 50\% and 95\% credible intervals of the model the row is scaled to (GENIE above, \texttt{hhh4} below). Only samples containing a $p$-th peak contribute so these estimates are conditional upon the forecast containing a peak; the horizontal axis is as in Figure \ref{fig:peak_week_accuracy}.}
    \label{fig:peak_height_rel_error}
\end{figure*}

Figure~\ref{fig:peak_height_rel_error} shows the median MRE in predicting the peak number of hospitalisations in a single week at the LTLA level for GENIE (top row) and \texttt{hhh4} (bottom row) indicating that the median MRE is lower for GENIE across available points of comparison for each of the three peaks.
For the first peak it falls from $1.44$ seven weeks ahead, an error as large as the peak itself, to $0.32$ two weeks ahead and $0.05$ on the day of the peak, whereas that of \texttt{hhh4} remains close to $1.0$ until roughly two weeks before the peak, reaching $0.13$ on the day itself.
The errors of GENIE on the second and third peaks are lower still ($0.19$ and $0.17$ two weeks ahead), while those of \texttt{hhh4} are still $0.86$ and $0.78$ at that lead time and stay above $0.7$ until a week before the peak.
We note here, however that the 95\% CI on the errors for peak 2 predictions are larger for GENIE than for \text{hhh4} despite the consistent better performance at the 50\% CI.
Analogous results with the absolute error are given in Appendix Section~\ref{sec:appendix_peaks_main}.

To assess the sensitivity in these results to our definition of the peak, we repeated them with the stricter peak definition in which \(\rho=2\).
Under this alternative definition of the peak all results remained qualitatively unchanged.
Full details of this sensitivity analysis are given in Appendix Section~\ref{sec:appendix_peaks_sensitivity}.

\FloatBarrier

\subsection{Ablation Study}\label{sec:results_ablation}

To answer Question (iv) from Section~\ref{sec:methods_evaluation} about the importance of accounting for geography when generating the forecasts, we carried out an ablation study in which parts of the neural network were removed to learn about their contribution.
We considered ablation of two components: the LPE, through which socio-demographic context is provided, (resulting in what we refer to as the ``No-LPE GENIE''); and the graph \(\mathcal{G}\) connecting the MSOAs, meaning that each MSOA is forecast independently, (resulting in what we refer to as the ``MLP GENIE'').
For both ablations, we calculated the percentage of simulations in which the full GENIE model outperformed the ablated version across various start dates and forecast horizons, similarly to the comparisons with Mantis and \texttt{hhh4} reported above.

Table~\ref{tab:results_ablation} shows the percentage of simulations in which GENIE outperformed each ablated version based on the ES. The full version of GENIE performed best in predicting hospitalisations and deaths relative to the No-LPE and MLP GENIE.
These results demonstrate that integrating location-specific information to GENIE's architecture enhances overall forecasting performance.
Further details of these results are given in Supplementary Figure \ref{fig:score_ablation_comparison}.

\begin{table}[h]
\centering
\begin{tabular}{@{}l C{3.5cm} C{2cm} @{}}
\toprule
\textbf{} & \textbf{Daily hospitalisations} & \textbf{Deaths} \\
\midrule
No-LPE GENIE & 76.00\% & 79.31\% \\
\midrule
MLP GENIE & 71.13\% & 65.60\% \\
\bottomrule
\end{tabular}
\caption{{Percentage of simulations where GENIE achieved the best score across different forecasting parameters compared to its ablated versions, based on the energy score.} 
}
\label{tab:results_ablation}
\end{table}

\FloatBarrier

\section{Discussion}\label{sec:discussion}

This work presents GENIE as a proof-of-concept framework for fine-scale, uncertainty-aware infectious disease forecasting that combines spatio-temporal modelling with simulation-based training.
We demonstrate that GENIE can (i) generate geographically granular probabilistic forecasts of multiple disease burdens across the whole epidemic duration, (ii) produce predictions during emergence that outperform existing state-of-the-art models, (iii) forecast the timing and magnitude of hospitalisation peaks, and (iv) provide a model whose essential components have been validated through ablation studies.
By training on synthetic, yet realistic datasets, the model internalises the complex interplay between the disease dynamics and localised socio-demographic factors, enabling accurate short and long term probabilistic forecasting.

A central motivating hypothesis of this work is that infectious disease dynamics are inherently spatio-temporal.
Local transmission processes create spatial dependence as infections spread through geographically proximate or otherwise connected areas, leading to correlated epidemic trajectories. 
We compare GENIE against two established models: Mantis, a non-spatial neural-network-based \cite{dudley2025mantis}, and \texttt{hhh4}, a widely used statistical model for spatio-temporal disease forecasting \cite{meyer2017spatio}.
GENIE outperforms Mantis consistently, and achieves superior performance against \texttt{hhh4}, even when evaluated using the Energy Score demonstrating that our spatial approach outperforms existing methods.
GENIE's forecasting performance improves as additional observations are included into its context window, it does well by approximately one week after the first hospitalisation, when transmission is still highly localised yet sufficiently data-rich to inform predictions.
GENIE outperforms both Mantis and \texttt{hhh4} on the majority of simulations, verifying the value of modelling spatial interactions at this early stage. 

In the simulated epidemics used to train our model, the basic reproduction number $R_0$ is predominantly concentrated between 2 and 9, implying that transmission typically accelerates rapidly following initial seeding events.
Early in an outbreak, when observational data are extremely sparse, GENIE tends to rely more heavily on its learned prior, which favours early growth in disease burden.
In contrast, Mantis, was trained on a broader range of conditions enabling it to better handle low burden settings. 
In these settings, conservative predictions can yield lower short-term error by matching the absence of recorded cases.
This observation highlights the sensitivity of emergence phase forecasting to training data assumptions and suggests that a wider simulated data range may improve the emergence phase, forecasting robustness.
More generally, this demonstrates fundamental difficulties in the use of neural methods in inference: 
it is not clear if a neural network will be able to generalise from its training data;
and for a finite sample of training data, outside of heavily constrained problems, it is not clear whether a new observation will lie near to that training data.
Developing robust general diagnostics for detecting when these difficulties have been encountered is an open problem for amortized inference.

Our results demonstrate that GENIE is capable of using of location-specific context when making forecasts. 
GENIE's architecture explicitly separates static socio-demographic information, encoded via the Local Profile Encoder (LPE), from dynamic inter-MSOA interactions, encoded by the Local Interaction Encoder (LIE).
The ablation study shows that removing the LPE degrades forecasting performance, confirming that persistent local characteristics, such as population structure, deprivation, and residential composition provide meaningful signals for predicting disease burden.
As epidemic dynamics are governed not only by shared biological mechanisms, but also by heterogeneity in local population and environmental conditions, we hypothesise this is a useful property for epidemic forecasting systems.

A key contribution of this work is the further demonstration and use of ABMs as a mechanism of generating simulated disease trajectories to train \emph{neural} epidemic forecasting models such as GENIE, extending prior work on the use of ABM generated data to train deep learning models \cite{dudley2025mantis,zarebski2026amortized,OutbreakFlow}.
ABMs enable the production of large volumes of spatially granular epidemic trajectories, while explicitly tracking and recording all major components of the simulation, this includes among other things the location of each agent over time and the full \emph{transmission tree} which describes who infected whom, and where transmission occurred.
This level of detail provides access to latent epidemiological quantities that are unobservable in routine surveillance data, such as true infection incidence and reproduction numbers.
This flexibility allows GENIE to learn relationships between observable disease burdens such as hospitalisations and deaths and unobserved epidemic states (infections and the case reproduction number \(R_t\)).
In this work, we compute and associate \(R_t\) based on the infector's MSOA of residence rather than the location of the infection event.
We adopt this definition as it enables the identification of geographic areas responsible for transmission.
An example of the strength of the ABM framework is the ease of $R_t$ redefinition, as the complete transmission tree is directly accessible within the simulation.

For healthcare planners, accurately anticipating when an epidemic will peak and the magnitude of this peak is important for inform resource allocation, hospital capacity management, and the timing of interventions \cite{Morbey2023, Morbey2025, Fong2024}. 
The results in Section~\ref{sec:results_peaks} demonstrate that GENIE does a better job of predicting the timing and magnitude of the peak than the \texttt{hhh4} model, despite \texttt{hhh4} having the epidemic's dominant period when used, an advantage that likely wouldn't exist in practise outside of seasonal pathogens.
This result is robust, as demonstrated across several sensitivity analyses to the peak definition and the specific metrics.

We acknowledge a number of limitations of this study.
First, model evaluation is conducted on a held-out subset of the agent-based simulations.
While this design ensures a separation between training, validation, and testing data, it does not validate the model against real-world surveillance data.
A preferable evaluation would involve training GENIE on synthetic data and assessing its performance on real-world datasets.
This would provide a clearer assessment of the model's ability to forecast real epidemics.

Second, the geographic scope of this study is limited to a subset of the North East of England.
While the selected study region includes a mixture of urban, suburban, and rural areas, and therefore provides a reasonable geography for a proof of concept, it does not capture the full heterogeneity present across the United Kingdom.
Scaling to larger and more diverse geographies will introduce additional challenges, both computational and modelling related, and the performance of GENIE under such conditions remains to be established.
Although the GNN structure of GENIE allows it to be applied to geographies unseen during training, its forecasting capability on novel geographies degrades.
Appendix \ref{sec:appendix_other_geos}, presents results of applying GENIE to simulations from a novel geography in which \texttt{hhh4} outperforms GENIE.
This may be in part due to the novel geographies considered being substantially smaller; when we consider only the 10\% of simulations with the most hospitalisations, GENIE again outperformed \texttt{hhh4}.
Again, this illustrates the importance of matching training data to the problem at hand, and the need for sufficient computational resources to simulate a large amount of training data.
There are efforts to overcome these computational challenges using surrogate models~\cite{scheurer2025uncertainty}.

Thirdly, the simulated outbreaks only consider a limited region of parameter space, resulting in a range of basic reproduction numbers and generation times that are broadly representative of COVID-like respiratory pathogens.
As a result, the trained GENIE may be less effective for pathogens with transmission dynamics that fall outside this parameter space.
This limitation is particularly relevant during disease emergence, where the model relies more heavily on the specific patterns of emergence exhibited in the training simulations.
Furthermore, the random sampling of parameters used to generate the synthetic dataset may have introduced unintended biases in the outbreak dynamics represented by the dataset, potentially affecting the generalizlisability of GENIE to outbreak scenarios under-represented in this parameter sampling scheme.

Finally, although GENIE uses a highly flexible neural architecture to condition forecasts useful context, the form of the predictive distribution is constrained to a given parametric family (as shown in Equation~\eqref{eq:distribution_q}).
It remains to answer: to what extent is predictive performance limited by the choice of parametric family rather than by the representation learned by the neural network?
One specific limitation here is that in equation~\ref{eq:distribution_q}, we treat each burden's predictive distributions as independent conditional on the context which is a restriction on the joint predictive distribution.
The limited flexibility of the parametric family may constrain the ability of the model to represent the conditional marginal distributions.
Non-parametric models, such as normalising flows, learn transformations of a simple base distribution and can represent substantially more complex distributions, including arbitrarily multimodal, skewed and heavy-tailed densities~\cite{Kobyzev2021}. 
Whether this additional flexibility translates into improved predictive performance for epidemic forecasting remains an unanswered empirical question.
The literature has demonstrated improved predictive performance from more flexible predictive distributions in other autoregressive time series settings~\cite{Rasul2021}.

There are a number of possible extensions for this work.
A natural extension is to broaden the range of simulated epidemics used for training by expanding the agent-based model parameter space.
This would enable GENIE to learn from a more diverse set of disease dynamics, improving robustness during disease emergence and enhancing cross-pathogen generalisability.
From a computational perspective, since simulating the training data is the main computational cost, the construction of an emulator or surrogate model for the agent-based simulator could substantially accelerate data generation. 
This would also facilitate training on larger and more heterogeneous geographies and disease dynamics. 
An alternative scaling strategy could involve training GENIE on multiple disjoint subgraphs or regional subsets and evaluating its ability to generalise across a full geography at test time.

This work shows that interpretability from ABM simulations and predictive performance from machine learning do not need to be treated as separate approaches. By training the model on simulation-generated data, we combine the strengths of both. The resulting framework supports surveillance systems that can anticipate future epidemic waves while providing forecasts that are timely for policy decisions and sufficiently detailed to guide local interventions.

\section{Data availability}

An implementation of the GENIE model along with the code to train the neural networks and evaluate their performance is available at \url{https://github.com/Enantiodromis/GENIE}.
Code to simulate the training data is available from \url{https://github.com/lauraguzmanrincon/JUNE4GENIE}.

\section*{Acknowledgments}

We thank Martha Correa-Delval for her assistance in the use and customisation of JUNE. GREB, LMG and DDA were funded by the UK Medical Research Council (MRC) programme MRC 
$\text{MC}\_\text{UU}\_00040/04$, which also part funded PJB.
GREB and JK acknowledge support from AI4CI (EPSRC) [grant number G937902].
PJB also acknowledges support from the Alan Turing Institute Foundations/Fundamental research programme.
AEZ and PJB acknowledge support from the Wellcome Trust [grant number 227438/Z/23/Z].
The authors acknowledge the use of ChatGPT and Claude for text editing and assistance with selected coding tasks.
The authors retain full responsibility for the manuscript's content.

\bibliographystyle{unsrt}  
\bibliography{referencesCombined}  

\clearpage

\appendix \setcounter{page}{1}

\begin{center}

{\Huge\bfseries{GENIE: Generative Neural Inference for Epidemics}} 

\end{center}

{\Large\bfseries Laura M Guzm\'{a}n-Rinc\'{o}n, George R E Bradley, Joel Kandiah, Kyriakos Flouris, Pietro Li\`{o}, Paul J. Birrell, Alexander E. Zarebski, Daniela De Angelis}

\begin{center}

{\Large\bfseries Supplementary Material} 

\end{center}

\vspace{1cm}

\addcontentsline{toc}{section}{Appendix}

\section{Data Construction}\label{sec:appendix_data}

\subsection{Scenario-Defining Parameters for the Synthetic Epidemic Dataset}\label{sec:appendix_data_parameters}

\juneold provides a built-in functionality for seeding infections within specified regions of England, a coarse geographic unit and the highest tier of sub-national division \cite{ONS_StatisticalRegions}. It also includes parameters controlling transmission intensity and the rates at which individuals progress to varying levels of severity. Building on this structure we extended \juneold in several ways. To generate heterogeneous scenarios, we introduced randomisation of model parameters at code level and enabled seeding at the finest available geographical scale (Output Area level in England). We also implemented a waning immunity mechanism allowing individuals to lose protection after recovery, a feature absent from the original model.

\textbf{Seeding events:} We randomise the timing, location, and initial magnitude of infections to initiate outbreaks under a diverse range of starting conditions. Seeding events are distributed over a randomly sized time window at the start of the epidemic, with a mean window length of 45 days across simulations.

\textbf{Infection transmission intensity:} We vary the global contact intensity defined in \juneold \cite{JUNE} to simulate different levels of population-wide infectiousness and interaction frequency.

\textbf{Degree of waning immunity:} We adjust the rate at which immunity fades to reflect different potential durations of protection after an individual's recovery. This rate was not incorporated in \juneold.

\textbf{Hospitalisation and death rate:} We modify the probabilities of hospitalisation and death outcomes to represent varying levels of disease severity across simulations. To account for the impact of socio-economic deprivation on severity, we link these variations to the Index of Multiple Deprivation (IMD) of each MSOA, as detailed in Section \ref{sec:sims_outcomes}.

\begin{table}[ht]
    \centering
    \begin{tabular}{ll}
    \toprule
        \textbf{Scenario-defining parameter} & \textbf{Sampling}\\
    \midrule
        Number of seeds & LHS \\
    \midrule
        Location of each seed $i$ & IS \\
    \midrule
        Day of each seed & IS \\
    \midrule
        Time window for seeding & LHS \\
    \midrule
        Degree of waning immunity & LHS \\
    \midrule
        Infection transmission intensity & LHS\\
    \midrule
        Hospitalisation and death rates & LHS \\
    \bottomrule
    \end{tabular}
    \caption{\textbf{Scenario-defining parameters used to generate simulations from \texttt{JUNE26}.} The second column shows the parameter name, and the third the method from which the samples are drawn: individual sampling (IS) or Latin Hypercube sampling (LHS).}
    \label{tab:june_parameters}
\end{table}

\subsubsection{Random Seeding}\label{sec:sims_seeding}

Each simulation begins by introducing infections into an initially healthy population.  
Since there is a day-of-the-week structure in JUNE, we need to randomise the day of the week in which the seeding occurs.
Seeding events are distributed across time and space as described below.
\begin{itemize}
    \item The \textit{first seed} occurs on a randomly chosen day of the year, $t_{0} \sim \Uniform{1}{365}$, where \(t_{0}=1\) corresponds to the seeding occurring on 1~January and day 365 corresponds to 31~December (ignoring leap years). The location of the infection seeding is drawn uniformly from all areas in the geography.
    \item The number of \textit{secondary seed} locations is drawn from $n \sim \Poisson{6}$, and their timing is spread over a window $w \sim \text{Poisson}(45)$ days. Each secondary seed $d_i$, for $i \in \{1,\dots,n\}$, occurs on a unique day sampled from $\{1,\dots,w\}$ days after the first seed, ensuring no two seeding events fall on the same day. The location of each secondary seed is drawn uniformly from the set of areas in the geography.
    \item For each seeding event, $i$, $i\in \{0,...,n\}$, the magnitude of the seeding event is an emergent property determined by assuming individuals of the corresponding area are infected with probability $p_i \sim \Beta{1}{1000}$. If no infections occur, one randomly selected individual is infected to guarantee at least one infection per seed.
\end{itemize}

\subsubsection{Infection Transmission Intensity}\label{sec:sims_transmission}

We first describe the original configuration of infection transmission in \juneold. We then detail the modifications implemented to vary the transmission rates for each simulation. Let $s$ be a susceptible individual interacting in a group $g$ (e.g. "school 1", "school 2") on a location of type $G$ (e.g. home, school, workplace). The number of individuals in the group is denoted by $N_g$ and the set of infectious individuals in the group is denoted by $I_g$. The susceptibility of individual $s$ to infection is denoted $\psi_s$.

The probability of a susceptible individual becoming infected in \juneold is defined in \cite{JUNE}. We adapt this formulation to include additional terms, $\psi^{(l)}$, $\beta$, to give adjusted infection probabilities:
\begin{equation*}
    \mathcal{P}_s(t, t + \Delta t) \approx 1 - \exp\left[-\psi^{(l)}\psi_s \Delta t \sum_{i \in I_g} \beta\beta_{si}^{(G, g)}(t) {\mathcal{I}_i(t)} \right],
\end{equation*}
and we vary these new parameters between simulation to cover a range of infection dynamics. The 
newly introduced parameters and are defined to be:

\begin{itemize}
    \item \textbf{Local susceptibility factor $\psi^{(l)}$}: modifies the individual’s susceptibility based on their residence location $l$.
    
    \item[] The factor $\psi^{(l)}$ is calculated from a function $z(l)$ that depends   on $N$ socio-demographic features of the location, denoted as $x_1^{(l)},\ldots,x_N^{(l)}$ (i.e. population density, index of deprivation). The function $z$ is defined to take values in the interval $(0,1)$, and these values are then transformed to map $\psi^l$ into a convenient range. We choose:
    
    $$\psi^{(l)} = 1 - \log \frac{z(l)}{0.5}.$$

    \item[] Assume that there exists $L$ locations in the geography indexed by $l_1,\ldots,l_L$. The feature values across these locations are usually unevenly distributed. To address this, we use the empirical cumulative distribution function (ECDF) to map feature values into a more uniform scale. For a given location $L$ and the feature indexed by $i$,

    \begin{equation*}
        \Phi\left(x_i^{\left(l\right)}\right) = \mathrm{ECDF}_{\left\{x_i^{\left(l_1\right)}, \ldots, x_i^{\left(l_L\right)}\right\}}\left(x_i^{\left(l\right)}\right)
    \end{equation*}

    \item[] denote the ECDF of feature $i$ across all locations, evaluated at $l$. We then compute an intermediate score\vspace{-\baselineskip}

    \begin{equation*}
        \sigma\left(l\right) = \log \frac{1}{\prod_i \Phi\left(x_i^{\left(l\right)}\right)} .
    \end{equation*}

    \item[] Finally, the function $z(l)$ is defined as the ECDF of the values $\{\sigma(l_1), \ldots, \sigma(l_L)\}$ across all locations, evaluated at $l$:\vspace{-\baselineskip}

    \begin{equation*}\label{eq:local_susceptibility}
        z(l) = \mathrm{ECDF}_{\left\{ s(l_1), \ldots, s(l_L) \right\}}\!\left( \sigma(l) \right).
    \end{equation*}

Figure \ref{fig:local_susceptibility} shows values of $\psi^{(l)}$ for all locations in a geography made of 56 MSOAs in England, using $N=2$ with the index of deprivation and population density as features. 

\begin{figure}[H]
    \centering
    \includegraphics[width=0.5\linewidth]{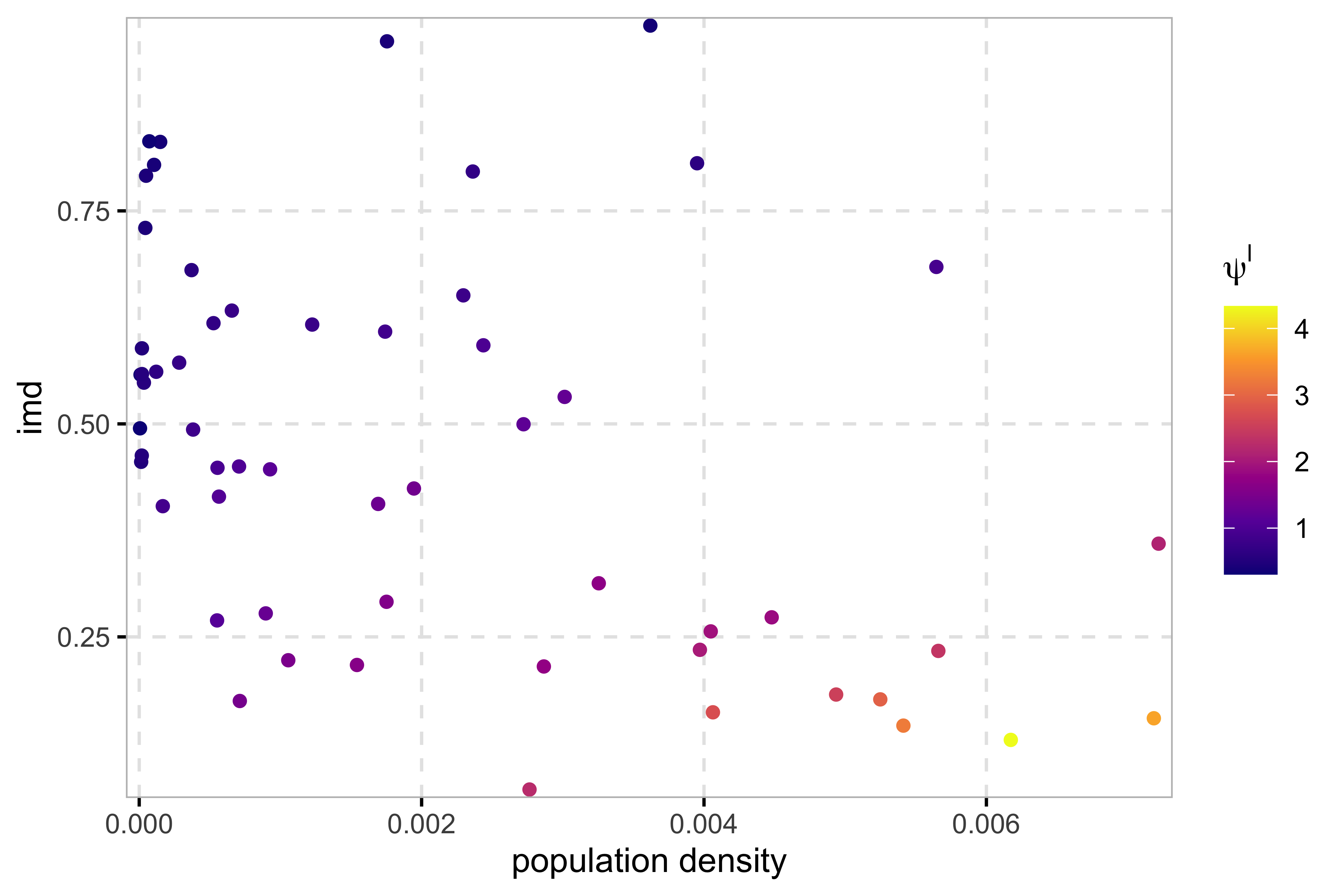}
    \caption{Local susceptibility $\psi^{(l)}$ (colour) for all MSOAs in Gateshead and Northumberland. The index of deprivation (IMD) is an ordinal quantity calculated from centiles that ranges from 0 to 1, where 0 representing to the most deprived and 1 representing the least deprived.}
    \label{fig:local_susceptibility}
\end{figure}

    \item \textbf{Global contact intensity factor $\beta$}: adjusts the overall contact intensity between susceptible and infectious individuals within the same group.

    \item[] We sample values of $\beta$ in the range $(0,2)$. Directly selecting a sampling distribution for $\beta$ is non-trivial, as its effect on outbreak dynamics is complex. To address this, we construct a heuristic transformation that maps a uniform variable $s \sim \text{Unif}(0,1)$ to $\beta$, such that the resulting distribution of epidemic curve maxima is approximately uniform. To derive the transformation, we generated a dataset of outbreak curves across a range of $\beta$ and susceptibility values, and observed empirically that $\exp(1.5 \:\hat{y})$ is approximately linear in $\beta$, where $\hat{y}$ denotes the maximum of the epidemic curve. Fitting a linear regression $\exp(1.5 \cdot \hat{y}) \approx m\beta + d$ and inverting this relationship yields a transformation of the form
    $$\beta(s) = \frac{\exp(bs) + c}{a},$$
    with values $a=30.574$, $b=1.5$, $c=-2.022$, so drawing $s$ uniformly produces values of $\beta$ whose corresponding epidemic peaks are approximately uniformly distributed.

\end{itemize}

\subsubsection{Waning immunity}\label{sec:sims_waining}

Individuals who have recently recovered from infection remain immune in \juneold. Alternatively, waning immunity can be simulated by setting susceptibility to $0.0$ after recovery (immune individual) and allowing it to gradually return to $1.0$ (susceptible individual). Let $s(t)$ denote the susceptibility of an individual, where $t$ is the time since recovery. A simple model is $s(t) = 1 - a^t$, with $a < 1$.

\begin{figure}[H]
    \centering
    \includegraphics[width=0.5\linewidth]{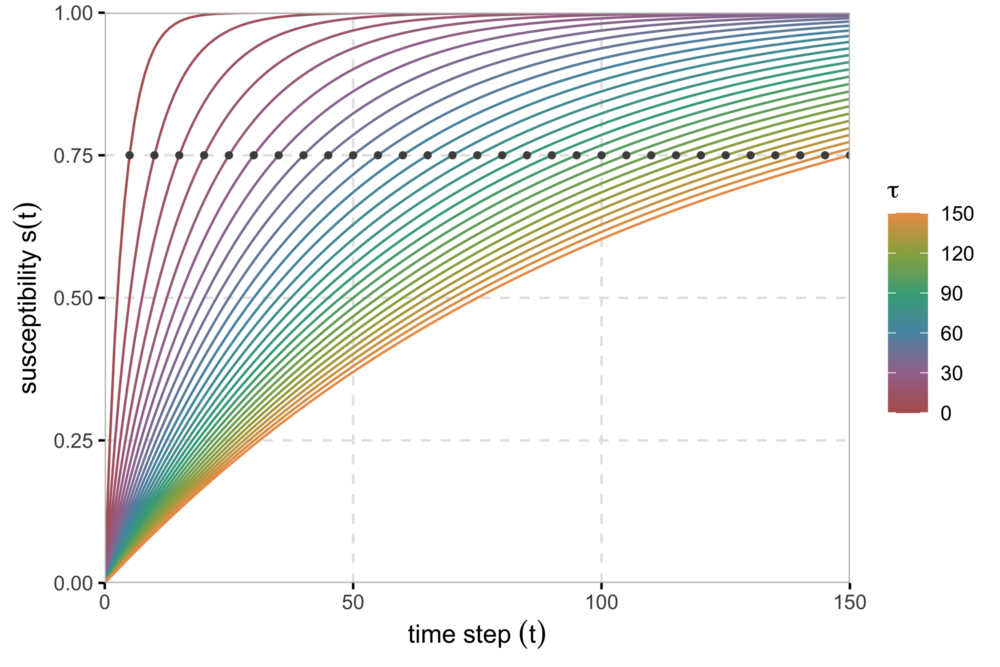}
    \caption{Waning immunity scenarios parameterised with $\tau$. Each line represents a different scenario. The grey dots mark the time when susceptibility reaches $0.75$.}
    \label{fig:waning}
\end{figure}

To introduce variability across simulations, different waning scenarios can be generated by changing $a$. For a more \textit{uniformly looking sampling}, we can reparametrise $a$ in terms of $\tau$: the time at which susceptibility reaches $0.75$. The resulting formulation is shown in Equation (\ref{eq:waning}). Figure \ref{fig:waning} illustrates $s(t)$ for several values of $\tau$.

\begin{equation}\label{eq:waning}
    s(t) = 1 - \exp \left[\frac{\log(0.25)}{\tau} \; t \right]
\end{equation}

We then can sample $\tau$, i.e. $\tau \sim \text{unif}(\tau_\text{min}, \tau_\text{max})$.

\subsubsection{Probabilities of Infection Outcomes}\label{sec:sims_outcomes}

For each individual, a symptom trajectory is randomly assigned at the onset of infection, based on their age $a$, gender $g$ (female \texttt{f}, male \texttt{m}) and population type $k$ (general population \texttt{gp} or care home population \texttt{ch}) (see \juneold configuration in \cite{JUNE}).
\begin{align*}
p_\text{asym} (a,g,k)      &= \texttt{asymptomatic rate} \\
p_\text{mild} (a,g,k)       &= \texttt{mild rate} \\
p_\text{ward} (a,g,k)       &= \texttt{recovers in ward rate} \\
p_{\text{icu}} (a,g,k)       &= \texttt{recovers in ICU rate} \\
p_{\text{death home}} (a,g,k) &= \texttt{death at home rate} \\
p_{\text{death ward}} (a,g,k) &= \texttt{ward death rate} \\
p_{\text{death icu}} (a,g,k)  &= \texttt{ICU death rate} \\
p_\text{severe} (a,g,k) &= \max\left(0, 1 - [p_\text{asym}(a,g,k) + p_\text{mild}(a,g,k) + p_\text{hosp}(a,g,k) + p_{\text{death home}} (a,g,k)]\right) \\
\end{align*}
Where $p_\text{hosp}(a,g,k) = p_\text{ward} (a,g,k) + p_{\text{icu}} (a,g,k) + p_{\text{death ward}} (a,g,k) + p_{\text{death icu}} (a,g,k)$. The probabilities can be adjusted to reflect lower or higher mortality and hospitalisation rates. For every trajectory $j$ in $\{\text{ward, icu, death home, death ward, death icu}\}$, we construct unnormalised values (interpretable as rates) $r_j$ by offsetting the original probability $p_j$ by $\delta$ and multiplying by a location-dependent factor $\alpha_l$.
\begin{align*}
    r_j(a,g,k,l) &= \alpha_l \; [p_j(a,g,k) + \delta]. \\
    \intertext{We then normalise all rates to obtain updated probabilities.}
    \tilde{p}_j(a,g,k,l) &= \frac{r_j(a,g,k,l)}{\sum_j r_j(a,g,k,l)}.
\end{align*}

Values of $\alpha_l$ can be chosen depending on local features. For the synthetic epidemic dataset used in this project, we use $\delta=0.02$ and $\alpha_l = a\: \text{imd}_l + b$, where $\text{imd}_l$ is the IMD of MSOA $l$, and $a$ and $b$ are chosen such that $\alpha_l=0$ for the lowest IMD (least deprived) and $\alpha_l=1$ for the highest IMD in the geography.

\begin{figure}[H]
     \centering
     \begin{subfigure}[b]{0.48\linewidth}
         \centering
         \includegraphics[width=\linewidth]{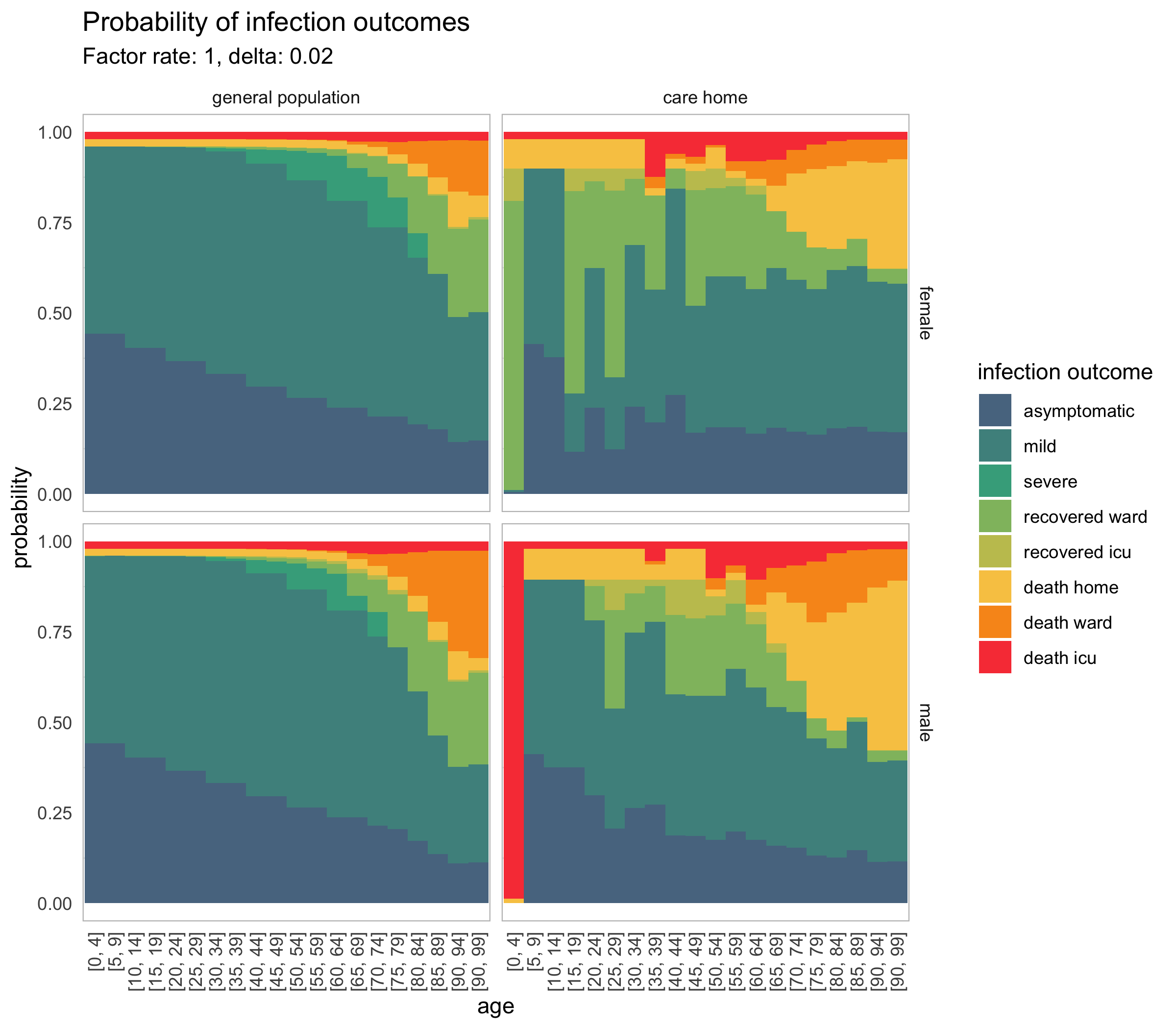}
         \caption{Example: "lowest" hospitalisation/mortality rate.}
         \label{fig:probabilities_01}
     \end{subfigure}
     \hfill
     \begin{subfigure}[b]{0.48\linewidth}
         \centering
         \includegraphics[width=\linewidth]{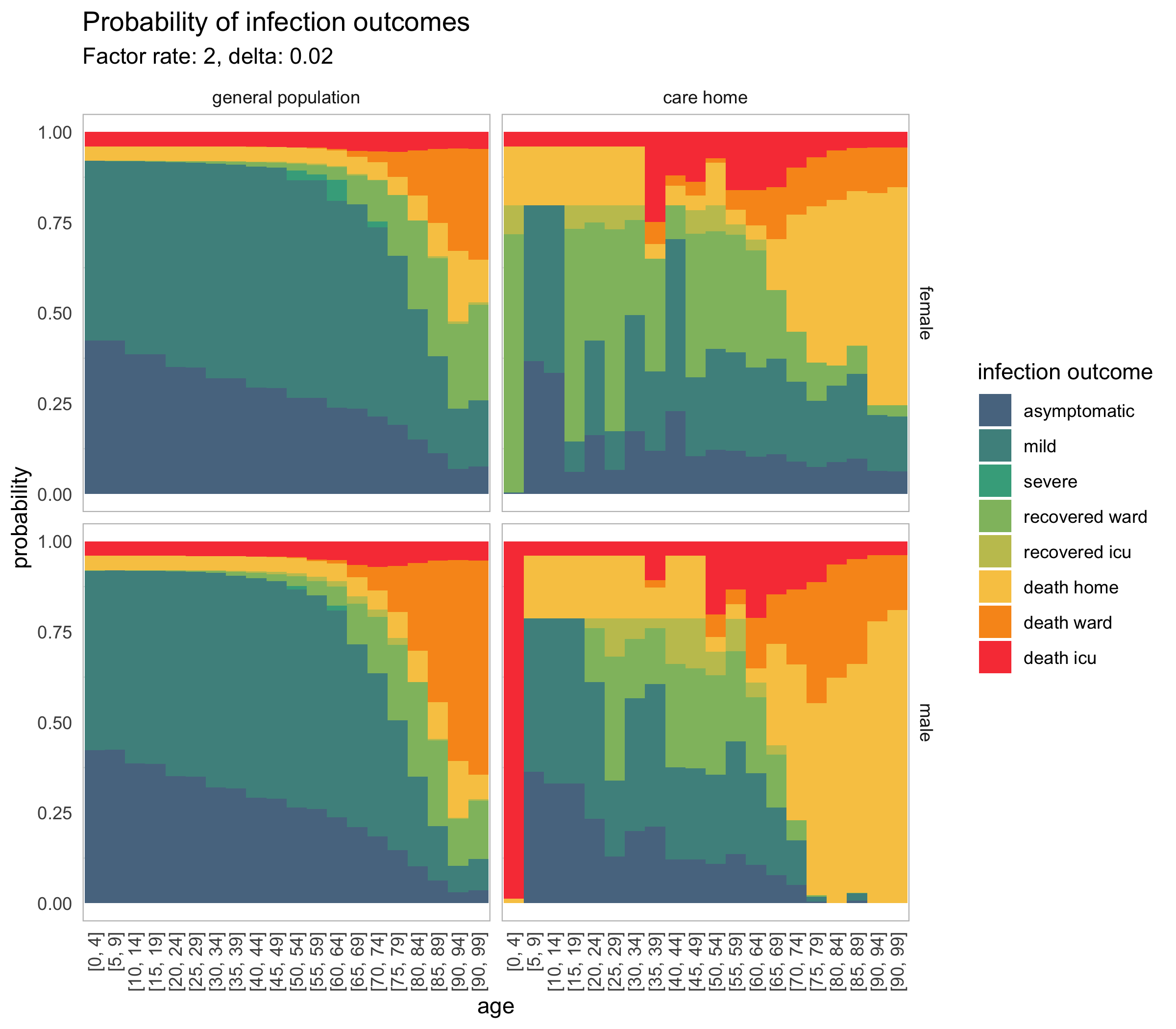}
         \caption{Example: ``highest" hospitalisation/mortality rate.}
         \label{fig:probabilities_11}
     \end{subfigure}
     \caption{Comparison of probabilities for hospitalisation and mortality rates.}
     \label{fig:probabilities_combined}
\end{figure}

Figure \ref{fig:probabilities_01} shows the adjusted probabilities for a factor rate $\alpha=1$ and $\delta = 0.02$. Figure \ref{fig:probabilities_11} shows the adjusted probabilities for a factor rate $\alpha=2$ and $\delta = 0.02$, reflecting a scenario with a high mortality and hospitalisation rates.

\subsection{Estimation of Transmission Parameters}\label{sec:appendix_data_R0}

For each simulation, we reconstructed the transmission chain from the recorded infection events, each comprising an infector, an infectee,
and the time of transmission. From these events, we derived the following daily quantities for every simulation: the number of new infection events occurring on day $t$, the total number of secondary infections
attributed to cases with index day $t$, and the generation time of all infector-infectee pairs, defined as the elapsed time between a primary
infection and the secondary infections it produced.

The \textit{mean generation time} $\bar{g}$ was estimated by averaging generation times across all transmission events where the primary infection occurred before day 150. The truncation at day 150 excludes the latter portion of the epidemic, where depletion of susceptibles begins to suppress transmission and would otherwise bias the estimate downward.

The \textit{basic reproduction number} $R_0$ was estimated from the early exponential growth phase of the epidemic, restricted to the first 8 days ($t < 8$), before susceptible depletion or behavioural changes could deflate transmission rates. For each simulation, $R_0$ was computed as the ratio of total secondary infections to total primary cases over this window:

\begin{equation*}
    R_0 = \frac{\sum_{t=0}^{7} S_t}{\sum_{t=0}^{7} I_t},
\end{equation*}

\noindent where $I_t$ is the number of newly infected individuals on day $t$ and $S_t$ is the total number of secondary infections they collectively produced over the remainder of the simulation.

The \emph{case reproduction number} $R_t^i$ on day \(t\) for MSOA \(i\) was estimated at the MSOA-level, attributing the cases to the MSOA of residence of the infector rather than the location of the infection.
For each simulation, $R_t^i$ was computed as the ratio of all secondary infections caused by primary cases in MSOA $i$ (counted through day 180) to the number of primary cases:
\begin{equation*}
    R_t^i = \frac{S^i_t}{I^i_t},
\end{equation*}
\noindent where $I_t^i$ is the number of individuals infected on day $t$ who reside in MSOA $i$, and $S^i_t$ is the total number of secondary infections they collectively produced in the whole geography over the remainder of the simulation.

\subsection{Data Overview}

\begin{figure}[h]
    \centering
    \includegraphics[width=0.8\textwidth]{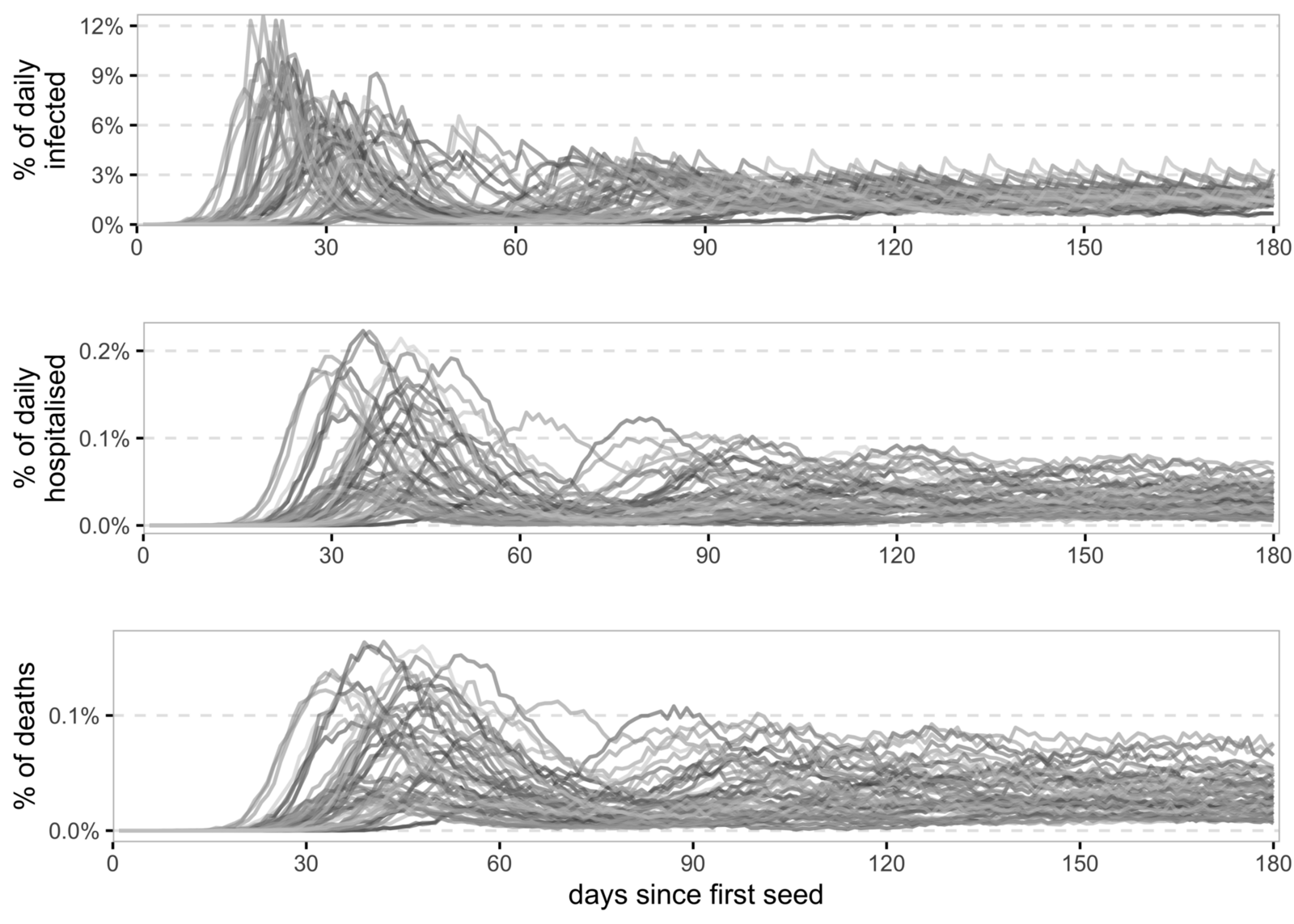}
    \caption{\textbf{Overview of 5\% of all simulations randomly sampled.} The top panel shows the percentage of the total population infected, the middle panel shows hospitalisations, and the bottom panel shows deaths. All figures present values aggregated across the 84 MSOAs.}
    \label{fig:simulations}
\end{figure}

\begin{figure}[h]
    \centering
    \includegraphics[width=\textwidth]{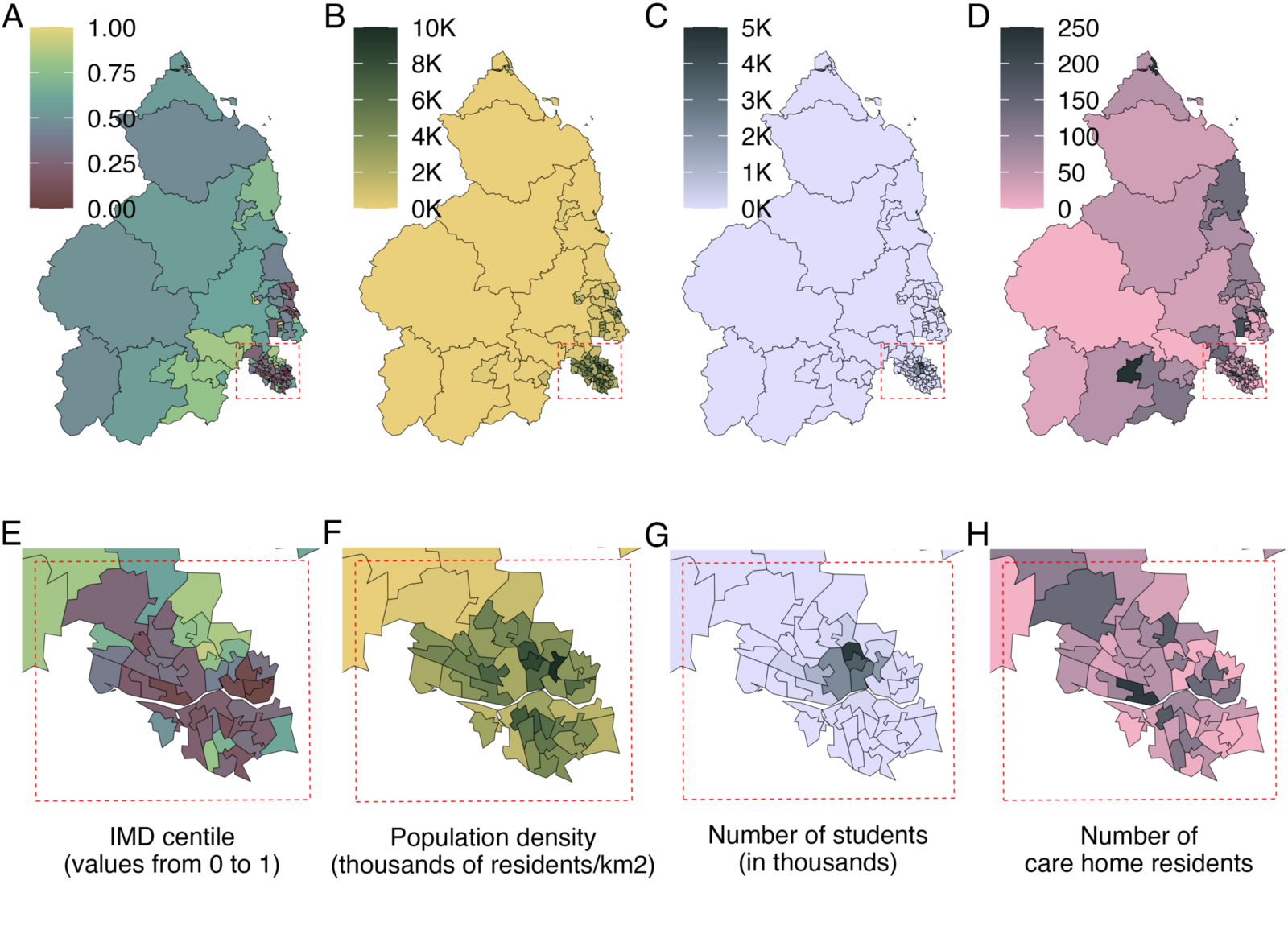}
    \caption{\textbf{Spatial variation of static features.} Geographic distribution of Index of Multiple Deprivation (IMD) centile (Panel A, E), population density (Panel B, F), number of students (Panel C, G), and number care home residents (Panel D, H) by Middle Layer Super Output Areas (MSOA). (Top) Full study area. (Bottom) Detailed view of the Newcastle-Gateshead urban core.}
    \label{fig:features}
\end{figure}

Care-home and household size distributions reflect potential contact intensity and vulnerability within residential settings. The Index of Multiple Deprivation provides a proxy for socio-economic inequality that may influence infection exposure, access to healthcare, and disease outcomes. Student, resident, and care-home populations jointly reflect differences in population mobility and age composition, which together influence both the frequency of social contact and the likelihood of severe clinical outcomes following infection. Population density quantifies potential contact opportunities across the area, while the latitude and longitude of each MSOA centroid introduces additional spatial information to the model. Figure \ref{fig:features} illustrates the spatial distribution of four of the location-specific features across the study region.

\clearpage

\section{Graph Construction}\label{sec:appendix_graphs}

\begin{figure}[h]
    \centering
    \includegraphics[width=\textwidth]{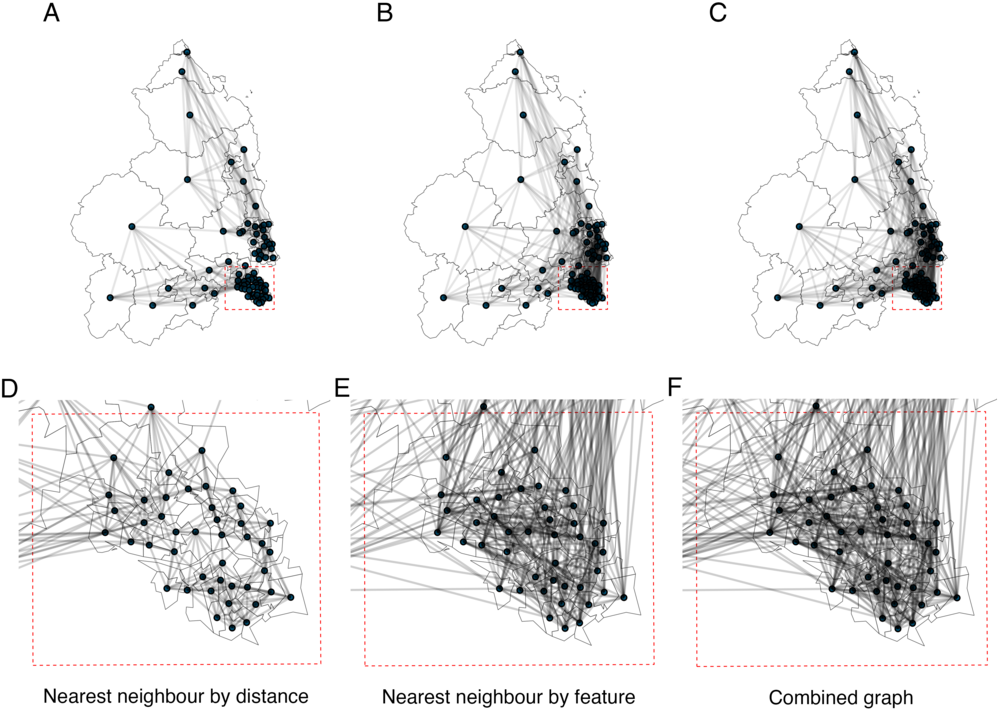}

    \caption{\textbf{Comparison of network connectivities.} Visualisation of different graphs constructed by spatial proximity (Panel A, D), feature-space similarity (Panel B, E), and the resulting combined graph formed by their union (Panel C, F). Detail of the most densely populated area in the geography is shown in Panels D, E, F.}
    \label{fig:Graph_Consruction}
\end{figure}

Figure \ref{fig:Graph_Consruction} shows the graphs constructed in Section \ref{sec:network_architecture}.

\clearpage

\section{Further Details on Model Training}\label{sec:appendix_model_training}

\subsection{Data Preprocessing and Scaling}\label{sec:appendix_data_preprocessing}
Prior to model training and testing, data inputs are preprocessed and partitioned into training, validation, and hold-out test sets. The location-specific socio-demographic features \(\mathbf{X}\) are rescaled using min–max normalisation to the unit interval for each feature across all MSOAs in the training set.
I.e. the columns are scaled onto the interval \([0,1]\).
The same scaling parameters are then applied to the validation and test sets to prevent data leakage.
Similarly, the columns of the \(\mathbf{Y}^{(t)}\) and \(\matrixt{W}\) are standardised to zero mean and unit variance within the training set. 
These transformation statistics are then reused to standardise the validation and test data.

\subsection{Numerical Stability Considerations}
\label{sec:appendix_numerical_stability}
To ensure stable training and prevent numerical instabilities when parameterising the NB distribution, we introduce small adjustments to the network outputs and probability values.

\subsubsection*{Preventing Underflow} Although the \(\mathrm{softplus}\) activation is strictly positive, it asymptotically approaches 0 for large negative inputs. Under finite-precision arithmetic, this can lead to \emph{underflow}, where very small values are effectively rounded to 0 during. A near-zero \(\mu\) destabilises the parametrisation of the NB distribution in the calculation of \(r\) which becomes unstable as \(\mu \to 0\). To mitigate this, we add a small constant offset of \(10^{-3}\) to \(\mu\) after applying \(\mathrm{softplus}\), ensuring that \(\mu\) remains above zero with a \emph{safety buffer} preventing underflow while preserving its positivity.

\subsubsection*{Preventing Residual Collapse}
The NB formulation requires that \(\sigma^2 > \mu\) to capture overdispersion. As we define \(\sigma^2 = \mu + \delta\), it is important that \(\delta\) does not collapse to values close to 0 as this would violate the overdispersion property and make the dispersion parameter \(r\) unstable. To enforce a sufficient separation between \(\sigma^2\) and \(\mu\), we add a constant offset of \(1.0\) to the \(\delta\) output after applying \(\mathrm{softplus}\), guaranteeing that the variance is always markedly greater than the mean.

\subsubsection*{Probability Bounds}
When parametrising the NB distribution, the probability parameter \(p\) is constrained within \([10^{-8}, 1 - 10^{-8}]\), preventing \(p\) from approaching extreme values of exactly \(0\) or \(1\).

\clearpage

\section{Further Details on Model Evaluation}\label{sec:appendix_model_comparison}

\subsection{Scoring Rules}

\subsubsection{Continuous Ranked Probability Score}

The Continuous Ranked Probability Score (CRPS) introduced by \cite{matheson_scoring_1976} is a scoring rule widely used to evaluate the quality of probabilistic forecasts against observed outcomes. We adopt the definition of CRPS given by \cite{gneiting_strictly_2007}, but drop the minus sign so that it is treated as an error to be minimised (taking non-negative values, with 0 corresponding to a perfect forecast). For a forecast represented by a cumulative distribution function (CDF) $F$ and a scalar observation $y$, the CRPS is given by:
\begin{equation*}
    \text{CRPS}(F, y) = \int_{-\infty}^{\infty} \left( F(x) - \mathbf{1}(x \geq y) \right)^2 dx,
\end{equation*}
where $\mathbf{1}(\cdot)$ denotes the indicator function.

Several approximations to CRPS have been proposed when the exact CDF is not available. We use the following approximation, identified as the optimal choice when samples from the probability distribution are available \cite{zamo_estimation_2018}. Given $S$ forecast samples $x_1, \ldots, x_S$, the CRPS can be estimated as (compare with equation~\ref{eq:crps_kernel}):
\begin{equation}\label{eq:crps_approx}
    \text{CRPS}(F, y) \approx \frac{1}{S} \sum_{s=1}^{S} |x_s - y| - \frac{1}{2S^2} \sum_{s=1}^{S} \sum_{s'=1}^{S} |x_s - x_{s'}|.
\end{equation}

This approximation remains valid when, instead of samples, we use the values of $F$ at the $K$ quantile levels of the form:
\begin{equation}\label{eq:quantile_form}
    \tau_j=\tfrac{j - 0.5}{K},\:\:j=1,\dots,K, 
\end{equation}
\cite{brocker_evaluating_2012}. This ensemble of quantiles has been shown to improve the accuracy of the CRPS approximation \cite{zamo_estimation_2018}.

\subsubsection{Energy Score}

The Energy Score (ES) is a scoring rule for evaluating multivariate probabilistic forecasts. We adopt this measure when evaluating GENIE's performance at a spatial level. Let $\mathbf{y} \in \mathbb{R}^M$ denote the observed burden across $M$ MSOAs at a time $t$, and let $\mathbf{X},\mathbf{X}'\in \mathbb{R}^M$ be independent random vectors drawn from the forecast with multivariate CDF $\mathbf{F}$. The ES is defined by:
\begin{equation*}
\text{ES}(\mathbf{F},\mathbf{y}) = \mathbb{E}_{\mathbf{F}}\|\mathbf{X}-\mathbf{y}\|_2 - \frac{1}{2}\mathbb{E}_{\mathbf{F}}\|\mathbf{X}-\mathbf{X}'\|_2,
\end{equation*}
where $\|\cdot\|_2$ denotes the Euclidean distance \cite{gneiting_assessing_2008}. For $S$ forecast samples $\mathbf{x}_1, \dots, \mathbf{x}_S\in \mathbb{R}^M$, the ES is approximated as \cite{gneiting_assessing_2008}:
\begin{equation*}
\text{ES}(\mathbf{F},\mathbf{y}) \approx \frac{1}{S} \sum_{s=1}^{S} \|\mathbf{x}_s - \mathbf{y}\|_2 - \frac{1}{2S^2} \sum_{s=1}^{S} \sum_{s'=1}^{S} \|\mathbf{x}_s - \mathbf{x}_{s'}\|_2.
\end{equation*}
The first term measures the mean distance between forecast samples and the observation, while the second term accounts for the internal dispersion among samples. Lower ES values indicate forecasts that are both accurate and well calibrated in multivariate space.

\subsection{Benchmark Models}

\subsubsection{The Mantis Model}

\subsubsection*{Implementation of Mantis for Daily MSOA-Level Forecasting}

We used the Mantis foundation model for disease forecasting \citep{dudley2025mantis}\footnote{Source code available at \url{https://github.com/carsondudley1/Mantis}.} to generate probabilistic forecasts of disease burden for each MSOA (Middle Layer Super Output Area) within the study area. Mantis uses ``a sequence-to-sequence neural network with convolutional and transformer-based components'' to perform time series forecasting, trained on weekly-aggregated epidemic data. It produces quantile predictions across nine probability levels and supports both univariate and covariate-conditioned inference. Full architectural and training details are given in the original paper; here we describe our specific experimental setup.

\subsubsection*{Experimental Configurations}

We ran two experiments for each simulation and each MSOA within the hold-out testing dataset using Mantis' pre-trained 8-step forecasting model with covariate support (\texttt{mantis\_8w\_cov.pt}). Each simulation contains a time series of daily deaths and daily hospitalisations for every MSOA in the study region. For these experiments we maintained the daily resolution; no weekly aggregation was applied. We did this as our primary focus is the development and evaluation of forecasting models at a daily temporal resolution. In the first experiment, the target variable was daily deaths with the covariate set to each MSOA's own daily hospitalisations. In the second, the target variable was daily hospitalisations with the covariate set to each MSOA's own daily deaths.

The historical target series was taken as all daily observations up to (but not including) the first timestep of the prediction horizon. The covariate history was constructed identically from the MSOA's own covariate column, covering the same temporal window. These were used to initialise the Mantis model object. The \texttt{.predict()} function was then called on this model object to produce a 9-quantile forecast across 8 timesteps. The output tensor is of shape $[8 \times 9]$, representing 8 forecast steps $\times$ 9 quantile levels $\tau \in \{0.05, 0.10, 0.25, 0.40, 0.50, 0.60, 0.75, 0.90, 0.95\}$ \citep[Appendix B.5.2]{dudley2025mantis}. For each prediction step $h \in \{0, \ldots, 7\}$, the 9 quantile predictions were stored alongside the ground truth value at that timestep, the simulation ID, the MSOA identifier, and the window start index.

\subsubsection*{Sliding Window Evaluation}

To obtain comprehensive coverage of Mantis forecasting across the entire data length, we employed a sliding window evaluation with stride 1. For each simulation, the set of valid prediction origins was computed as every timestep $t$ such that $t \geq 1$ (ensuring at least one historical observation) and $t + 8 \leq T$ (ensuring the full 8-step forecast horizon falls within the observed data). This produced a set of overlapping evaluation windows, one for each possible prediction origin.

Each simulation in the testing dataset has a length of $T = 180$ daily timesteps. With a forecast horizon of 8 steps and a minimum history requirement of 1 timestep, the valid prediction origins range from $t = 1$ to $t = 172$, yielding 172 evaluation windows per MSOA per simulation.

\subsubsection*{Probabilistic Scoring}

Each forecast was assessed at every individual prediction step using the Continuous Ranked Probability Score (CRPS), computed via a quantile-based approximation in (\ref{eq:crps_approx}). The quantiles provided by Mantis are unevenly spaced, a format unsuitable for existing CRPS approximations \cite{zamo_estimation_2018}. To obtain an ensemble of quantiles as in (\ref{eq:quantile_form}), we linearly interpolate the quantiles provided by Mantis. With $K=10$, the lowest and highest quantiles in the ensemble correspond to $0.05$ and $0.95$, the minimum and maximum provided by Mantis, so no extrapolation is needed. To broaden the range of quantiles and improve the approximation, we alternatively use $K=30$. In this case, the minimum and maximum quantiles fall outside the range provided by Mantis, so we estimate them by linearly extrapolating from the first and last two Mantis quantiles, respectively. Since this extrapolation relies on strong assumptions about the tails of the distribution, we compute the CRPS using both approaches and report whichever is more favourable to Mantis.

\subsubsection{The \texttt{hhh4} Model}

Forecasts are generated using the \texttt{hhh4} modelling framework \cite{meyer2017spatio} provided in the \texttt{R} package \texttt{surveillance}. It assumes that the case counts $Y_{i,t}$ for MSOA $i$ at time $t$ follow a Negative Binomial distribution with mean $\mu_{it}$ and overdispersion parameter $\psi$, where the mean decomposes additively into an endemic component $e_i\nu_t$, an autoregressive component $\lambda_i Y_{i,t-1}$, and a spatio-temporal neighbourhood component $\phi_i \sum_{j \neq i} w_{ji} Y_{j,t-1}$:
\begin{equation*}
    \mu_{it} = e_i \nu_t + \lambda_i Y_{i,t-1} + \phi_i \sum_{j \neq i} w_{ji} Y_{j,t-1}.
\end{equation*}
with an offset $e_i$ in the endemic component. The weights $w_{ji}$ are distance-decaying neighbourhood weights between locations $i$ and $j$, with a maximum neighbouring order of 5, beyond which the weights are set to 0. They are defined in terms of the adjacency order $o_{ji}$ in the neighbourhood graph of the regions, $w_{ji} = o_{ji}^{-d}$, where $d$ is estimated directly from the data. The graph is constructed so that only neighbouring regions are connected.

The term $\nu_t$ is a log-linear predictor incorporating an intercept shared across locations, and a time seasonality term with a given frequency $\omega$:

\begin{align*}
    \log(\nu_{t}) &= \alpha^{(\nu)} + \gamma \sin(\omega t) + \delta \cos(\omega t).
\end{align*}

The endemic component $\lambda_i$ and neighbourhood component $\phi_i$ are also log-linear predictors with an intercept and MSOA-level covariates selected from the JUNE parameters listed in Table \ref{tab:static_features}. We exclude those related to number of households, longitude, and latitude, resulting in four features being included: deprivation $x_{\text{imd}}$, population density $x_{\text{den}}$, number of students $x_{\text{st}}$, and number of care home residents $x_{\text{ch}}$.

\begin{align*}
    \log(\lambda_i) &= \alpha^{(\lambda)} + \beta_{\text{imd}}^{(\lambda)} \text{logit}(x_{\text{imd}})
    + \beta_{\text{den}}^{(\lambda)} \log(x_{\text{den}}) + \beta_{\text{st}}^{(\lambda)} \log(x_{\text{st}}) + \beta_{\text{ch}}^{(\lambda)} \log(x_{\text{ch}}),\\
    \log(\phi_i) &= \alpha^{(\phi)} + \beta_{\text{imd}}^{(\phi)} \text{logit}(x_{\text{imd}})
    + \beta_{\text{den}}^{(\phi)} \log(x_{\text{den}}) + \beta_{\text{st}}^{(\phi)} \log(x_{\text{st}}) + \beta_{\text{ch}}^{(\phi)} \log(x_{\text{ch}}),
\end{align*}

Inference is performed using the quasi-Newton algorithm to maximise the log-likelihood, as implemented in the \texttt{surveillance} package. The tolerance for convergence is set to 1e-5, and the optimiser is allowed a maximum of 100 iterations. Forecasting is carried out by refitting the model for every simulation and every timestep, using only the data observed up to that timestep. Once the model has converged, forecasts are generated for a 60-day window ahead.

Since it is unclear which of these MSOA-level parameters would favour \texttt{hhh4}, we perform a forward variable selection previous to forecasting: starting from the four candidate variables, we added the one that produced the best AIC improvement, then repeated the process on the remaining variables until no further addition improved the AIC. This process is performed using the whole time series, before performing forecasts. The periodicity of the seasonal component $\omega$ is likewise inferred from the whole time series, by aggregating observations across the entire geography into a single time series and applying a fast Fourier transform to determine the dominant frequency (period). Performing these analyses on the whole time series gives \texttt{hhh4} an advantage over GENIE, as it effectively provides a view of future observations when forecasting. We include this advantage for \texttt{hhh4}, since without it, it is difficult to determine which features to incorporate into the model.

\clearpage

\section{Peak prediction}

\subsection{Spatial and temporal aggregation}\label{sec:methods_peak_aggregation}

Let $\mathcal{L}$ denote the set of LTLAs and $\mathcal{M}_{l}\subseteq\{1,\dots,M\}$ the set of MSOAs contained in the LTLA $l$; the $\{\mathcal{M}_{l}\}_{l\in\mathcal{L}}$ partition the MSOAs. 
The day $t$ burden for LTLA $l$ is
\begin{equation}\label{eq:ltla_aggregation}
   z^{(t)}_{l}=\sum_{m\in\mathcal{M}_l}(\matrixt{Y})_{m,1},\qquad l\in\mathcal{L}.
\end{equation}
\noindent
where \((\matrixt{Y})_{m,1}\) indicates the number of hospitalisations in MSOA \(m\) on day \(t\).
Each day index $t$ corresponds to a calendar date.
Write $\omega(t)$ for the ISO week containing the date \(t\), and $\mathcal{T}_w=\{t:\omega(t)=w\}$ for the set of days of week $w$ present in the series.
\emph{We use the weekly median to smooth the stochasticity in \(z_{l}^{(t)}\) over time, and at this geographic scale, also day of week effects}:
\begin{equation}\label{eq:weekly_median}
    m^{(w)}_{l}=\operatorname{median}\left\{z^{(t)}_{l}:t\in\mathcal{T}_w\right\}.
\end{equation}
The magnitude reported for a peak is defined as the weekly sum of the number hospitalisations in that week,
\begin{equation}\label{eq:weekly_sum}
   s^{(w)}_{l}=\sum_{t\in\mathcal{T}_w}z^{(t)}_{l},
\end{equation}
Timing and height are always read from the same week $w$.

Our series do not naturally begin or end on a Monday, so a simulation's first and last ISO weeks may be only partially observed and statistics are therefore not comparable with those of complete weeks.
Writing $\Delta_w=\max\mathcal{T}_w-\min\mathcal{T}_w$ for the span in days of the observations falling in week $w$, we discard the first week if $\Delta_{w_{\min}}<6$ and the last if $\Delta_{w_{\max}}<6$.
The retained weeks are re-indexed $w=1,\dots,W$.

\subsection{Definition of a peak}\label{sec:methods_peak_definition}

Let $\delta_{l}^{(w)}=m^{(w)}_{l}-m^{(w-1)}_{l}$ for $w=2,\dots,W$ denote the change between consecutive weeks, and define the direction of the change
\begin{equation}\label{eq:direction}
   \sigma_{l}^{(w)}=
   \begin{cases}
        \text{sign}\left(\delta_{l}^{(w)}\right), & \delta_{l}^{(w)}\neq0,\\
       \sigma_{l}^{(w-1)}, & \delta_{l}^{(w)}=0,
   \end{cases}
\end{equation}
so that a plateau inherits the direction that led into it and (by choice) a turning point is placed 
at the last week of a plateau.

A turning point requires the time series to have gone from increasing to decreasing, or vice versa.
We allow the definition to demand that the approach direction is repeated over more than a single week or more.

Let $\left[a_w,w\right]$ be the longest interval of weeks over which the sign of change is constant and ending at $w$, that is $a_{w,l}=\min\{a\le w:\sigma_{l}^{(j)}=\sigma_{l}^{(w)}\ \forall j\in[a,w]\}$, and let
\begin{equation}\label{eq:run_length}
    r_{l}^{(w)}=\left|\left\{j\in\left[a_{w,l},w\right]:\text{sign}\left(\delta_{l}^{(j)}\right)=\sigma_{l}^{(w)}\right\}\right|
\end{equation}
count the \emph{strict} moves in the interval \([a_{w,l},w]\).
Note that because plateau weeks are absorbed into the run by \eqref{eq:direction} but, not being changes, they do not contribute to $r_{l}^{(w)}$.
I.e. the run counts the total number of changes in a particular direction, not just the number of weeks in which there hasn't been a reversal of direction, which is what would be captured by \(w-a_{w,l}\).

Fixing a minimum run length $\rho\ge1$, week $w$ is defined to be a peak if
\begin{equation}\label{eq:peak_def}
   \sigma_l^{(w)}=+1,\qquad \sigma_l^{(w+1)}=-1,\qquad r_l^{(w)}\ge\rho,\qquad 1<w<W,
\end{equation}
and defined to be a trough if $\sigma_l^{(w)}=-1$, $\sigma_l^{(w+1)}=+1$ and $r_l^{(w)}\ge\rho$ over the same range of $w$.
The restriction $1<w<W$ excludes the first and last retained weeks, for which the direction is undefined on one side.
Peaks are numbered $p=1,2,\dots$ chronologically within each (simulation and for each LTLA's) time series, and the height of peak $p$ is $s^{(w_p)}_{l}$, where $w_p$ is the week in which the peak occurs.

Two choices of $\rho$ are of particular interest.
Setting $\rho=1$ gives a simple decision rule: a single increase followed by a single decrease marks a peak.
Setting $\rho=2$ gives a rule which we use in sensitivity analysis, setting \(\rho=2\) requires two consecutive increases before a decrease that defines a peak, and correspondingly two decreases before the increase that defines a trough.
Using \(\rho=2\) suppresses any turning points of ``low-amplitude'' which the simple rule finds but which might plausibly be attributed to noise, at the cost of discarding potentially real but short-lived peaks and troughs.

\subsection{Application to forecast samples}\label{sec:methods_peak_samples}

As described in Section~\ref{sec:forecast-rollout}, a forecast initiated on day $t_0$ produces $J$ sampled trajectories over the horizon $t\in\{t_0,\dots,t_0+H\}$ with $H=59$ (producing a forecast over a \(60\) day interval).
Because a turning point requires information from at least one week before and after the peak week, a sample evaluated over the horizon alone could never identify a peak occurring near $t_0$, and the first weeks of every forecast would be impossible to evaluate consistently.
We therefore evaluate each sampled trajectory extended to include the observed history up to the day before forecast initiation and the sampled trajectory thereafter,
\begin{equation}\label{eq:spliced_series}
    \tilde{z}^{(t),j}_{l}=
    \begin{cases}
        \displaystyle\sum_{m\in\mathcal{M}_\ell}(\matrixt{Y})_{m,1}, & 1\le t<t_0,\\[4mm]
        \displaystyle\sum_{m\in\mathcal{M}_\ell}(\matrixt{Y}_j)_{m,1}, & t_0\le t\le t_0+H-1,
    \end{cases}
\end{equation}
where, recall that \((\matrixt{Y})_{m,1}\) is the observed number of hospitalisations in MSOA \(m\) on day \(t\), and \((\matrixt{Y}_j)_{m,1}\) is the number of predicted hospitalisations in MSOA \(m\) on day \(t\) in the \(j\)th sampled trajectory.
For each of the extended LTLA trajectories, \(\tilde{z}^{(t),j}_{l}\) we apply \eqref{eq:weekly_median}--\eqref{eq:peak_def} to get the peak statistics.
This gives us two additional useful properties.
First, a sample may legitimately place a peak in a week that begins before $t_0$: the week containing the initiation day is typically split between observation and forecast, and the model is credited with the turning point its trajectory implies there. 
Second, if the entire multi-week neighbourhood required to define a peak precedes $t_0$ then that peak is reconstructed from observed data alone and is recovered identically by every sample.

\subsection{Evaluation}\label{sec:methods_peak_evaluation}
Performance is reported separately for each peak index $p$, and only for the JUNE simulations in which there is a $p$-th peak to be predicted within the simulation window of 180 days.
For each LTLA $l$ we write
\begin{equation}\label{eq:peak_eval_set}
    \mathcal{S}_{p,l}=\left\{\text{simulations whose series }m^{(w)}_{l}\text{ has a }p\text{th peak for some week \(w\)}\right\},
\end{equation}
so that a simulation in which the epidemic in LTLA $l$ does not produce a $p$-th peak contributes nothing to the $p$-th peak results.

Consider a simulation in $\mathcal{S}_{p,l}$ and an forecast initiation day $t_0$.
Let $w_p$ be the week of the $p$-th simulated peak, $\tau_p$ the midpoint date of that week and $s_p=s^{(w_p)}_{l}$ its height.
Given a set of samples from the posterior predictive distribution, extended as described in Section~\ref{sec:methods_peak_samples}, let $\hat{w}^{\,j}_p$ and $\hat{s}^{\,j}_p$ denote the week and height of the $p$-th peak of the \(j\)th sample using Equations~\eqref{eq:weekly_median}--\eqref{eq:peak_def}.

The performance in peak prediction is reported against the offset
\begin{equation}\label{eq:days_after_peak}
    u_p=t_0-\tau_p,
\end{equation}
the number of days between the day the forecast was initiated and the day the peak actually occurred.
The offset, \(u_p\), is negative when the peak is in the future and positive once the peak is in the past, i.e. when it is before \(t_0\).
This means we show results  for $-H+11\leq u_p\leq 10$ days.
The limits are determined by the earliest date from which the peak can be predicted at all (i.e. One full week after the peak week as recognising a peak under \eqref{eq:peak_def} requires a week of decrease after it) and the first day we are guaranteed to determine it to have been a peak week (once we have observed data one full week after the peak).

Recall that we use four quantities to assess peak prediction:
\begin{enumerate}
    \item \(\Prob{\text{peak in horizon}}\), the predicted probability of a peak in the forecast horizon;
    \item \(\Prob{\text{correct week}}\), the predicted probability of a peak falling in the correct week;
    \item the absolute error (AE) and the  relative absolute error (RAE) in the peak number of hospitalisations, conditional upon there being a peak in the forecast;
    \item and the bias in the estimate of the peak week, conditional upon there being a peak in the forecast.
\end{enumerate}
When assessing the performance in predicting the \(p\)th peak, we consider each posterior sample $j$ of a forecast initiated at time $t_0$ in a single LTLA \(l\) against a JUNE simulation in $\mathcal{S}_{p,l}$, and collect all such posterior samples $u$ days from the $p$-th peak into the set
\begin{equation}\label{eq:record_set}
    \mathcal{D}_p(u)=\left\{\left(\text{simulation},l,t_0,j\right):\ \text{simulation}\in\mathcal{S}_{p,l},\ u_p=u,\ 1\le j\le J\right\}.
\end{equation}
We compute quantiles or proportions over this set which contains all the relevant simulations and posterior samples.
A simulation enters in proportion to the number of its forecast windows that sit $u$ days from the peak. 
We also write define a useful subset:
\begin{equation}\label{eq:record_set_with_peak}
    \mathcal{D}^{+}_p(u)=\left\{\left(\text{simulation},l,t_0,j\right)\in\mathcal{D}_p(u):\ \text{sample }j\text{ has a }p\text{th peak}\right\}.
\end{equation}
A posterior sample whose series contains no turning point at all reports no peak of any number, and so lies in $\mathcal{D}_p(u)$ but not in $\mathcal{D}^{+}_p(u)$.

The first quantity used to assess peak performance is the proportion of samples that contain the peak,
\begin{equation}\label{eq:peak_containment}
    \widehat{\Pr}\left(\text{peak }p\text{ present}\right)=\frac{\left|\mathcal{D}^{+}_p(u)\right|}{\left|\mathcal{D}_p(u)\right|}.
\end{equation}
We use this to see whether the model anticipates a $p$-th peak.

The second is the proportion of samples that place the peak in the correct ISO week,
\begin{equation}\label{eq:peak_week_accuracy}
    \widehat{\Pr}\left(\text{correct week}\right)=\frac{1}{\left|\mathcal{D}_p(u)\right|}\sum_{\mathcal{D}_p(u)}\mathbf{1}\left[\hat{w}^{\,j}_p=w_p\right].
\end{equation}
We take this over $\mathcal{D}_p(u)$, it counts a sample that forecast no $p$-th peak as incorrect and is a proportion of every forecast issued at $u$.

The third quantity is the error in the height of the peak. For each record in $\mathcal{D}^{+}_p(u)$, we compute either the absolute error (AE) or relative absolute error (RAE):
\begin{align}
    \mathcal{E}_{\mathrm{abs}}\left(\hat{s}^{\,j}_p, s_p\right) &= \left|\hat{s}^{\,j}_p - s_p\right|, \label{eq:height_abs}\\
    e^{\,j}_p = \mathcal{E}_{\mathrm{rel}}\!\left(\hat{s}^{\,j}_p, s_p\right) &= \frac{\left|\hat{s}^{\,j}_p - s_p\right|}{s_p}, \qquad s_p > 0. \label{eq:peak_height_error}
\end{align}
The RAE normalises the error across LTLAs and simulations, where a value of $1$ represents an error of the same magnitude as the true peak. (Simulated peak heights with $s_p = 0$ are excluded from this metric.)

The fourth quantity is the error in the timing of the peak for each record in $\mathcal{D}^{+}_p(u)$, defined as the signed difference in timing:
\begin{equation}\label{eq:peak_week_error}
    d^{\,j}_p = \hat{w}^{\,j}_p - w_p.
\end{equation}

To aggregate these error metrics across forecasts initiated $u$ days from the peak, we take the mean over all relevant posterior predictive samples in $\mathcal{D}^{+}_p(u)$. Specifically, we evaluate peak height accuracy using the \emph{mean relative absolute error} (MRE):
\begin{equation}\label{eq:mre_height}
    \mathrm{MRE}_p(u) = \frac{1}{\left|\mathcal{D}^{+}_p(u)\right|} \sum_{\mathcal{D}^{+}_p(u)} e^{\,j}_p,
\end{equation}
(with the unnormalised \emph{mean absolute error}, $\mathrm{MAE}$, reported in Appendix~X), and peak timing using the \emph{mean absolute error} (MAE):
\begin{equation}\label{eq:mae_timing}
    \mathrm{MAE}_p(u) = \frac{1}{\left|\mathcal{D}^{+}_p(u)\right|} \sum_{\mathcal{D}^{+}_p(u)} \left|d^{\,j}_p\right|.
\end{equation}
Crucially, these means are calculated directly across all posterior predictive samples in $\mathcal{D}^{+}_p(u)$.

\subsection{Peak Prediction Results}\label{sec:appendix_peaks}

This section collects the peak prediction results summarised in Section \ref{sec:results_peaks}. 
All quantities are those defined in Section \ref{sec:methods_peak_evaluation}, evaluated on hospitalisations aggregated to LTLA and ISO week, and are plotted against the offset $u_p$ of \eqref{eq:days_after_peak} between the forecast initiation day and the day of the peak.
Section \ref{sec:appendix_peaks_availability} explains the range over which \texttt{hhh4} can be evaluated on the first peak, Section \ref{sec:appendix_peaks_main} gives the results for the peak definition used throughout the main text ($\rho=1$), and Section \ref{sec:appendix_peaks_sensitivity} repeats the analysis under the stricter definition ($\rho=2$) as a sensitivity analysis.

\subsubsection{Availability of \texttt{hhh4} Forecasts Before the First Peak}\label{sec:appendix_peaks_availability}

\begin{figure*}[h]
    \centering
    \includegraphics[width=1.0\textwidth]{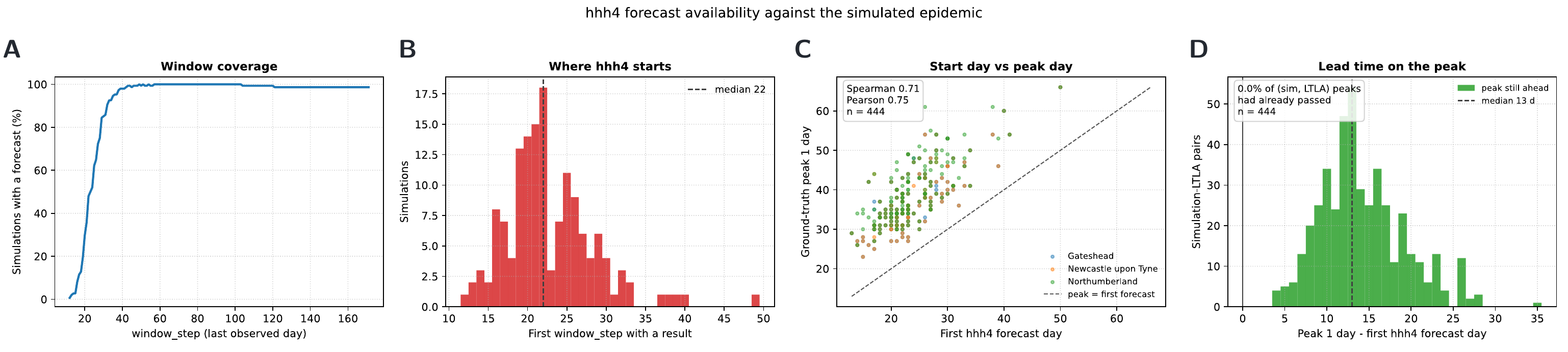}
    \caption{\textbf{Convergence of \texttt{hhh4} estimation relative to the first peak.} (Panel A) Percentage of simulations for which \texttt{hhh4} returns a forecast, as a function of the last observed day of the fitting window. (Panel B) Distribution of the first day on which \texttt{hhh4} returns a forecast, across simulations; the dashed line is the median of 22 days. (Panel C) Day of the first \texttt{hhh4} forecast against the day of the first simulated peak, one point per (simulation, LTLA) pair and coloured by LTLA; the dashed line marks equality, so points above it are pairs in which the peak is still ahead when \texttt{hhh4} first becomes available. (Panel D) Distribution of the resulting lead time, the number of days between the first \texttt{hhh4} forecast and the first peak; the dashed line is the median of 13 days and no pair has a negative lead time.}
    \label{fig:hhh4_availability}
\end{figure*}

When generating forecasts with \texttt{hhh4}, it is refitted at each forecast initiation day using only the data observed up to that day, and early in an epidemic there is a lack of convergence within the iteration budget of Section \ref{sec:appendix_model_comparison}, so no forecast can be generated.
Figure \ref{fig:hhh4_availability} presents the consequence. A forecast is available for half of the simulations only from day 22 onwards, for $95\%$ of them from day 36, and for essentially all of them from day 43, whereas GENIE produces a forecast from the first day of every simulation. The first forecast always precedes the first peak, by a median of 13 days and by at most 35 days. From the second peak onwards, sufficient data has accumulated for \texttt{hhh4} to be available at every offset of interest, and the two models are evaluated over identical ranges.

\subsubsection{Results for the Simple Peak Definition ($\rho=1$)}\label{sec:appendix_peaks_main}

\begin{figure*}[h]
    \centering
    \includegraphics[width=0.95\textwidth]{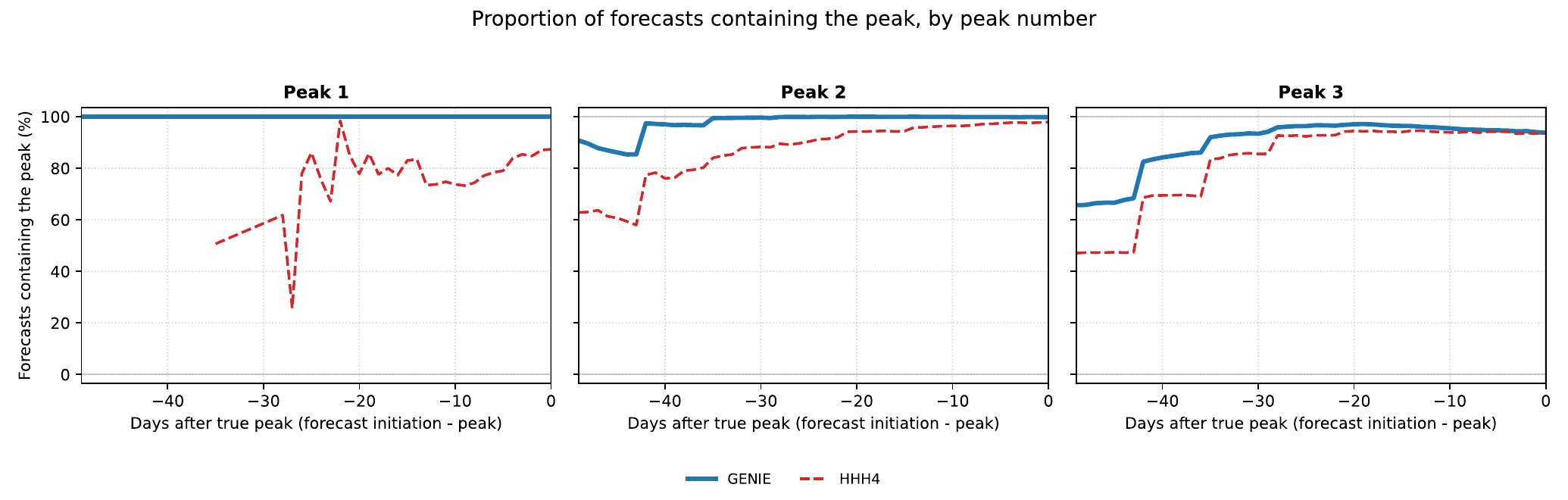}
    \caption{\textbf{Proportion of samples in the forecasts containing the peak.} Percentage of samples whose extended trajectories (as defined in Equation~\eqref{eq:spliced_series}) contains a $p$th peak within the forecast horizon, for GENIE (solid blue) and \text{hhh4} (dashed red), for the first three peaks.}
    \label{fig:peak_containment}
\end{figure*}

Figure \ref{fig:peak_containment} reports the proportion of samples that contain the peak \eqref{eq:peak_containment}, i.e. whether a model anticipates the peak at all, irrespective of when it places it.
Later peaks are harder to anticipate for both models and the gap between them persists: seven weeks ahead of the second peak $90.7\%$ of GENIE samples contain it against $62.8\%$ for \texttt{hhh4}, and for the third peak $65.7\%$ against $47.0\%$. GENIE contains the third peak in $95\%$ of samples four weeks ahead of it, a level \texttt{hhh4} attains only once that peak has passed.

\begin{figure*}[h]
    \centering
    \includegraphics[width=0.95\textwidth]{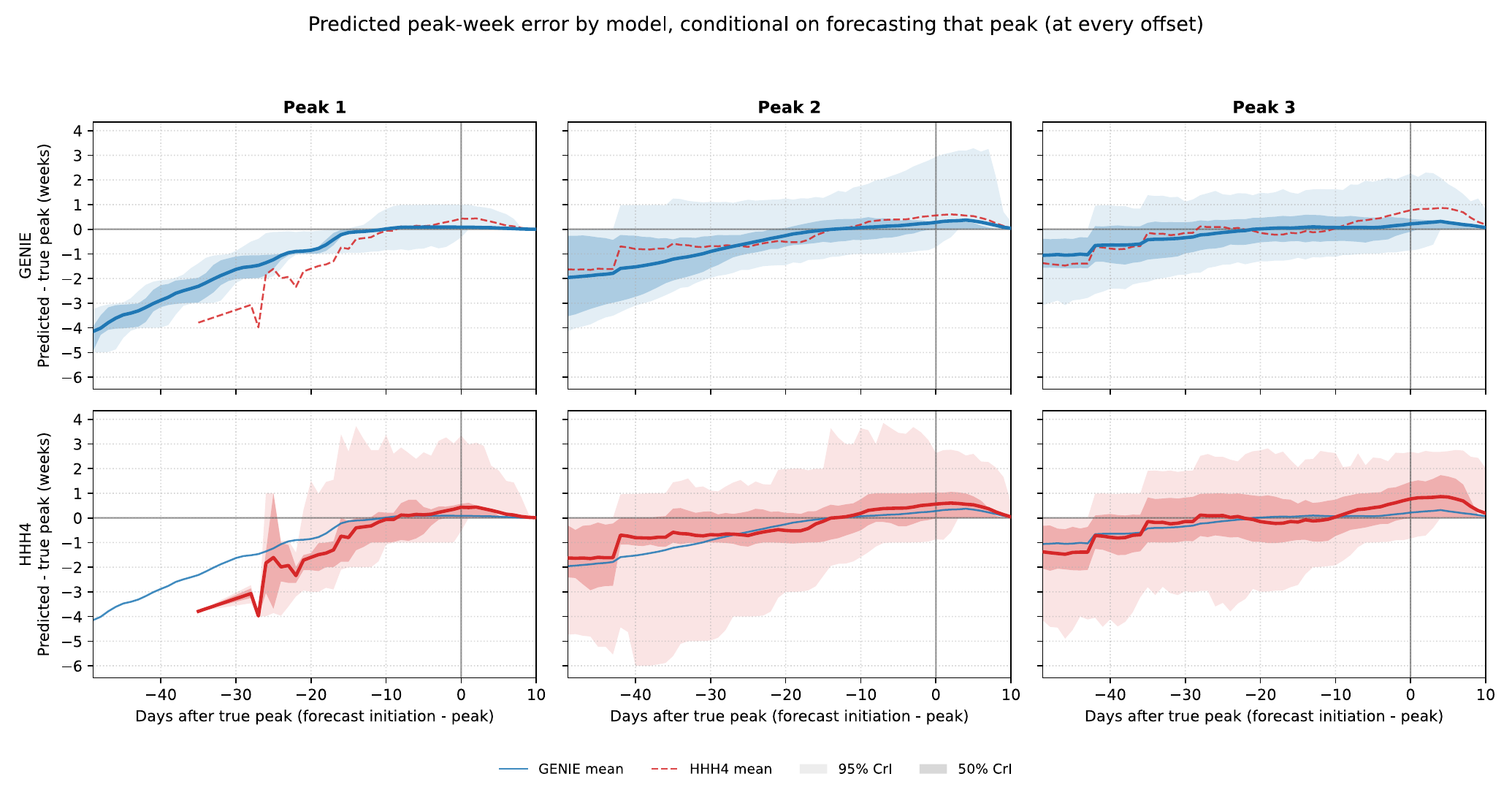}
    \caption{\textbf{Error in the prediction of the peak week.} The number of ISO weeks between the week in which a sample trajectory places peak $p$ and the week in which it occurs --- see Equation~\eqref{eq:peak_week_error} --- so negative values are predictions in which the peak is too early. The top row shows GENIE's bias and the bottom row that of \texttt{hhh4}; both rows show both medians, together with the 50\% and 95\% credible intervals of the model of that row. Only samples containing a $p$-th peak contribute, i.e. this is the bias conditioning upon there being a peak.}
    \label{fig:peak_timing_error}
\end{figure*}

Figure \ref{fig:peak_timing_error} shows the bias in the estimate of the peak timing (defined in Equation~\eqref{eq:peak_week_error}) over the samples that do contain the peak. 
From 14 days before the first peak the median GENIE sample places it on exactly the right day, with a central $50\%$ spanning less than a day ($-0.5$ to $0.0$); the median \texttt{hhh4} sample is still $5.4$ days early at that lead time, its central $50\%$ spanning $-9.2$ to $0.0$ days, and does not reach zero until 8 days before the peak.

\begin{figure*}[h]
    \centering
    \includegraphics[width=0.95\textwidth]{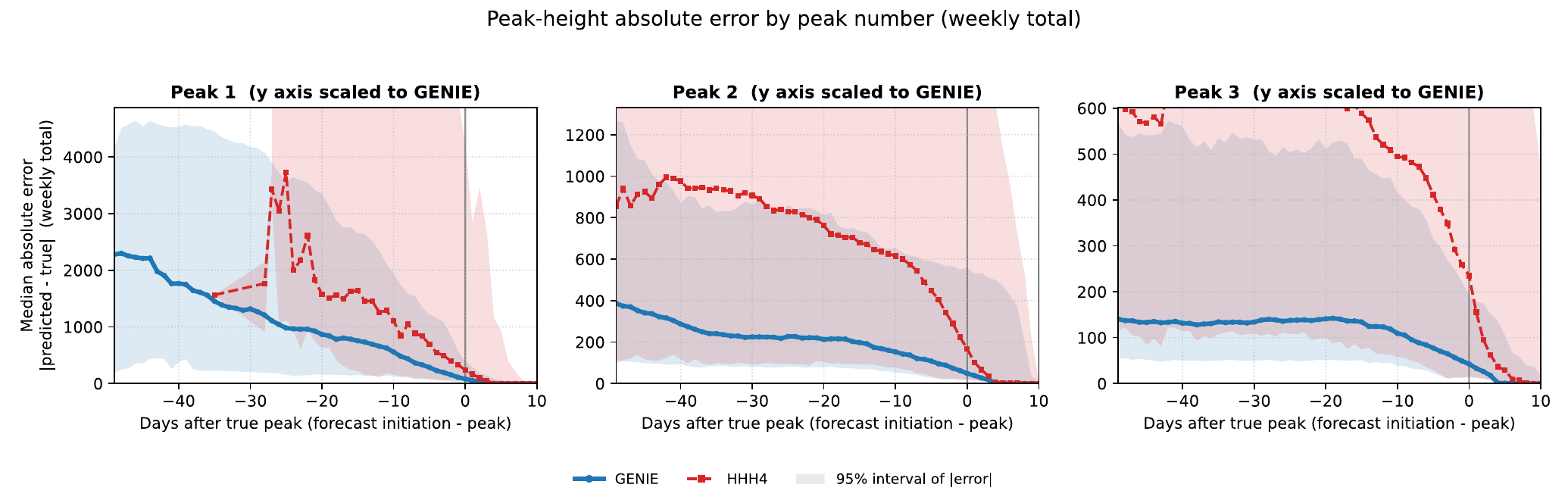}
    \caption{\textbf{Absolute error in the predicted peak height.} Median absolute error (as shown in Equation~\eqref{eq:height_abs}) between the weekly total of hospitalisations at the peak of a sample and at the peak of the simulation, with the shaded band giving the 95\% credible interval of the errors. The vertical axis of each panel is scaled to contain the output of GENIE. Only samples containing a $p$-th peak contribute.}
    \label{fig:peak_height_abs_error}
\end{figure*}

Figure \ref{fig:peak_height_abs_error} reports the absolute error (defined in Equation~\eqref{eq:height_abs}) in the height of the peak, in hospitalisations per week. Because this is pooled across LTLAs of between 115,000 and 316,000 residents it is dominated by the largest of them, which is why the relative error (from Equation~\eqref{eq:mre_height}) of Figure \ref{fig:peak_height_rel_error} is used in the main text; the ordering of the two models is the same under both.

\clearpage

\subsubsection{Sensitivity to the Peak Definition ($\rho=2$)}\label{sec:appendix_peaks_sensitivity}

\begin{figure*}[h]
    \centering
    \includegraphics[width=0.95\textwidth]{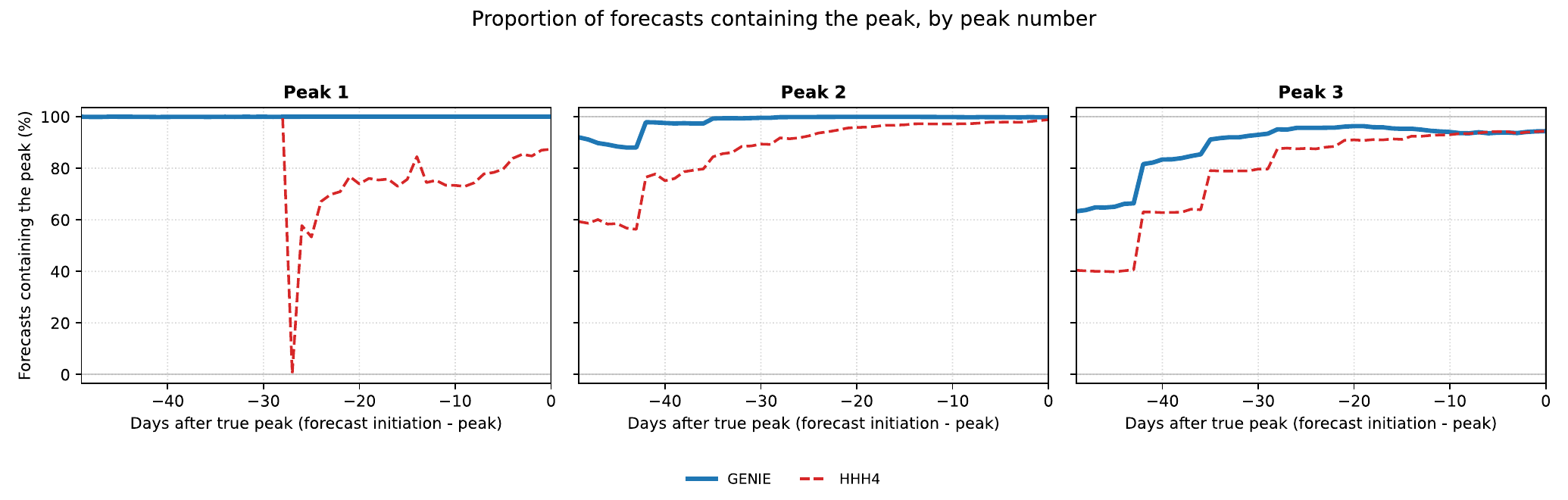}
    \caption{\textbf{Proportion of forecasts containing the peak, $\rho=2$.} As Figure \ref{fig:peak_containment}, with peaks defined by two consecutive weekly increases before the decrease.}
    \label{fig:peak_containment_rho2}
\end{figure*}

\begin{figure*}[h]
    \centering
    \includegraphics[width=0.95\textwidth]{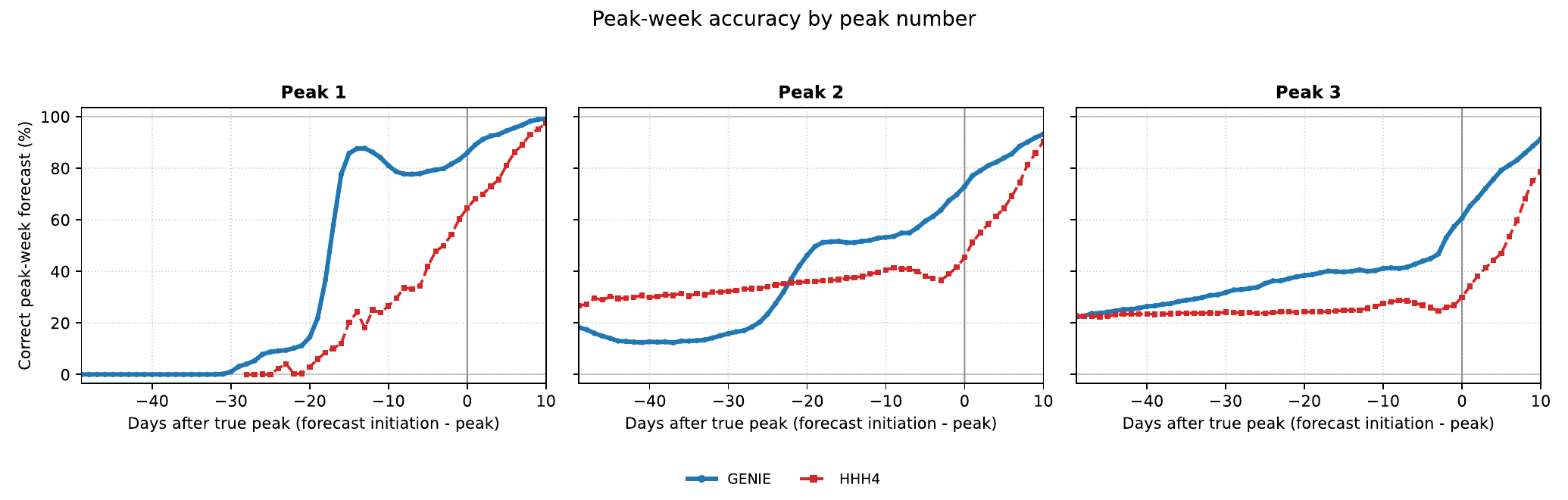}
    \caption{\textbf{Accuracy of the predicted peak week, $\rho=2$.} As Figure \ref{fig:peak_week_accuracy}, with peaks defined by two consecutive weekly increases before the decrease. The \texttt{hhh4} curve for the first peak now begins 28 days before the peak.}
    \label{fig:peak_week_accuracy_rho2}
\end{figure*}

\begin{figure*}[h]
    \centering
    \includegraphics[width=0.95\textwidth]{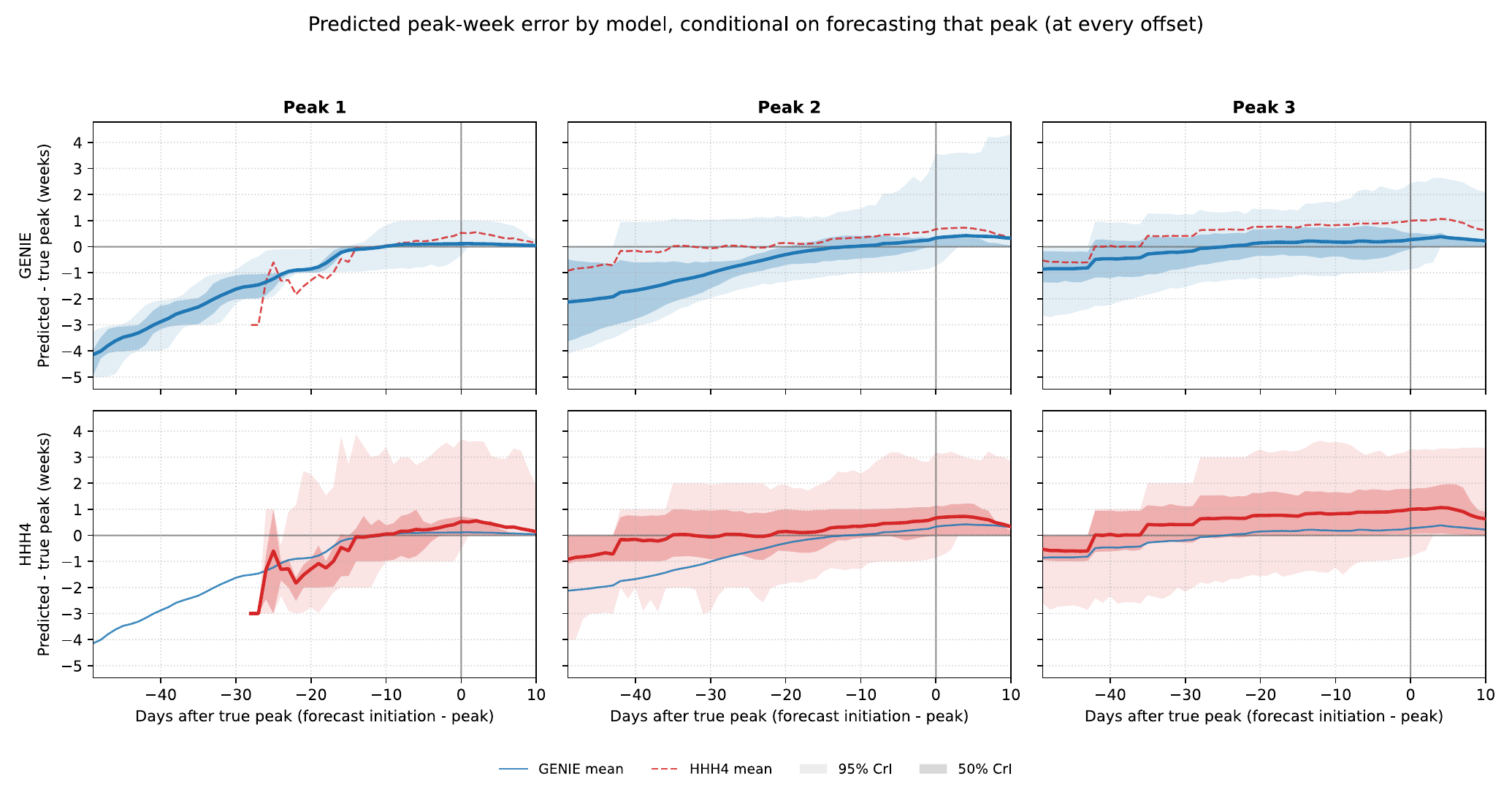}
    \caption{\textbf{Error in the predicted peak week, $\rho=2$.} As Figure \ref{fig:peak_timing_error}, with peaks defined by two consecutive weekly increases before the decrease.}
    \label{fig:peak_timing_error_rho2}
\end{figure*}

\begin{figure*}[h]
    \centering
    \includegraphics[width=0.95\textwidth]{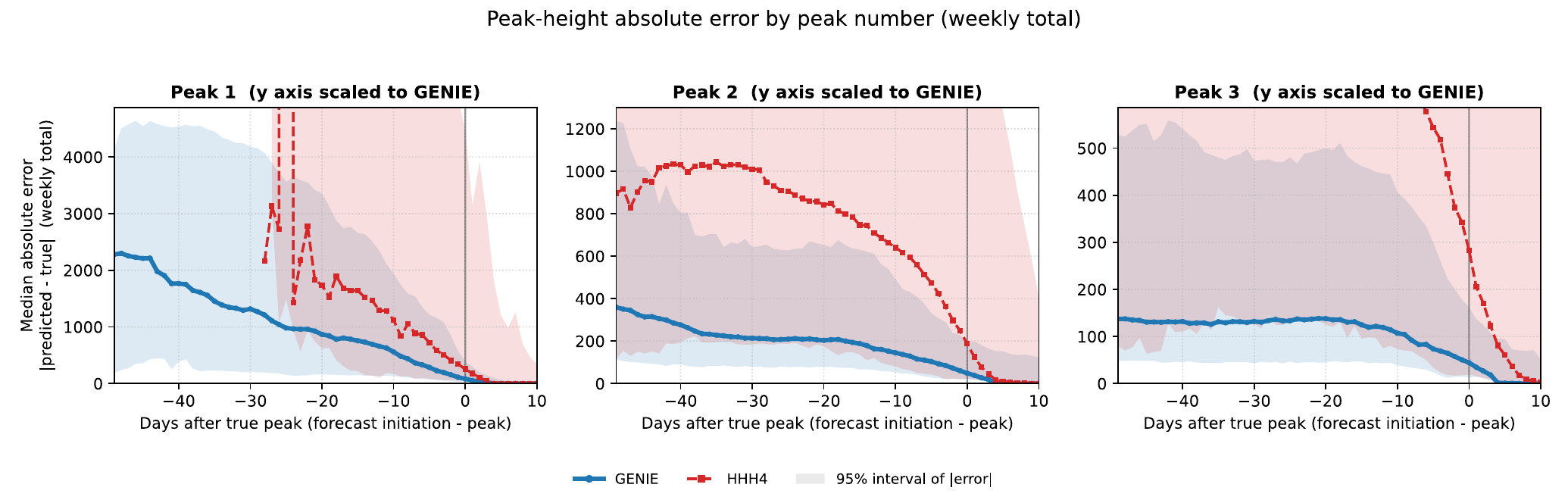}
    \caption{\textbf{Absolute error in the predicted peak height, $\rho=2$.} As Figure \ref{fig:peak_height_abs_error}, with peaks defined by two consecutive weekly increases before the decrease.}
    \label{fig:peak_height_abs_error_rho2}
\end{figure*}

\begin{figure*}[h]
    \centering
    \includegraphics[width=0.95\textwidth]{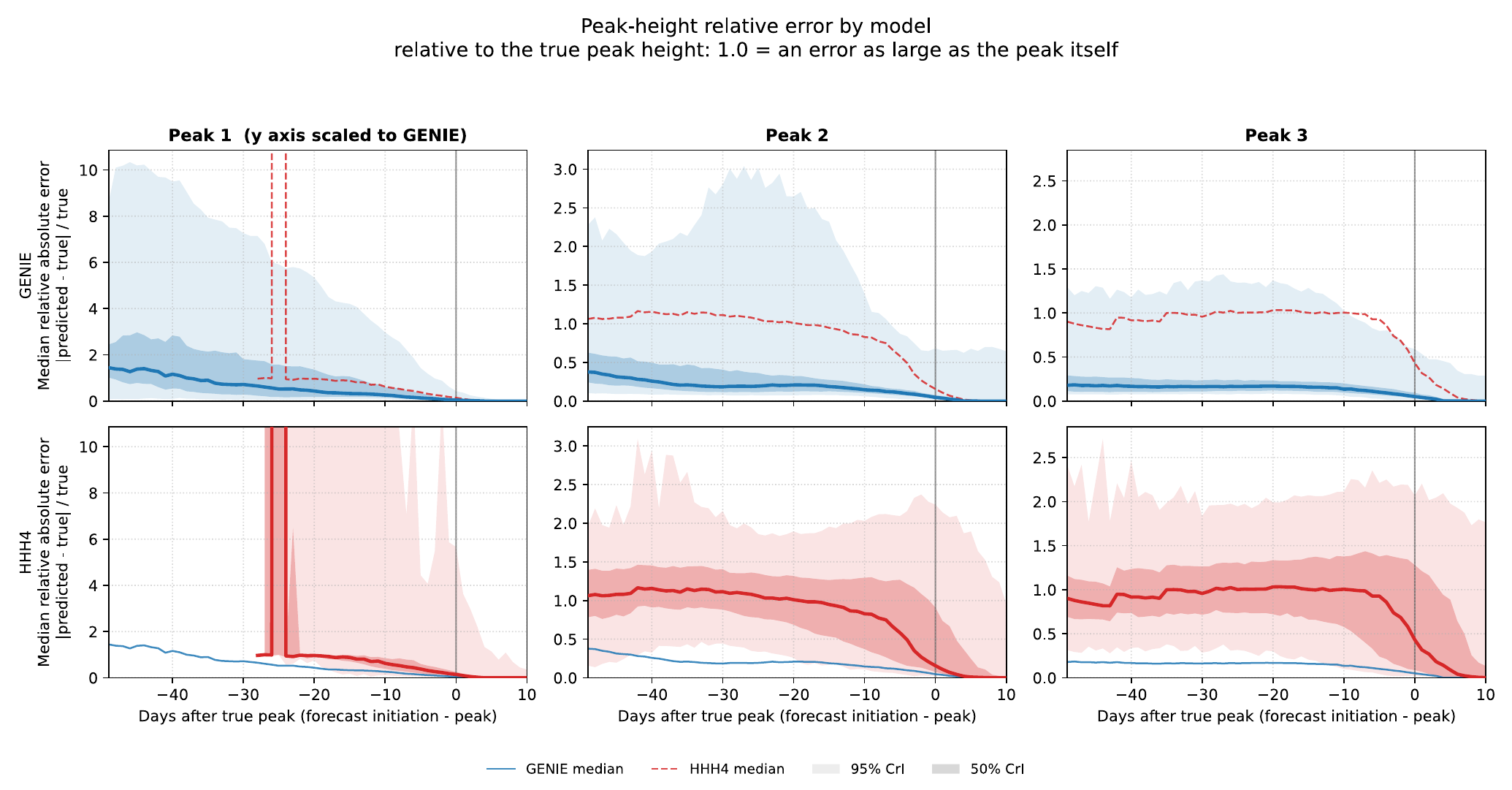}
    \caption{\textbf{Relative error in the predicted peak height, $\rho=2$.} As Figure \ref{fig:peak_height_rel_error}, with peaks defined by two consecutive weekly increases before the decrease.}
    \label{fig:peak_height_rel_error_rho2}
\end{figure*}

The results in the main text pertaining to prediction of peak hospitalisations use the simple peak definition, $\rho=1$, in which a single weekly increase followed by a single weekly decrease marks a peak.
As described in Section \ref{sec:methods_peak_definition}, requiring two consecutive increases instead ($\rho=2$) suppresses low-amplitude turning points that may be artefacts of stochasticity, at the cost of discarding genuine but short-lived peaks.
As a form of sensitivity analysis to this choice of \(\rho=1\), here in Figures \ref{fig:peak_containment_rho2}--\ref{fig:peak_height_rel_error_rho2} we show the results of repeating the whole analysis under this stricter definition in which \(\rho=2\).

The conclusions are unchanged. Across all offsets, the peak-week accuracy of GENIE moves by at most $0.8$ percentage points for the first peak and $4.9$ for the second and third, and its median relative height error by at most $0.003$ and $0.02$ respectively; the peak-week accuracy of GENIE for the first peak is $87.6\%$ two weeks ahead under either definition. The median relative height error of GENIE remains below that of \texttt{hhh4} at every lead time up to the peak and for each of the three peaks, as it is under $\rho=1$, and the timing results are ordered as before. The visible differences are confined to the earliest offsets, where the stricter rule discards a small number of series: the range over which \texttt{hhh4} can be evaluated on the first peak shortens from 35 to 28 days ahead of the peak, and the proportion of samples containing the third peak falls slightly for both models seven weeks ahead (from $65.7\%$ to $63.3\%$ for GENIE and from $47.0\%$ to $40.4\%$ for \texttt{hhh4}), widening rather than narrowing the gap between them.

\clearpage

\section{Further Results}

\subsection{Results on Model Evaluation}

\begin{figure*}[h]
    \centering
    \includegraphics[width=0.95\textwidth]{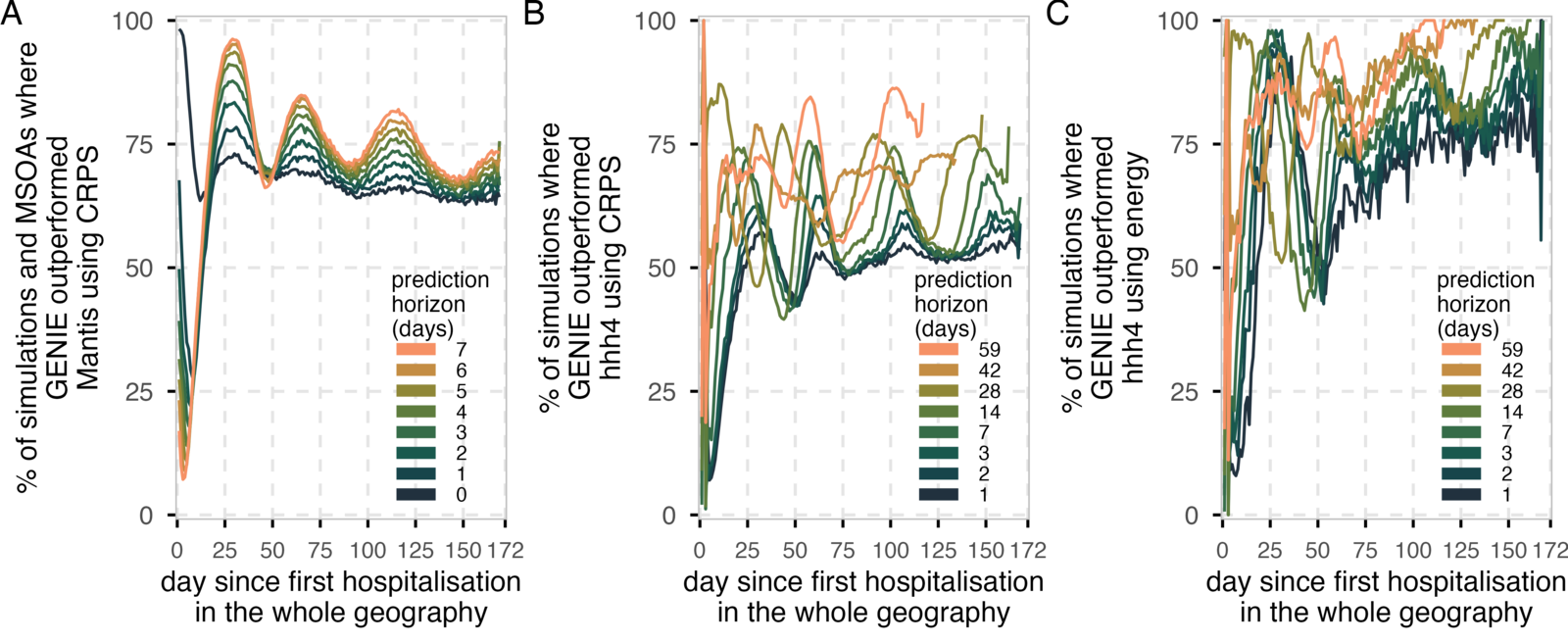}
    
    \caption{\textbf{Model comparison for daily hospitalisation prediction.} Curves showing the percentage of simulations where GENIE achieved the best (lowest) score across different forecasting parameters compared to Mantis (using CRPS, Panel A) and \texttt{hhh4} (using CRPS, Panel B, and energy score, Panel C). The horizontal axis indicates the forecast initiation day; each curve corresponds to different forecast window horizon. Higher values represent a higher percentage of simulations where the GENIE outperformed the others for that day/horizon combination.}
    \label{fig:comparison_models_hosp}
\end{figure*}

\begin{figure*}[h]
    \centering
    \includegraphics[width=0.95\textwidth]{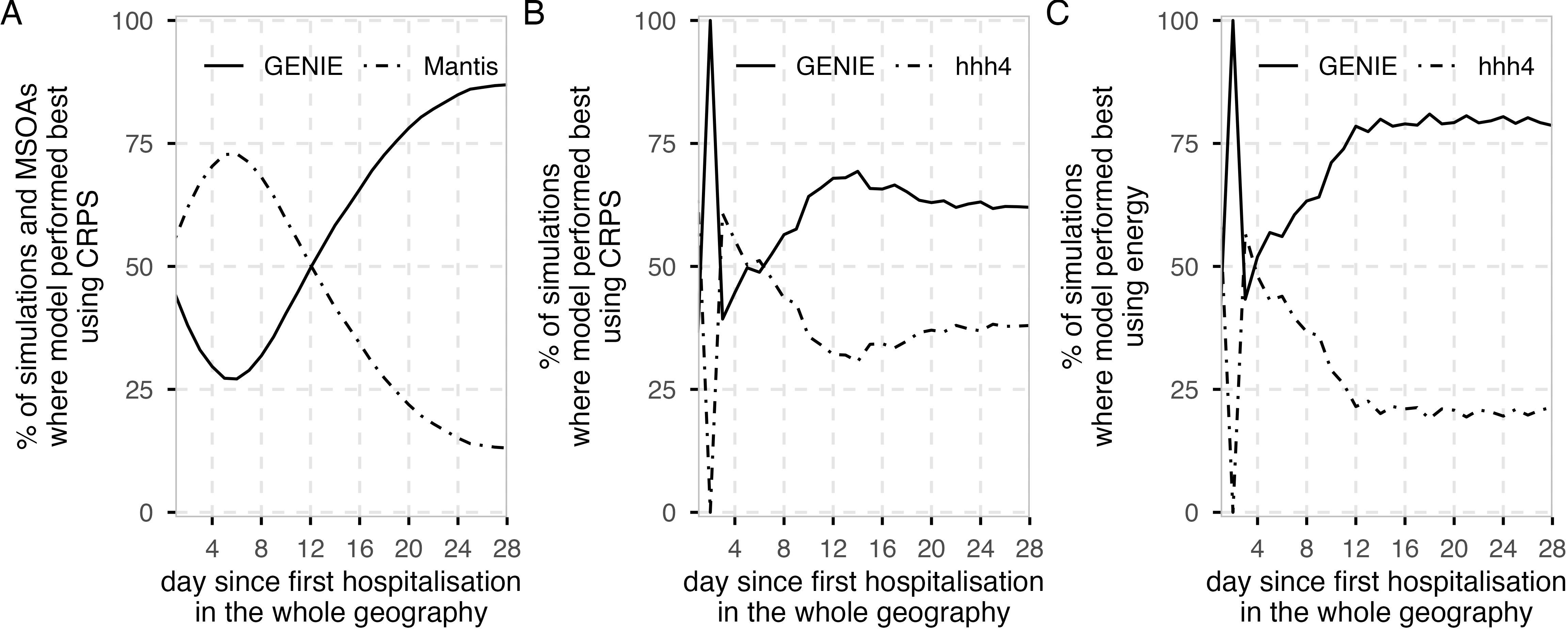}
    
    \caption{\textbf{Comparison of model performance during the emergence phase.} Percentage of simulations where GENIE (solid line) and the benchmark model (dotted line) achieved the best score in hospitalisation forecasting, for Mantis (CRPS, Panel A), \texttt{hhh4} (CRPS, Panel B) and \texttt{hhh4} (energy score, Panel C). The x-axis indicates the forecast initiation day; the y-axis represents the percentage of simulations where each model achieved the best score.}
    \label{fig:score_emergence_comparison}
\end{figure*}

\begin{figure*}[h]
    \centering
    \includegraphics[width=0.95\textwidth]{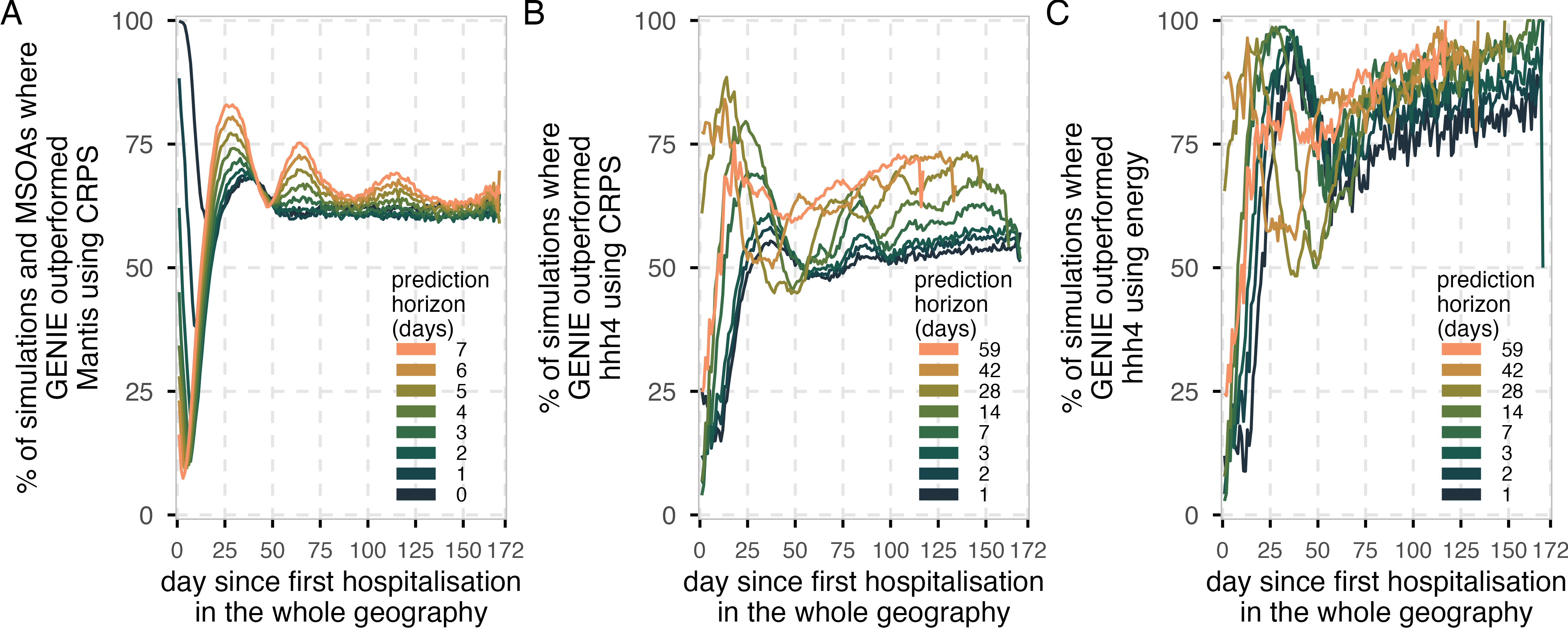}
    
    \caption{\textbf{Model comparison for deaths prediction.} Curves showing the percentage of simulations where GENIE achieved the best (lowest) score across different forecasting parameters compared to Mantis (using CRPS, Panel A) and \texttt{hhh4} (using CRPS, Panel B, and energy score, Panel C). The horizontal axis indicates the forecast initiation day; each curve corresponds to different forecast window horizon. Higher values represent a higher percentage of simulations where the GENIE outperformed the others for that day/horizon combination.}
    \label{fig:comparison_models_death}
\end{figure*}

\clearpage

\subsection{Ablation}\label{sec:appendix_ablation}

\begin{figure*}[h]
    \centering
    \includegraphics[width=0.95\textwidth]{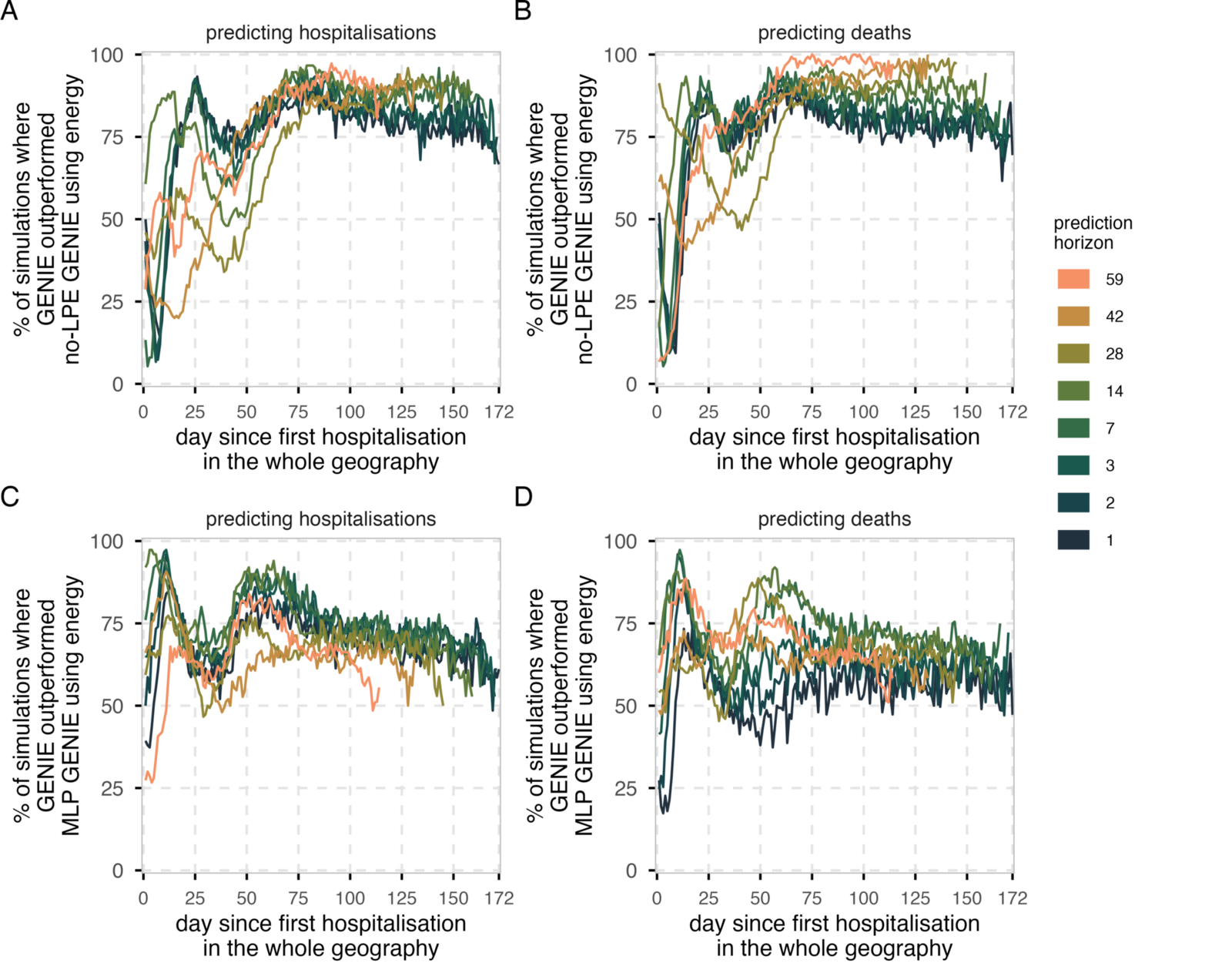}
    \caption{\textbf{Ablation analysis of GENIE.} Curves showing the percentage of simulations where GENIE achieved the best (lowest) score across different forecasting parameters using the energy score, compared to no-LPE GENIE (daily hospitalisations, Panel A; deaths, Panel B) and MLP GENIE (daily hospitalisations, Panel C; deaths, Panel D). The horizontal axis indicates the forecast initiation day; each curve corresponds to different forecast window horizon. Higher values represent a higher percentage of simulations where the GENIE outperformed the others for that day/horizon combination.}
    \label{fig:score_ablation_comparison}
\end{figure*}

Figure \ref{fig:score_ablation_comparison} shows the percentage of simulations in which GENIE outperformed its ablated versions.

\subsection{Application to Other Geographies}\label{sec:appendix_other_geos}

\begin{figure}[h]
    \centering
    \subfloat[Map showing the location of the new geographies relative to the area used for model training.]{\includegraphics[width=0.32\textwidth]{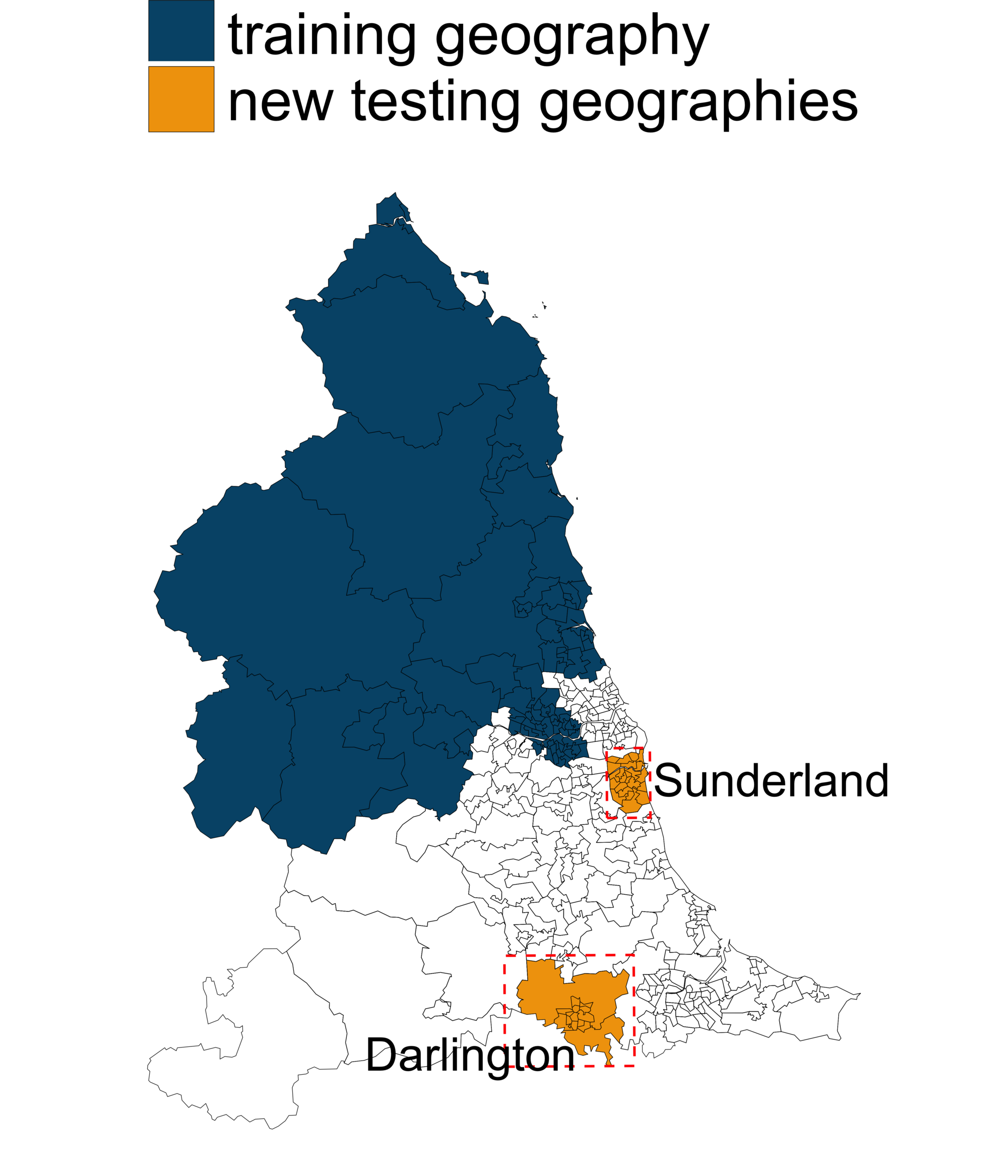}\label{fig:map_new_geos}}
    \hfill
    \subfloat[Map of Middle Layer Super Output Areas (MSOAs) in Sunderland.]{\includegraphics[width=0.32\textwidth]{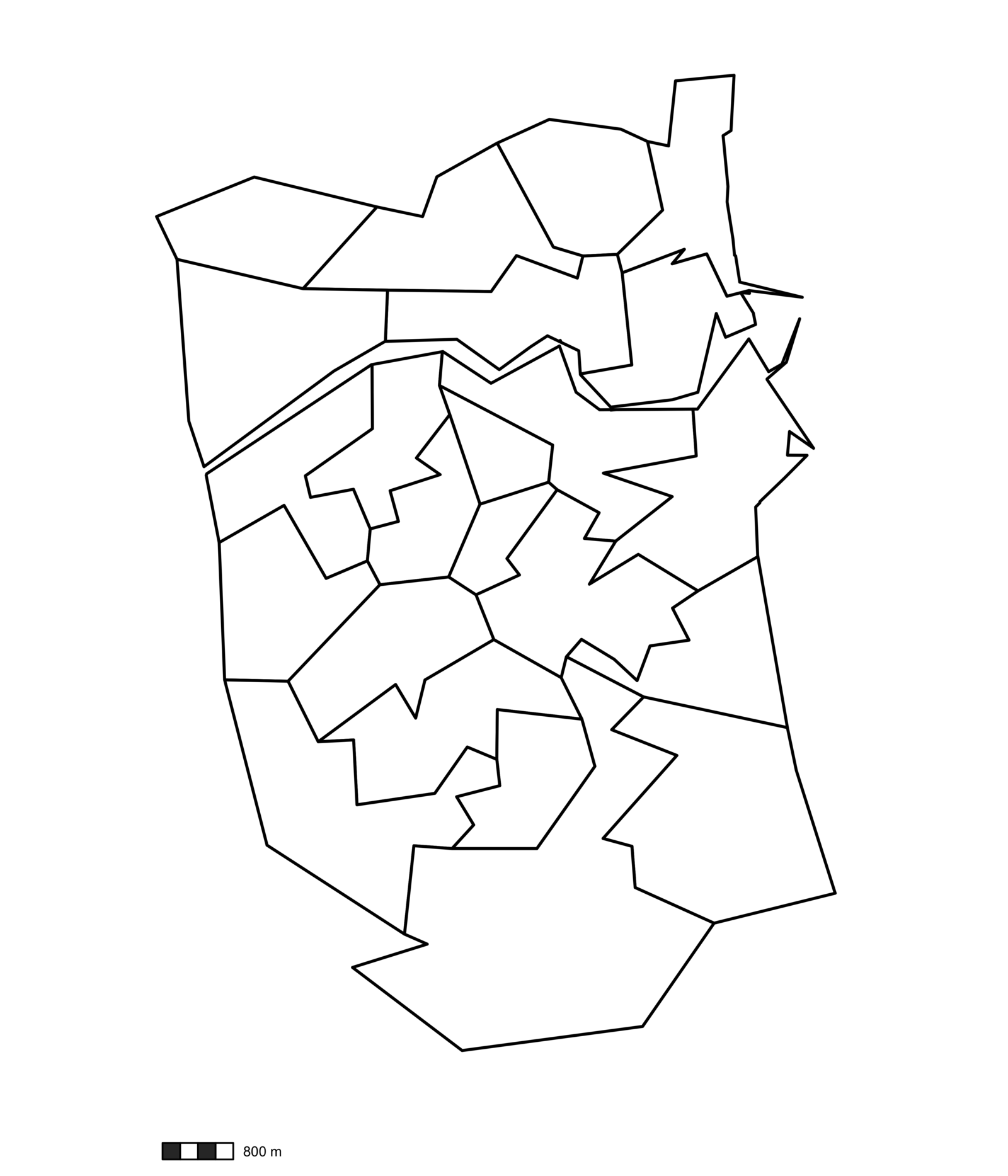}\label{fig:map_sunderland}}
    \hfill
    \subfloat[Map of Middle Layer Super Output Areas (MSOAs) in Darlington.]{\includegraphics[width=0.32\textwidth]{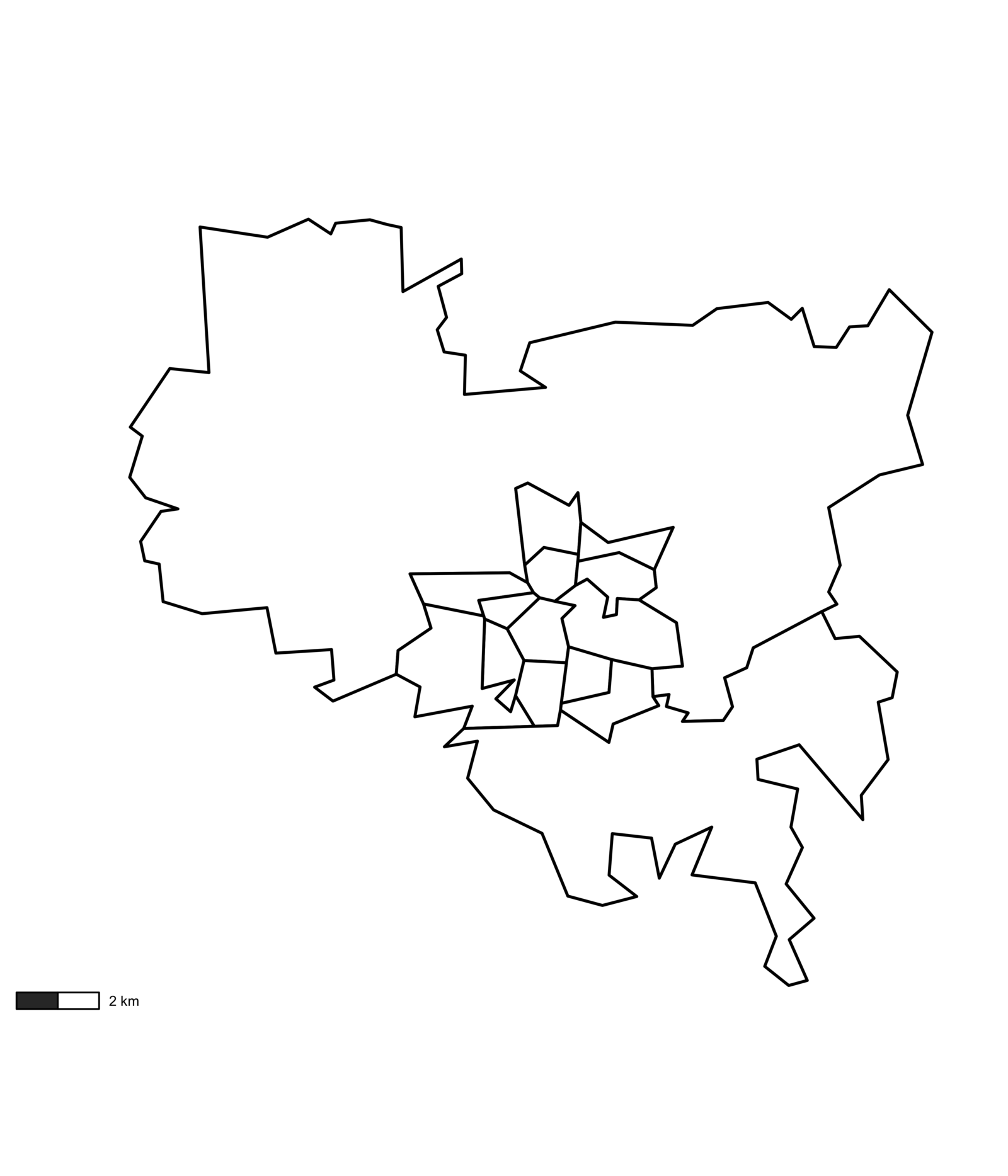}\label{fig:map_darlington}}
    
    \caption{\textbf{Geographic representation of the study areas.} Subfigure (a) illustrates the spatial relationship between the training geography (dark shade) and the two new testing geographies (Sunderland and Darlington, light shade), located in the North East of England. Subfigures (b) and (c) provide a detailed view of the Middle Layer Super Output Areas (MSOAs) within Sunderland and Darlington, respectively, which serve as the spatial units for analysis.}
    \label{fig:new_geographic_maps}
\end{figure}

We test whether the model supports \emph{geographic transfer} by generating forecasts on an unseen geography using our trained model. To evaluate out-of-distribution performance, we applied the trained model to 150 simulations on the cities of Sunderland and Darlington, areas entirely absent from the training data. For testing, the model requires only a reconstructed spatial graph and a consistent set of socio-demographic features to operate in unseen regions. The relative positioning of these new geographies is illustrated in Figure \ref{fig:new_geographic_maps}.

\begin{figure}[h!]
    \centering
    \includegraphics[width=0.95\textwidth]{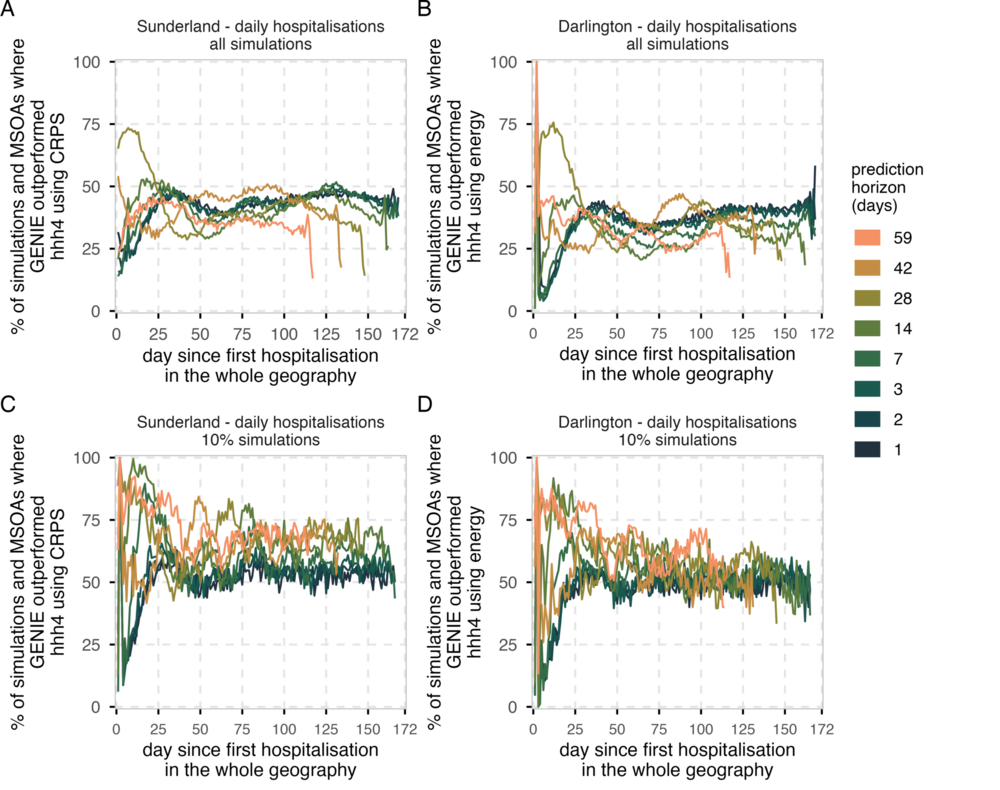}
    
    \caption{\textbf{Model comparison for additional geographies.} Curves showing the percentage of simulations where GENIE achieved the best (lowest) score across different forecasting parameters compared to \texttt{hhh4}, for Sunderland (Panel A) and Darlington (Panel B), based on all 150 simulations. A similar comparison is calculated for the 10\% simulations that cumulated the highest number of hospitalisations during the whole epidemic, for Sunderland (Panel C) and Darlington (Panel D).}
    \label{fig:appendix_newgeo_hospitalisations_scoring}
\end{figure}

We generated forecasts for every day following the first recorded hospitalisation, using  rolling 60-day forecast horizons across all simulations. Performance was then evaluated for each day, window, simulation, and MSOA against \texttt{hhh4} using CRPS. Figure \ref{fig:appendix_newgeo_hospitalisations_scoring} shows the percentage of simulations in which GENIE outperformed \texttt{hhh4} for Sunderland (Panel A) and Darlington (Panel B), where GENIE outperformed \texttt{hhh4} in 41.86\% and 35.26\% of simulations/MSOAs, respectively. Simulations in the new geographies had substantially fewer recorder hospitalisations compared to the primary geography presented in the main text. We repeated the same analysis on the top 10\% simulations, ordered by number of hospitalisations, for Sunderland (Panel C) and Darlington (Panel D), where GENIE displays a better performance, outperforming \texttt{hhh4} in 60.54\% of simulations/MSOAs for Sunderland and 54.29\% for Darlington.

\end{document}